\documentclass[fleqn,usenatbib]{mnras}

\usepackage{newtxtext,newtxmath}

\usepackage[T1]{fontenc}
\usepackage{ae,aecompl}

\usepackage{graphicx}	
\usepackage{amsmath}	
\usepackage[dvipsnames]{xcolor}
\usepackage{lscape}
\usepackage{stfloats}

\title[Scaling relations for GCS II]{Scaling relations for globular cluster systems in early-type galaxies. \\
II. Is there an environmental dependence?}
\author[De B\'ortoli et al.]{Bruno J. De B\'ortoli$^{~1,2}$, Juan P. Caso$^{~1,2}$, Ana I. Ennis$^{~1,2}$ and Lilia P. Bassino$^{~1,2}$\thanks{E-mail:
brudebo.444@gmail.com}\\
$^{1}$Facultad de Ciencias Astron\'omicas y Geof\'isicas de la Universidad Nacional de La Plata,    
and \\ Instituto de Astrof\'isica de La Plata (CCT La Plata -- CONICET, UNLP), Paseo del Bosque S/N,  
B1900FWA La Plata, Argentina\\   
$^{2}$Consejo Nacional de Investigaciones Cient\'ificas y T\'ecnicas, Godoy Cruz 2290, C1425FQB,  
Ciudad Aut\'onoma de Buenos Aires, Argentina}

\date{Released 2002 Xxxxx XX}

\pagerange{\pageref{firstpage}--\pageref{lastpage}} \pubyear{2021}

\begin{document}

\label{firstpage}

\maketitle

\begin{abstract}
The current properties of globular cluster
systems (GCSs) are the result of the evolution experienced by their host 
galaxies, which shape the richness of the GCS as well as its spatial
distribution, among other features. We carry out an analysis of the  projected radial distribution of globular clusters for a sample of almost 30 early-type galaxies (ETGs) of intermediate and low luminosity, located in cluster environments (Virgo, Fornax and Coma).
We also include in the study six ETGs, for which the  parameters of their GCS radial profiles are publicly  available. The final analysis is performed on an enlarged sample ($\sim 100$ GCSs), by adding the GCSs of ETGs from our previous paper (Paper\,I).
Scaling relations involving different parameters of the GCSs are obtained for the whole sample and complement those obtained in Paper\,I. Several of such relations point to a second-order dependence on the environmental density.
Finally, the results are analysed in the literature context.
\end{abstract}


\section{Introduction} 
\label{sec.intro}

Globular clusters (GCs) are compact stellar systems, typically considered to be among the oldest objects in the Universe \citep{han13,ush19,fah20}. The properties of globular cluster systems (GCSs) are highly influenced by the evolutionary processes experienced by their host galaxies. 
In the current paradigm, the GCSs in bright early-type galaxies (ETGs) are built through a two-phases process \citep{for11,boy18,cho19,elb19,rei20}. First, {\it in situ} formation occurs at high redshift, with merger episodes playing a main role in the formation and early-survival of GCs \citep{li14,kru15}. Then, the accretion of GCs from satellite galaxies largely contributes to the growth of the GCS, and particularly to its outer regions. This is supported by numerical studies \citep[e.g.][]{ton13,ram18} as well as observational evidence from bright ETGs \citep[e.g.][]{par13,coc13,cas17,bea18}, and from satellite galaxies in dense environments \citep[e.g.][]{pen08,liu19}. In this latter case, the population of the GCSs depends on the distance to the central galaxy.

On past decades, GCs studies have been biased to massive ETGs, usually characterised by very populated and spatially extended GCSs \citep[e.g.][]{bro00,har00,ric04,for06,har06,for07,har16b,har17a}. In recent years, a few surveys based on ACS/HST observations on both low \citep{geo10} and high-density \citep{jor04,jor07} environments have enlarged the sample towards lower galaxy masses. Studies of GCSs focused on moderately bright ETGs in low-density environments \citep[e.g.][]{spi08,cho12,sal15} have complemented our understanding of GCSs, although a more complete and homogeneous sample is desirable. These systems present rather less populated GCSs, although some exceptions exist \citep[e.g.][]{enn20}. Several isolated bright ellipticals also present less populated GCSs than their counterparts in clusters \citep[e.g.][]{cas13b,lan13,ric15,bas17}, pointing to the relevance of the environment in the build-up of the GCS. It is also interesting the case of relic galaxies, where the lack of accretion processes leads to poor GCSs \citep{ala21}.

Regarding the radial profiles of GCSs in dense environments, \citet{bas06b} analyse three satellite galaxies in the vicinity of NGC\,1399, the dominant galaxy in the Fornax cluster, which present poor and compact GCSs. A similar scenario seems to occur with NGC\,3311 and NGC\,3309 \citep{weh08} in the Hydra cluster. \citet{coe09} fit the projected radial profile of the GCSs for a small sample of galaxies from the Virgo cluster. They note the lack of correlation between the slope of the radial profiles and the distance to M\,87, but the sample contains galaxies with a wide range of luminosities and the analysis does not take this into consideration. On the contrary, the existence of an environmental dependence on parameters of the radial profile has already been suggested by \citet{hud18} on the basis of comparing the halo mass and the effective radius of the GCS for a sample of galaxies. They consider an underlying effect which is pulling massive (mainly central) galaxies in a different direction than those presumably satellites.

From the assumption of coeval evolution of the GCSs and their host galaxies, it becomes a natural step to look for scaling relations that provide evidence about the physical processes ruling the current properties of GCSs. The first efforts have been made by \citet{spi09} from a mixed sample of early and late-type galaxies. From a larger sample that spans a wide range in stellar masses, \citet{har13} explore the relation between the population of GCSs and several parameters of the host galaxies. The relative richness of the GCSs, represented by the specific frequency \citep[$S_N$,][]{har81}, versus the absolute magnitude of the host galaxy produces an U-shape relation, with intermediate mass galaxies at the bottom. 
On the basis of the same sample and relations derived from weak lensing techniques, \citet{hud14} expand the suggestion of \citet{bla97} of a  uniform GC production rate per unit available mass and find that the mass enclosed in GCs correlates with halo mass with a large scatter. \citet{har15} revisit the previous scaling relations, including the fraction of red (metal-rich) GCs, to look for differences in the typical scenario of two subpopulations of GCs. \citet{for18} include halo mass for a sample of nearby dwarf galaxies, extending the correlation with the mass enclosed in GCs down to $M_{\rm vir}\approx 10^9\,M_{\odot}$. Regarding the parameters of the radial distribution, \citet{kar14,kar16} present scaling relations as a function of the stellar mass and effective radius of the host galaxy. Recently, \citet{for17} and \citet{hud18} provide improved scaling relations from larger samples, although some of their results disagree.

Stripping and accretion processes rule the late evolution of galaxy haloes in the last Gyrs, also affecting the halo populations, like the GCs. This motivates our focus on scaling relations derived from parameters of the GCSs radial profiles. In \citet[][hereafter Paper\,I]{cas19}, we fit the GCSs radial profiles for 24 ETGs with intermediate luminosities, residing in both low-density environments, and in the Fornax and Virgo galaxy clusters.  Besides, three ``stacked GCSs'' are built from Virgo dwarfs. The sample is supplemented with properties of GCSs from the literature. Several scaling relations from previous works are extended to lower stellar masses, and for some of them we find a change in the behaviour at a pivot mass of $\approx 5\times\,10^{10}\,M_{\odot}$. In the present paper, we add 23 intermediate luminosity ETGs from Virgo, Fornax and Coma clusters, plus four ``stacked GCSs'' built from Virgo dwarfs, with similar stellar masses each. Including GCSs from the literature, the updated sample achieves 100 GCSs, and constitutes the larger sample collected for this purpose. The objective in this paper is to explore the dependence on the environment of the scaling relations already presented in Paper\,I,  to provide clues about the processes experienced by satellite and central galaxies, and their haloes. In particular, we choose the richness of the GCSs and several parameters related with the radial profile, which should be affected by the environmental processes experienced by the galaxies in dense environments.

\smallskip
This paper is organised as follows. The observations and data reduction are described in Section\,\ref{sec.obs}, and observational 
catalogues used along the work are indicated in Section\,\ref{sec.lit}, where the set of environmental density parameters is also explained. In Sections\,\ref{sec.res} and \ref{sec.dis} we present the results and discussion of the GCSs scaling relations, respectively. Finally, in Section\,\ref{sec.sum} our conclusions are summarised.

\section{Observational data and reduction}
\label{sec.obs}

The sample of GCSs with radial profiles fitted in this paper consists of ETGs from the nearby clusters of Virgo \citep[$D\approx 17$\,Mpc,][]{mei07}, Fornax \citep[$D\approx 20$\,Mpc,][]{bla09} and Coma \citep[$D\approx$  100\,Mpc,][]{car08}. The data set is based on observations carried out with the HST/ACS (Hubble Space Telescope / Advanced Camera for Surveys) and available at the Mikulski Archive for Space Telescopes (MAST). The observed filters are $F475$ and $F814$ for Coma galaxies \citep[programme 10861][]{car08}, and $F475$ and $F850$ for Virgo \citep[programme 9401][]{cot04} and Fornax \citep[programme 10217,][]{jor07} ones. These have been widely used to select and analyse GC candidates. The fields are typically centred on the galaxies, presenting a field-of-view (FOV) of $202\times202\,{\rm arcsec^2}$ and a pixel scale of 0.05\,arcsec. 

New photometry of GC candidates is obtained only for galaxies in the Coma cluster. For GCSs in Virgo and Fornax, we use available GC photometry from \citet{jor09,jor15}, but selecting GC candidates on the basis of colour and brightness ranges, as explained in Section\,\ref{datlit}.

Regarding the surface brightness profiles of the ETGs, the parameters for single S\'ersic profiles are already available in the literature for  Coma \citep{hoy11} and most of the Virgo galaxies \citep{fer06b}. However, some ETGs from Virgo as well as those from Fornax lack fitted parameters for single S\'ersic profiles, and they are fitted in the present paper. Photometric procedures are described in the next subsections.

\subsection{Surface photometry for Virgo and Fornax galaxies} 
\label{sec.gals}
In order to analyse the surface brightness profiles of those galaxies in the sample that have no published S\'ersic fits, we use the task ELLIPSE within {\sc iraf}. Ellipticity and position angle are only calculated for the inner regions of the galaxies, typically up to $\sim 30-45\,\rm{arcsec}$. They are fixed for larger galactocentric distances to avoid fluctuations related with the low surface brightness level and the edges of the FOV. We obtain the profiles in filters $g$ and $z$ from the AB system applying the zero points calculated by \citet{sir05}, $ZP_{F475}= 26.068$ and $ZP_{F850}=24.862$.  Then, we apply corrections for Galactic extinction from NED, calculated through the \citet{sch11} calibration. The brightness profiles as well as the fitted parameters from the S\'ersic profile are presented in Section\,\ref{sec.briprof} and Table\,\ref{sersic}.

\subsection{GC candidates in Coma galaxies} 

\subsubsection{Photometry and selection of point sources}
\label{sec.fotcoma}
For the five galaxies in the Coma cluster PSF photometry of the GC candidates is performed in both filters. First, we use the tasks ELLIPSE and BMODEL within {\sc iraf} to model the diffuse brightness profile. Then,  we subtract it from the galaxy, favouring the detection of GC candidates in the inner regions. A preliminary catalogue of sources is built with SE{\sc xtractor} \citep{ber96}, considering as a positive identification of a source every detection of at least three connected pixels above a threshold of $3\sigma$ from the sky level. Following the procedure applied in Paper\,I and references therein, the catalogue is restricted to objects with elongation smaller than 2:1 and full width at half-maximum (FWHM) smaller than 3\,px to avoid extended sources.

The photometry for the GC candidates is performed by means of DAOPHOT \citep{ste87} within {\sc iraf}. At the Coma cluster distance, the mean effective radius of GCs of $\approx 3$\,pc   \citep[e.g.][]{pen08,cas14} corresponds to $\approx 10$\,\,per cent of the typical FWHM for the PSF of these images, FWHM\,$\approx 0.08-0.10$\,arcsec. Then, for our purposes GCs can be treated as point sources, in agreement with the analysis by \citet{pen11} for these same images. Although some objects in the range of extended clusters \citep[e.g.][]{bro11} might be marginally resolved, their number should be negligible in comparison with the general population of GCs, and this simplification does not affect our results. A spatially variable PSF is built for each filter from the selection of $40-50$ bright point sources, looking for an homogeneous spatial distribution, to account for PSF variations. The fitting radius is chosen at 3\,px. The PSF photometry is run with the task ALLSTAR, the parameters sharpness and $\chi^2$ are used to separate point-like from extended sources. For each filter, limits are chosen as the 95-percentile of the measurements of these parameters, for the artificial stars added for completeness analysis (see Section\,\ref{compl}). Aperture corrections are calculated, for each filter, from the same objects used to model the PSF.

\subsubsection{Calibration, extinction corrections and GC candidates selection}
\label{sec.calib}
For the point sources from the Coma cluster the instrumental magnitudes are calibrated based on the zero-point magnitudes from \citet{sir05}, already indicated in Section\,\ref{sec.gals}. The resulting magnitudes correspond to $g$ and $I$ bands in the AB system, respectively. Then, we apply corrections by foreground extinction from NED, based on the calibration by \citet{sch11}.

Finally, GC candidates are selected according to their colours and luminosity, choosing those objects that fulfill $0.5 < (g-I)_0 < 1.5$\,mag, in agreement with previous studies with these photometric data \citep[e.g.][]{pen11}, and $23 < I_0 < 26.5$\,mag. The fainter limit for our GC candidates is defined by the photometric completeness, set at  $I_0=26.5$\,mag (see Section\,\ref{compl}). The brighter limit restricts the inclusion of bright foreground stars and transitional objects, like UCDs \citep[e.g.][]{bru11}. It results from typical turn-over-magnitude for GCs, $M_{V,{\rm TOM}}\approx -7.4$\,mag \citep{ric03,jor07}, and the expected dispersion for the GC luminosity function (GCLF), usually lower than $1.3$\,mag in intermediate-mass galaxies \citep{har14}. Assuming a Gaussian GCLF and the $3\sigma$ criterion, it is reasonable to restrict the brightness of GCs to $M_V=-11.3$\,mag, i.e. $M_I\approx -12.3$\,mag. This latter value corresponds to the Vega system and, from the zero point difference \citep{sir05}, it results $M_I\approx-12$\,mag in the AB system.

\subsubsection{Completeness analysis}
\label{compl}   
\begin{figure}    
\includegraphics[width=85mm]{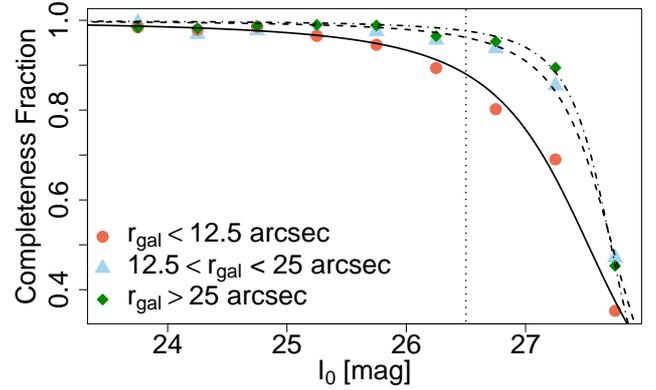}    
\caption{Completeness as a function of $I_0$ magnitude for NGC\,4906, in the Coma cluster. The solid, dashed and dash-dotted curves represent the completeness for three
increasing ranges of galactocentric distance (${\rm r_{gal} [arcsec]}$), according to Equation\,\ref{eq.compl}. The dotted vertical line at $I_0=26.5$\,mag 
indicates the adopted magnitude limit, at which the completeness is $\approx 90$\,\,per cent for an average ${\rm r_{gal}}$. Analogue analysis was performed for the other Coma galaxies.}
\label{compl1}
\end{figure}

The photometric completeness for each galaxy in the Coma cluster is obtained by the addition of artificial stars to the images in both bands, spanning the typical colour range adopted for GCs (see Section\,\ref{sec.calib}) and magnitudes $23 < I_0 < 28$. In order to avoid issues related with crowded regions, we add only 50 artificial stars per iteration, repeating the process $1\,200$ times to achieve a final sample of $60\,000$ objects. The PSF photometry is run in the same manner as for the science fields. The procedure is repeated in both filters, and then a unified catalogue is built with the artificial stars detected and measured in both filters. The completeness curves for different galactocentric ranges are represented in Fig.\,\ref{compl1} with different symbols. We select as limiting magnitude $I_0= 26.5$\,mag, which corresponds to approximately a 90\,\,per cent completeness, with full completeness for GCs brighter than $I_0=25$\,mag. These results do not differ from those by \citet{pen11} for the same data set. In order to apply differential completeness corrections to the GCS radial distributions, we fit the following analytic function:

\begin{equation}
\label{eq.compl}
f(m) = \frac{1}{2}\left[ 1 - \frac{\alpha(m-m_0)}{\sqrt{1+\alpha^2(m-m_0)^2}}\right] 
\end{equation}

\noindent with $\alpha$, and $m_0$ as free parameters. This has already been used in GC studies based on HST/ACS observations by \citet{har09c}. A generalisation of this expression, including a multiplicative factor that accounts for maximum completeness values below unity, is applied in Paper\,I. The resulting completeness corrections are
represented in Fig.\,\ref{compl1} by solid, dashed and dash-dotted 
curves, corresponding to intervals of increasing galactocentric
distance.

\section{Data from the literature and environment}
\label{sec.lit}

\subsection{Catalogues of GC candidates from Virgo and Fornax clusters}
\label{datlit}

We also take advantage of the publicly available photometry of GCs for a sample of ETGs with intermediate luminosity, namely 12 galaxies in Virgo ($-20.5<M_B<-17.9$) and 6 in Fornax ($-20.7<M_B<-17.6$), from  \citet{jor09} and \citet{jor15}, respectively. These ETGs are selected because they present GCSs populated enough to fit their radial profiles. The GC candidates are selected according to their colours and luminosity. We picked out those sources in the range $0.6 < (g-z)_0 < 1.7$\,mag for Fornax and Virgo galaxies, typical range that includes old GCs for early-type galaxies \citep[e.g.][]{pen06}. The choice of the brightness ranges is analogous to that described in Section\,\ref{sec.calib} for the Coma cluster. In Virgo and Fornax, the faint limit is set at $z_0=24$\,mag, to avoid the drop in photometric completeness (calculated in Paper\,I for the same dataset). The bright limit at $M_V=-11$\,mag, corresponds to $M_z\approx -12.3$\,mag, and the brightness ranges are $18.6 < z_0 < 24$ and $19 < z_0 < 24$ for GCs in Virgo and Fornax, respectively.

We apply completeness corrections to this Virgo/Fornax sample, following the procedure described in Section\,2.3 from Paper\,I. That is, we use NGC\,4621 from Virgo and NGC\,1340 from Fornax as model cases to obtain a detailed analysis of the photometric completeness. These galaxies are among the brightest in each cluster in our sample, and do not present prominent  underlying substructure. Their surface brightness profiles extend over a large radial range (i.e. corresponding to a large range in surface brightness), allowing for a direct comparison with the rest of the galaxies in each cluster. For both model galaxies, the completeness curves at different galactocentric annuli are obtained by adding 250\,000 artificial stars to the images in both filters. Then, the mean surface brightness of the galaxy in the $z$ band ($\mu_{mean,z}$) is calculated for each annulus (i.e. at different radial radius) and associated to its corresponding completeness curve. By means of those model cases, the completeness corrections for these GCSs were calculated in different radial regimes (typically four) from the numerical integration of its surface brightness profiles, fitted in this paper (see Section\,\ref{sec.briprof} and Table\,\ref{sersic}) or published in the literature \citep{fer06b}, to obtain the corresponding $\mu_{mean,z}$. Then, these latter values of $\mu_{mean,z}$ are used to select the appropriate completeness curves from those derived for the model galaxies. These completeness corrections are applied to the projected density distribution of the GCs in Section\,\ref{sec.radprof}.

Additionally, in order to extend our GC analysis to low luminosity host galaxies, we select 20 dwarfs from the Virgo cluster. As they have 
poorly populated GCSs, we split them into four different groups and stacked their projected GC spatial distributions. Details of the stacking and results of the analysis of their GC radial profiles will be presented in Section\,\ref{sec.radprof}.

\begin{figure} 
\centering
\includegraphics[width=70mm]{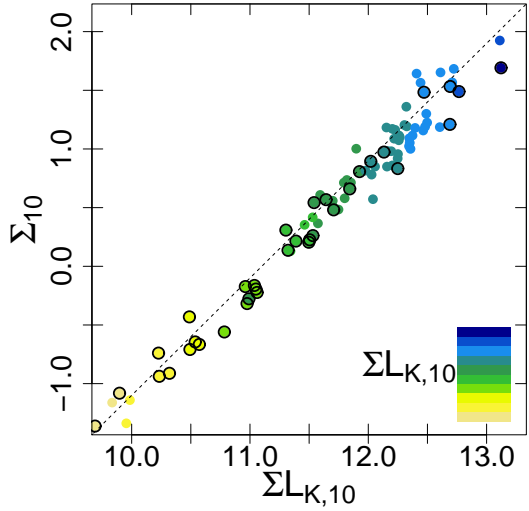} 
\caption{Comparison between the environmental density estimators applied in this paper. The environmental density up to the 10th nearest neighbour, $\Sigma_{10}$, as a function of the environmental density weighed by the luminosity in the $K$ filter of the 10th nearest neighbours, $\Sigma L_{K,10}$. The colour gradient represents this latter parameter, and framed symbols highlight central galaxies. The dashed line has slope 1, and its is arbitrary scaled for comparison purposes.}
\label{compdens}    
\end{figure} 

\begin{figure*}
\raggedright
\includegraphics[width=58mm]{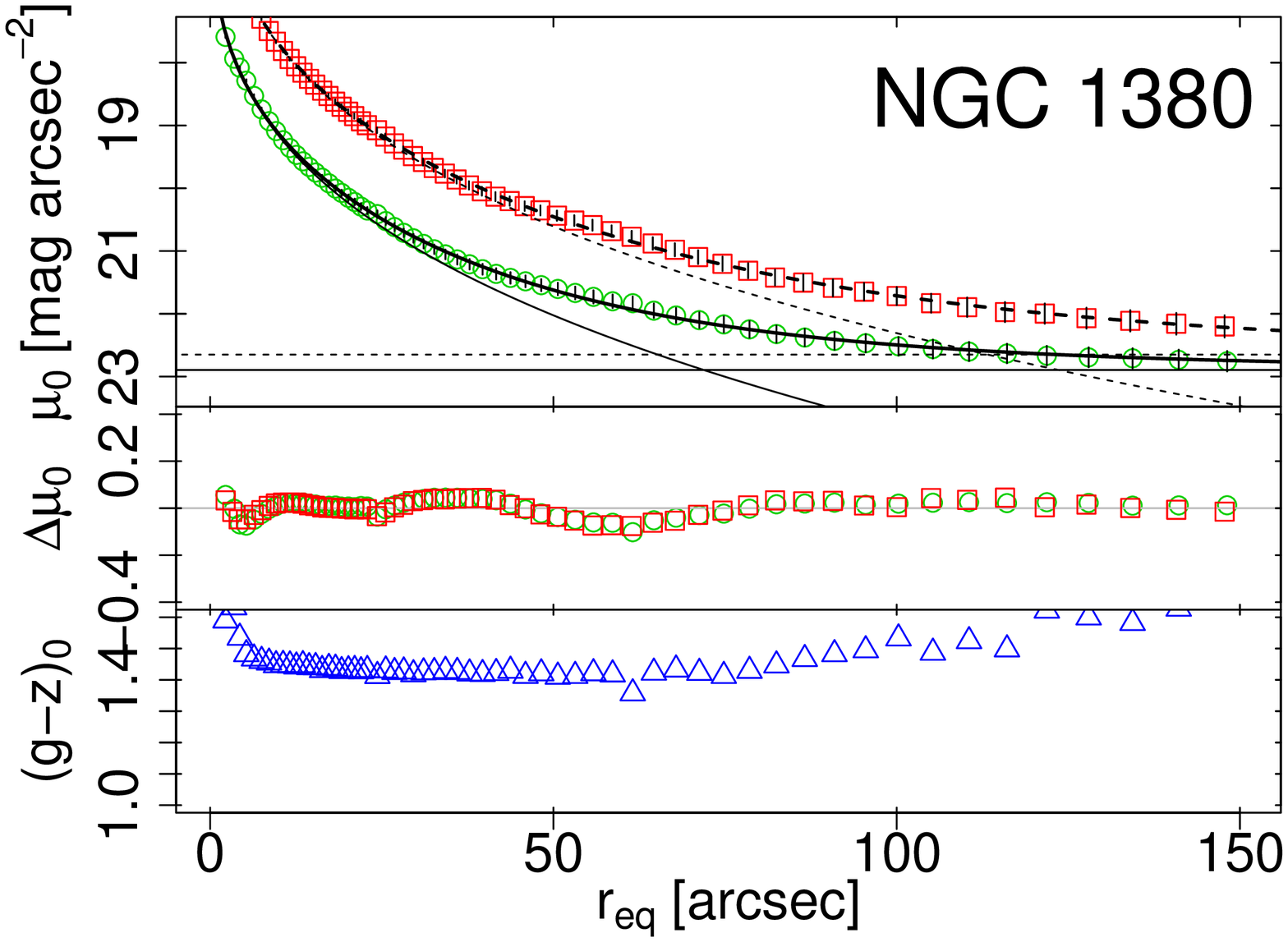}    
\includegraphics[width=58mm]{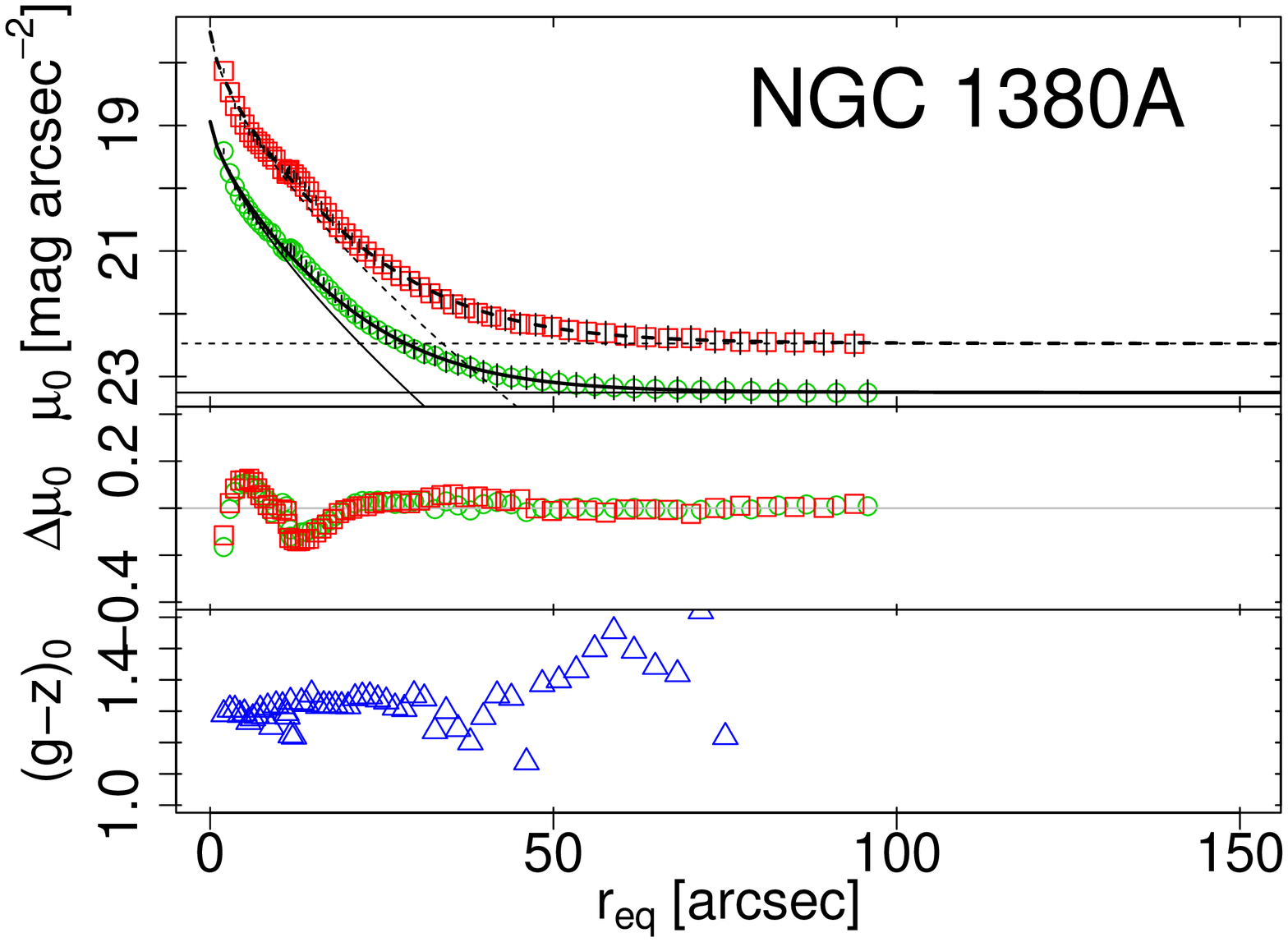}   
\includegraphics[width=58mm]{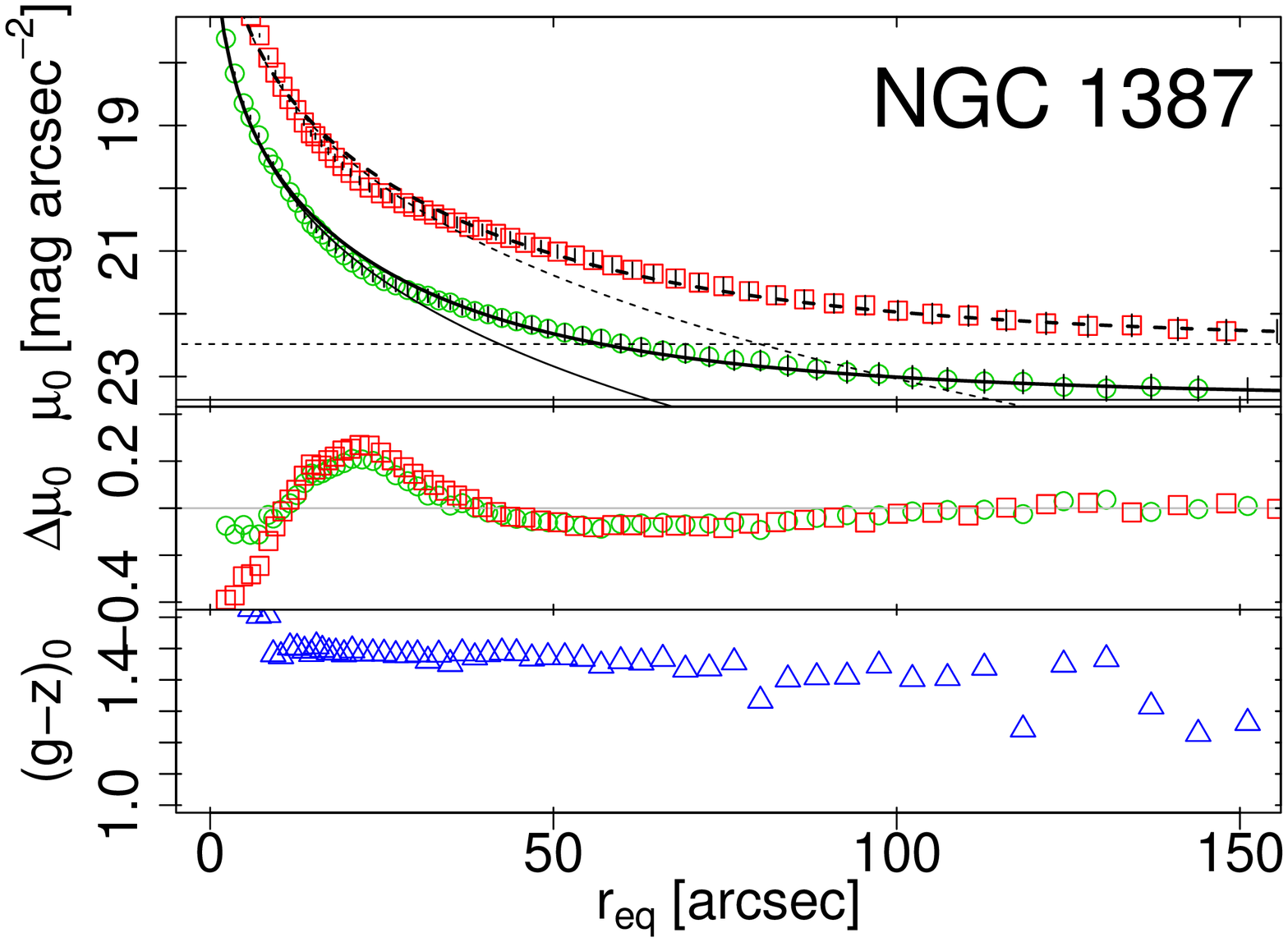}    
\includegraphics[width=58mm]{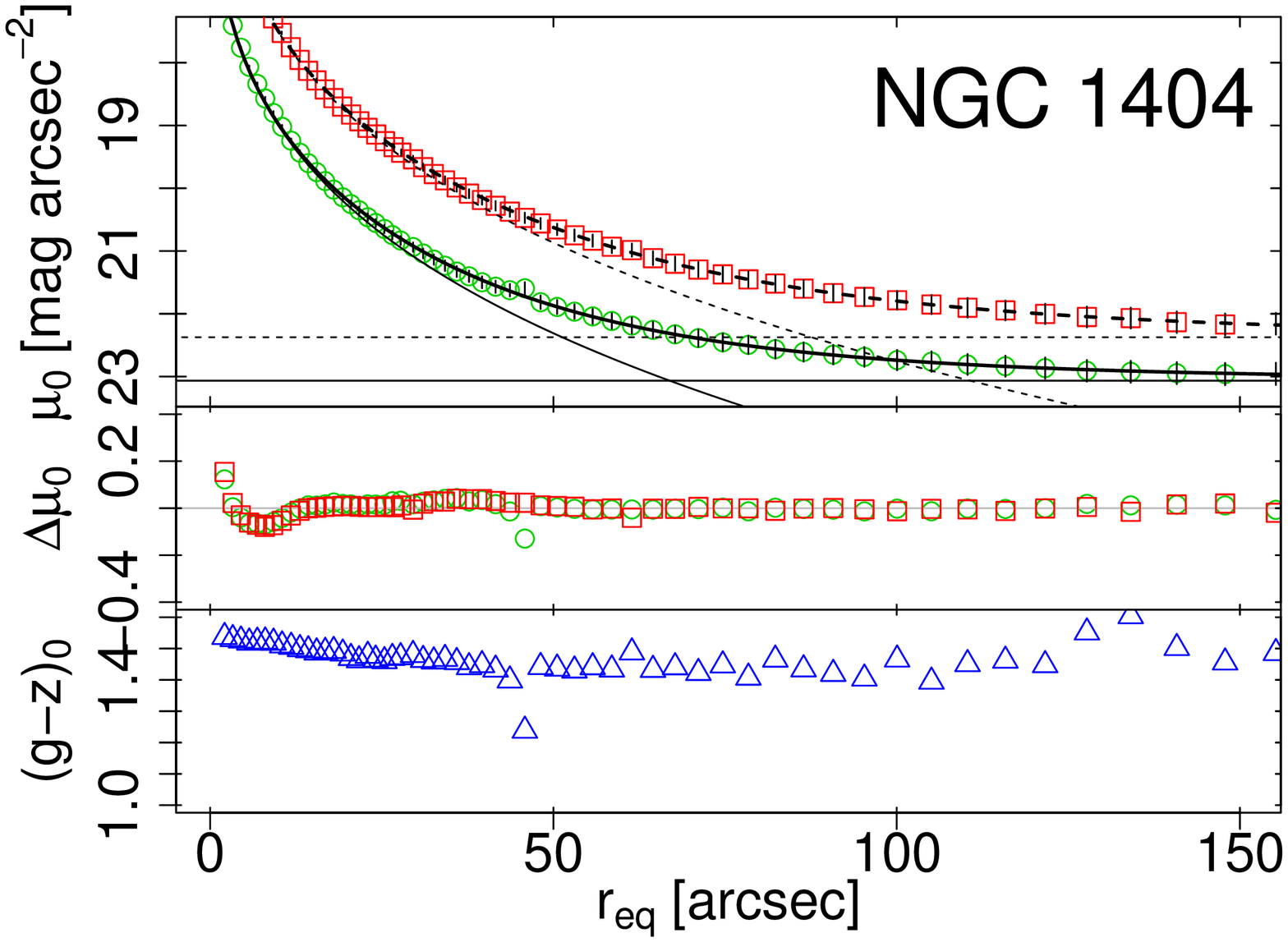}    
\includegraphics[width=58mm]{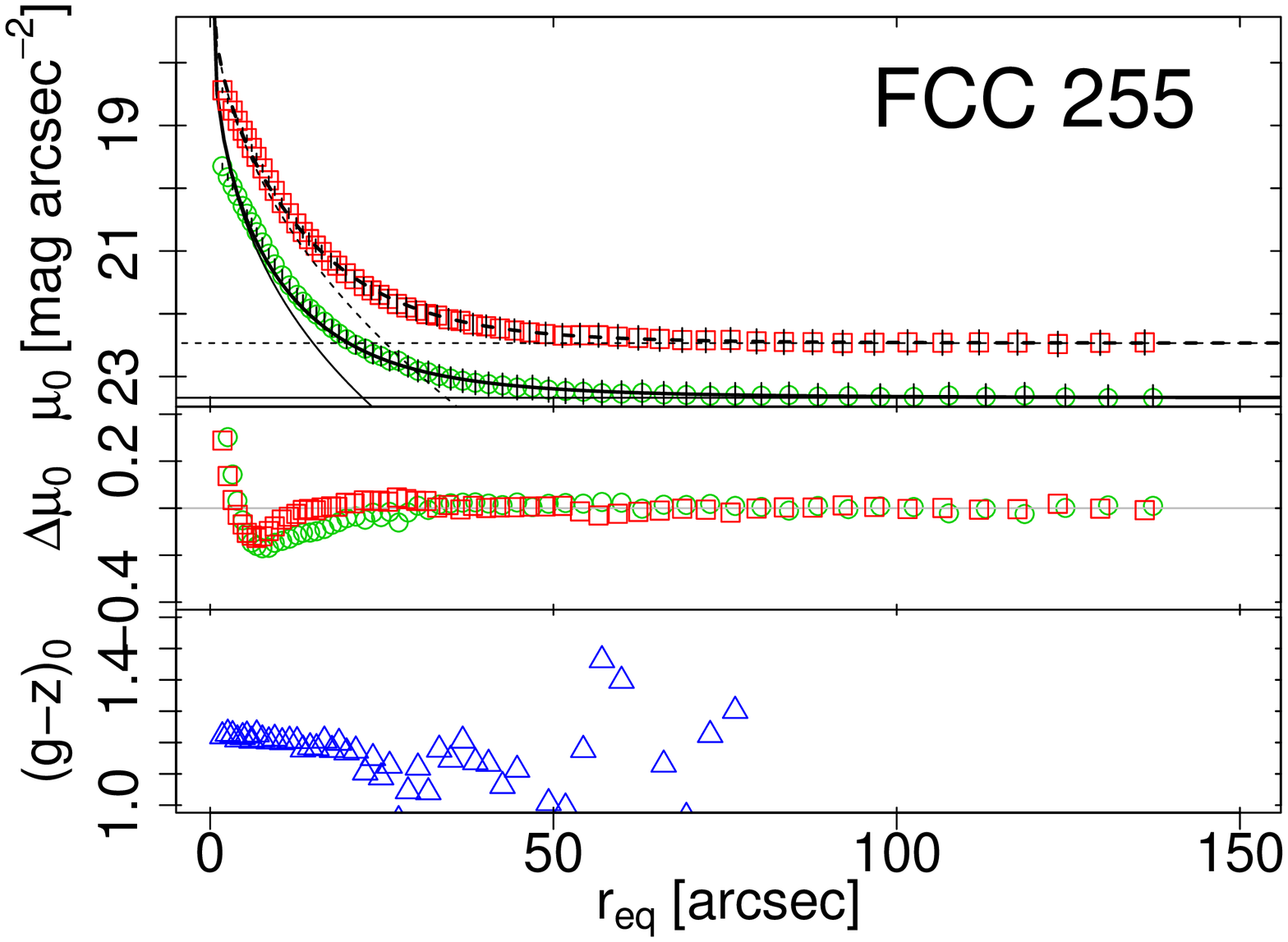}    
\includegraphics[width=58mm]{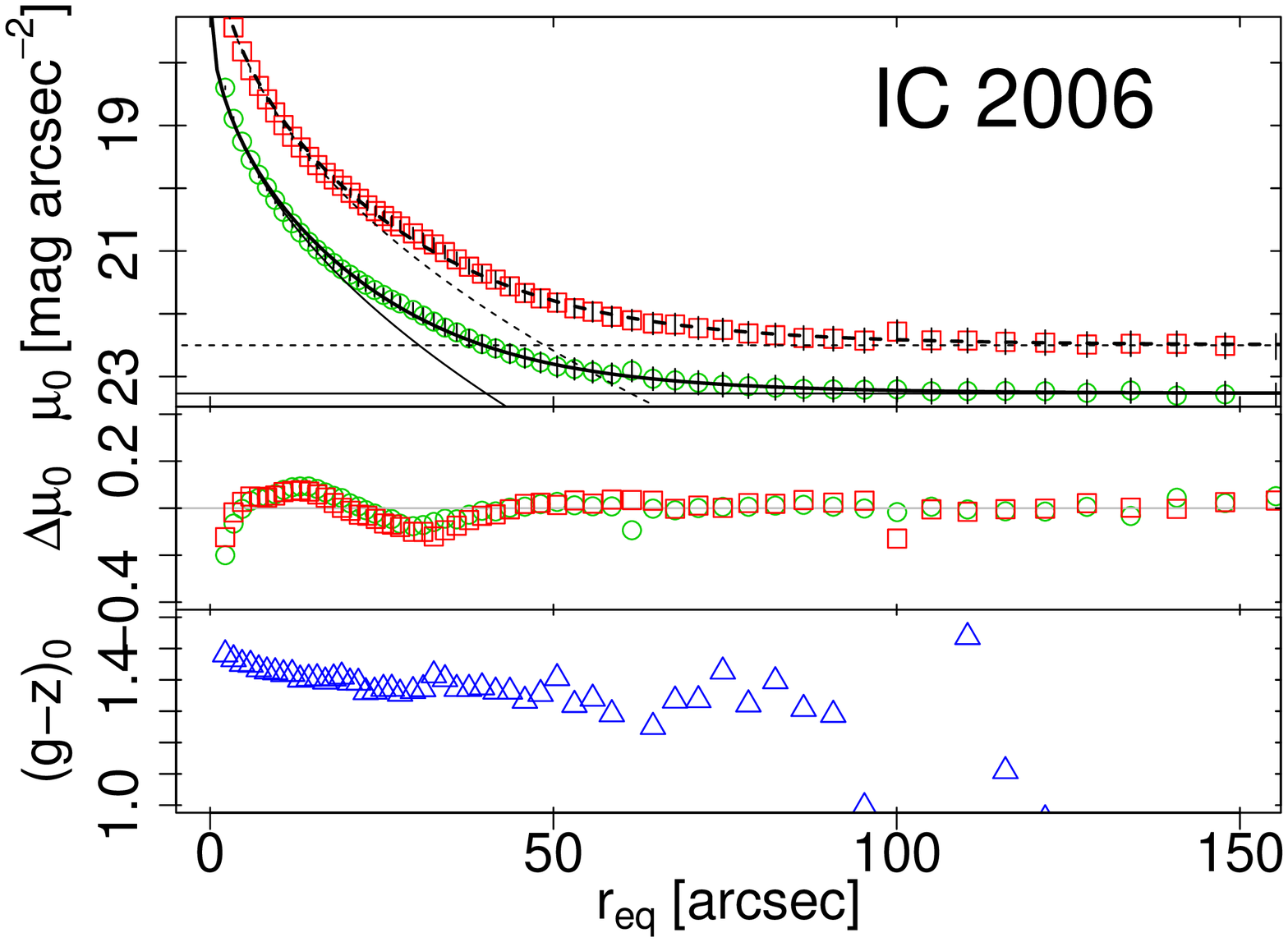}    
\includegraphics[width=58mm]{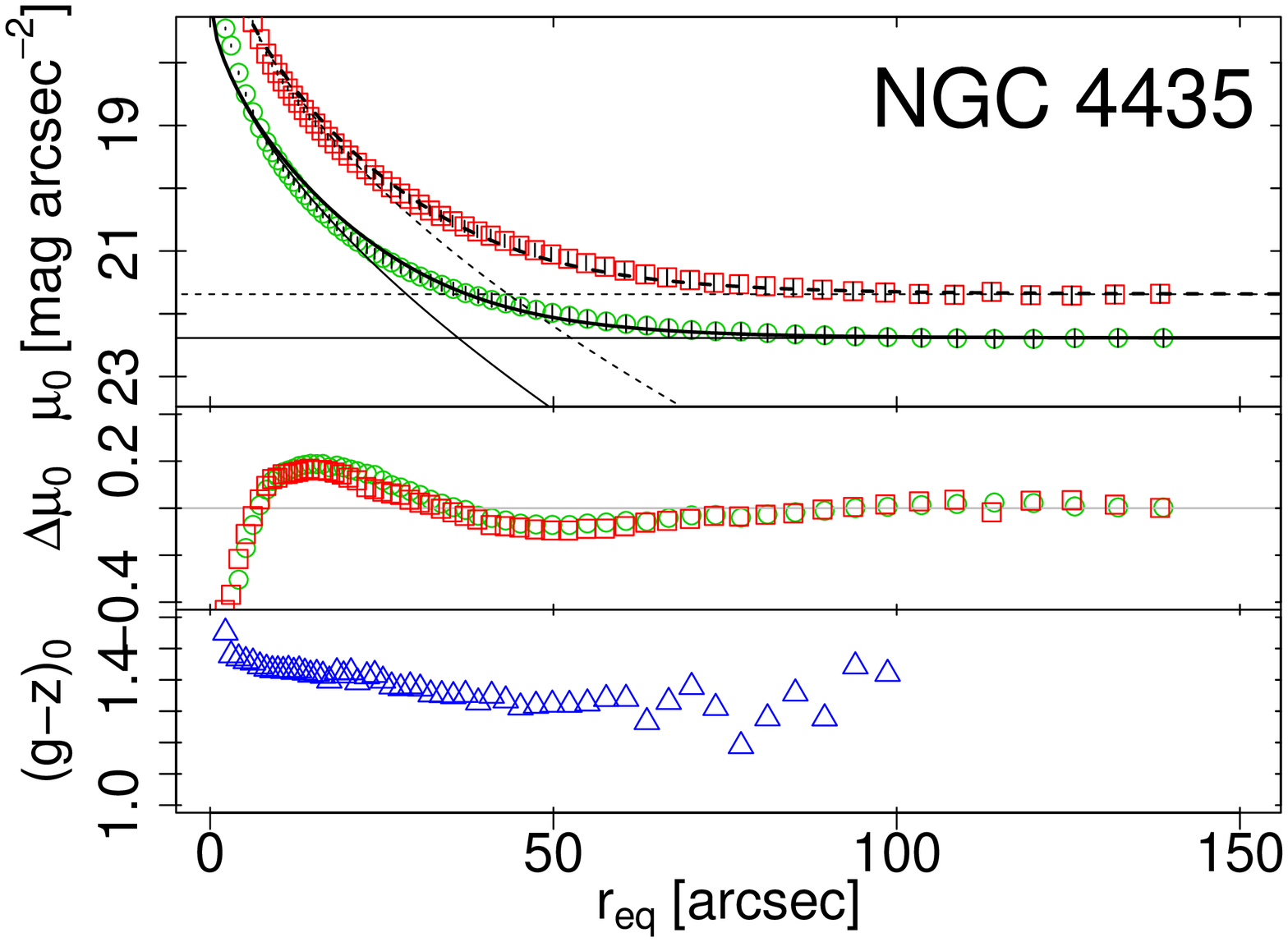}
\includegraphics[width=58mm]{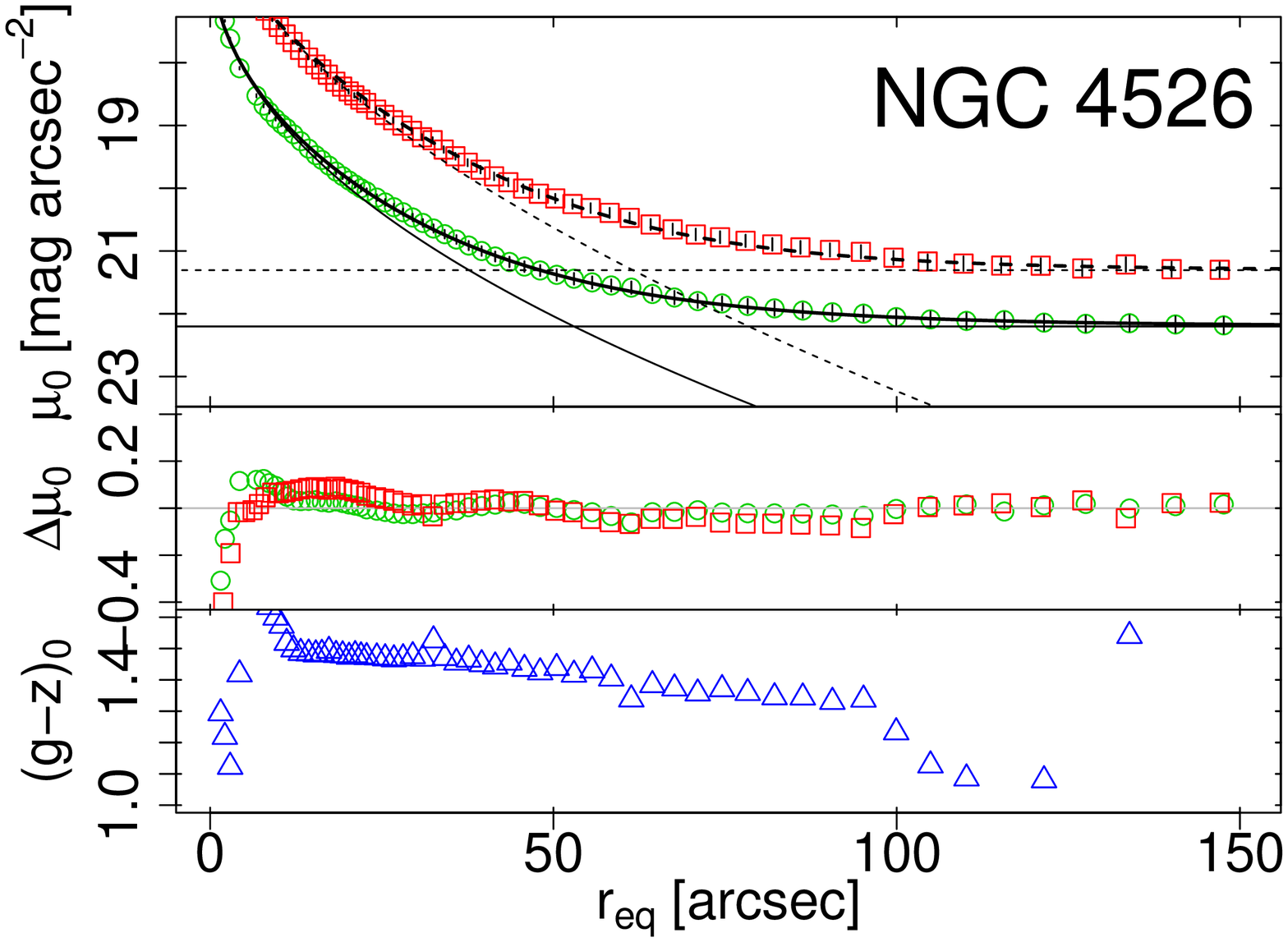}

\caption{Upper panels show the $(g,z)$  surface brightness profiles versus equivalent radius (${\rm r_{eq}}$), for a subsample of 2 Virgo and 6 Fornax intermediate luminosity ETGs that lack of published parameters from single S\'ersic fits. Green circles correspond to the $g$ filter and red squares to the $z$ one. Solid and dashed horizontal lines show the respective background levels. The fitted S\'ersic profiles are shown as thin curves, while thick curves represent the contribution of the galaxy plus the background. Fit residuals are shown in the middle panels, with similar symbols as above for each band. Lower panels present the respective colour profiles in $(g-z)_0$. Resulting parameters are listed in Table\,\ref{sersic}.} 
\label{perfilgal}    
\end{figure*}

\subsection{Estimating the local galaxy density}
\label{sec.dens}
 We calculate local environmental density parameters for the sample of galaxies analysed in this paper, plus those previously presented in Paper\,I and those compiled from the literature. To estimate such local galaxy density, we consider galaxies more luminous than $M_K=-21$\,mag from several surveys, with redshift distances obtained assuming a Planck cosmology \citep{pla13}. On account of distance uncertainties, we use a redshift cylindrical geometry with a limiting difference in radial velocities of $\Delta V_{\rm R}=600$\,km\,s$^{-1}$, which is of the same order as the velocity dispersion of the galaxy clusters involved in this study \citep[e.g.][]{con01}, and is thought to avoid biases in dense environments. Depending on spatial coverage and depth, the 2MASS Redshift Survey \citep{huc12}, the 6dF Galaxy Survey \citep{jon09}, the SDSS Spectroscopic Catalogue \citep{aba09}, and papers focused on the Coma cluster \citep{mob01,eis07} are used to calculate the environmental
density parameters. We are aware that redshifts as proxy of distances are uncertain due to peculiar velocities, and this is particularly important in cluster environments, but the main strength of such parameters is the higher completeness in comparison with other distance estimators, needed for environmental density measurements.

First, we propose a parameter to characterise the numerical density, calculated as $\Sigma_N = N/(\pi R^2_N)$ with $R_N$ being the radius of the cylinder centred on the galaxy, which contains the $N$ nearest neighbours in projected distance. In order to convert $R_N$ into metric units, we assumed the distances from Tables\,\ref{hubpar2}, \ref{tab.otros}, and Paper\,I for each galaxy of the sample. Similar estimators have proven to be useful for environmental analysis in the literature \citep[e.g.][]{dre80,cap11}. We choose the value $N=10$ to describe the density, $\Sigma_{10}$.

A second estimator shares the geometry with the previous one, but sums over luminosity in the $K$ filter for the 10 nearest neighbours, instead of galaxy counts. Its purpose is to assign different weights to those galaxies located close to massive ones, in the core of clusters. We worked with this estimator on a logarithmic scale. Fig.\,\ref{compdens} presents this parameter $\Sigma_{10}$ as a function of $\Sigma L_{K,10}$. In general, both environmental parameters are in agreement, and their results should not vary significantly. Framed symbols highlight central galaxies and reflect a bias in the environment of the GCSs analysed in the literature. Satellite galaxies are predominantly in intermediate/dense environments, while centrals span a large range. The dashed line has slope 1, and is arbitrarily scaled for comparison purposes. As expected, the parameter $\Sigma L_{K,10}$ slightly deviates to larger values for dense environments. Thus, in the following, just the parameter $\Sigma L_{K,10}$ will be used.

We note that projected distances might lead to uncertain environmental densities, but the typical errors in redshift-independent distance estimators, plus the lack of homogeneous distance determinations for the entire sample, prevent us from calculating spatial densities instead of projected ones. Tables\,\ref{morfodens} and \ref{morfodens2} at the Appendix list the density estimators for the entire sample (i.e. galaxies analysed in this paper and Paper\,I in the former table, and galaxies from the literature in this paper and Paper\,I in the latter), as well as the effective radius of the galaxy.

\begin{table}   
\begin{minipage}{85mm}   
\begin{center}   
\caption{Surface brightness corrected by extinction fitted for background levels ($\mu_{\rm backg,0}$), effective surface brightness for the galaxies ({$\mu_{\rm eff,0}$}), their effective radius (${\rm r_{eff,gal}}$), and S\'ersic index (n) from single S\'ersic laws fitted to the surface brightness profiles in $g$
and $z$ bands for a subsample of Virgo and Fornax galaxies without 
published parameters (Fig.\,\ref{perfilgal}). The last two columns correspond to the integrated galaxy colour and mean ellipticity.
 Galaxies are listed in decreasing $B$-band luminosity.}
\label{sersic}   
\scalebox{0.83}{
\begin{tabular}{@{}r@{}c@{}ccc@{}cc@{}}   
\hline   
\multicolumn{1}{@{}c}{Name}&\multicolumn{1}{c}{$\mu_{\rm backg,0}$}&\multicolumn{1}{c}{$\mu_{\rm eff,0}$}&\multicolumn{1}{c}{${\rm r_{eff,gal}}$}&\multicolumn{1}{c}{n}&\multicolumn{1}{c}{$(g-z)_{\rm 0,gal}$}&\multicolumn{1}{c}{<$\epsilon$>}\\
\multicolumn{1}{@{}c}{}&\multicolumn{2}{c}{${\rm mag\,arcsec^{-2}}$}&\multicolumn{1}{c}{${\rm arcsec}$}& &\multicolumn{1}{c}{mag}& \\
\hline   
NGC\,1404\,$g$ & $23.1$ & $20.7\pm0.1$ & $24.7\pm0.3$ & $4.1\pm0.1$ & 1.50 & 0.13\\ 
          $z$ & $22.4$ & $19.1\pm0.1$ & $23.3\pm0.2$ & $4.2\pm0.1$ & & 0.13\\
NGC\,4526\,$g$ & $22.2$ & $20.4\pm0.1$ & $24.8\pm0.4$ & $2.1\pm0.1$ & 1.42 & 0.40\\
          $z$ & $21.3$ & $18.7\pm0.1$ & $22.8\pm0.5$ & $1.8\pm0.1$ & & 0.39\\
NGC\,1380\,$g$ & $22.9$ & $21.4\pm0.1$ & $38.0\pm0.8$ & $3.3\pm0.1$ & 1.37 & 0.37\\ 
          $z$ & $22.4$ & $19.8\pm0.1$ & $33.4\pm0.6$ & $3.3\pm0.1$ & & 0.37\\ 
NGC\,1387\,$g$ & $23.4$ & $22.8\pm0.2$ & $50.0\pm5.7$ & $6.3\pm0.8$ & 1.53 & 0.17\\
          $z$ & $22.5$ & $20.8\pm0.2$ & $37.4\pm3.1$ & $5.0\pm0.6$ & & 0.19\\
NGC\,4435\,$g$ & $22.4$ & $20.3\pm0.2$ & $15.8\pm1.0$ & $1.7\pm0.2$ & 1.53 & 0.45\\
          $z$ & $21.7$ & $18.8\pm0.2$ & $14.5\pm1.0$ & $2.0\pm0.2$ & & 0.45\\
IC\,2006\,$g$ & $23.3$ & $21.3\pm0.1$ & $17.7\pm0.3$ & $2.1\pm0.1$ & 1.47 & 0.12\\
          $z$ & $22.5$ & $19.8\pm0.1$ & $16.8\pm0.6$ & $2.3\pm0.1$ & & 0.12\\
NGC\,1380A\,$g$ & $23.2$ & $21.7\pm0.1$ & $15.2\pm0.1$ & $1.4\pm0.1$ & 1.31 & 0.71\\ 
          $z$ & $22.5$ & $20.4\pm0.1$ &  $15.1\pm0.2$ & $1.5\pm0.1$ & & 0.72\\ 
FCC\,255\,$g$ & $23.3$ & $22.3\pm0.2$ & $14.2\pm1.1$ & $3.5\pm0.4$ & 1.48 & 0.52\\ 
          $z$ & $22.5$ & $20.8\pm0.2$ & $12.6\pm0.8$ & $2.3\pm0.2$ & & 0.52\\ 
\hline
\end{tabular}}  
\end{center}    
\end{minipage}   
\end{table}

\section{Results}
\label{sec.res}

\subsection{Galaxy surface brightness profiles}
\label{sec.briprof}

The parameters of single S\'ersic fits on the surface brightness profiles are available in the literature \citep{fer06b,cot07,gla11} for the majority of the ETGs in our sample. However, as 2 Virgo galaxies and the 6 Fornax ones have no published profiles, single S\'ersic models are fitted to their $g$ and $z$ surface brightness profiles obtained from the ACS images through the task ELLIPSE (see Section\,\ref{sec.gals}). In Fig.\,\ref{perfilgal} we show the surface brightness profiles measured for these 8 galaxies as a function of the equivalent radius in arcsec (${\rm r_{eq}} = \sqrt{ab}$). The $g$ and $z$ profiles and residuals are represented by green circles and red squares, respectively. The S\'ersic model fitted to each profile is described by the following equation:

\begin{equation}
\mu({\rm r_{eq}}) = \mu_{\rm eff} + 1.0857*{\rm b_n}\left[ \left( \frac{{\rm r_{eq}}}{\rm r_{ eff,gal}} \right)^{\frac{1}{\rm n}}-1 \right]
\end{equation}

\noindent with ${\rm r_{eq}}$ and $r_{\rm eff,gal }$ (galactic effective radius)  measured in arcsec, and $\mu({\rm r_{eq}})$ and $\mu_{\rm eff}$ in  mag\,arcsec$^{-2}$. We obtain $b_n$ from the expression in \citet{cio91}, and $n$ is the S\'ersic shape index. We achieve acceptable fits with a single component, taking into account the FOV of the ACS camera (residuals are shown in the middle panels of Fig.\,\ref{perfilgal}). Because of the reduced size of this FOV we can not estimate the background level accurately. Instead, it is handled as a free parameter and obtained  by fitting the count level at galactocentric distances larger than 100\,arcsec. Both the background level and the S\'ersic model are fitted interactively, and their corresponding contributions are subtracted at each step until the parameters converged, and the residuals for measurements further than 100\,arcsec from the galactic centre reached $\sim 10^{-2}$. In the upper panels of Fig.\,\ref{perfilgal}, the S\'ersic model for each band is shown in solid ($g$) and dashed ($z$) thin curves, the corresponding background levels are indicated with horizontal lines, and the contributions of the galaxy plus background are drawn as thick curves.

In Table\,\ref{sersic}, we present the S\'ersic parameters for these 8 galaxies, corrected by extinction, as well as their corresponding background levels ($\mu_{\rm backg,0}$ ). As an additional test, we check that our estimated backgrounds show negligible differences with respect to the values estimated using the ACS Exposure Time Calculator\footnote{http://etc.stsci.edu/etc/input/acs/imaging/}, in units of electrons per second, for similar dates, filters, and exposure times as the corresponding observations. Moreover, the fitted values for $\mu_{\rm backg,0}$ are in agreement with those  presented by \citet{jor04} for the Virgo galaxies, using a similar instrumental configuration.

In the last two columns of Table\,\ref{sersic}, we also list the galactic integrated colour $(g-z)_{\rm 0,gal}$ and the mean ellipticity for the galaxies. This colour results from integrating inwards the S\'ersic profiles. Though they are $\approx 0.1$\,mag bluer than those 
of Virgo galaxies within an analogous luminosity range \citep{smi13}, this difference is similar to that found for other ETGs in low-density environments \cite[e.g][]{lac16}. The lower panels of Fig.\,\ref{perfilgal} show the colour profiles in $(g-z)_0$ for each galaxy, where a negative colour gradient is clear in most galaxies. For some of them, the colours at large radii are missing, due to their surface brightness profiles falling quickly to the background level.

\begin{figure*}   
\caption{Projected radial distribution for each GCS, with the fitted modified Hubble profile shown in red solid lines. Grey regions show the variations resulting from  individual fits performed by shifting  the bin breaks (see text for further details)}.
\includegraphics[width=55mm]{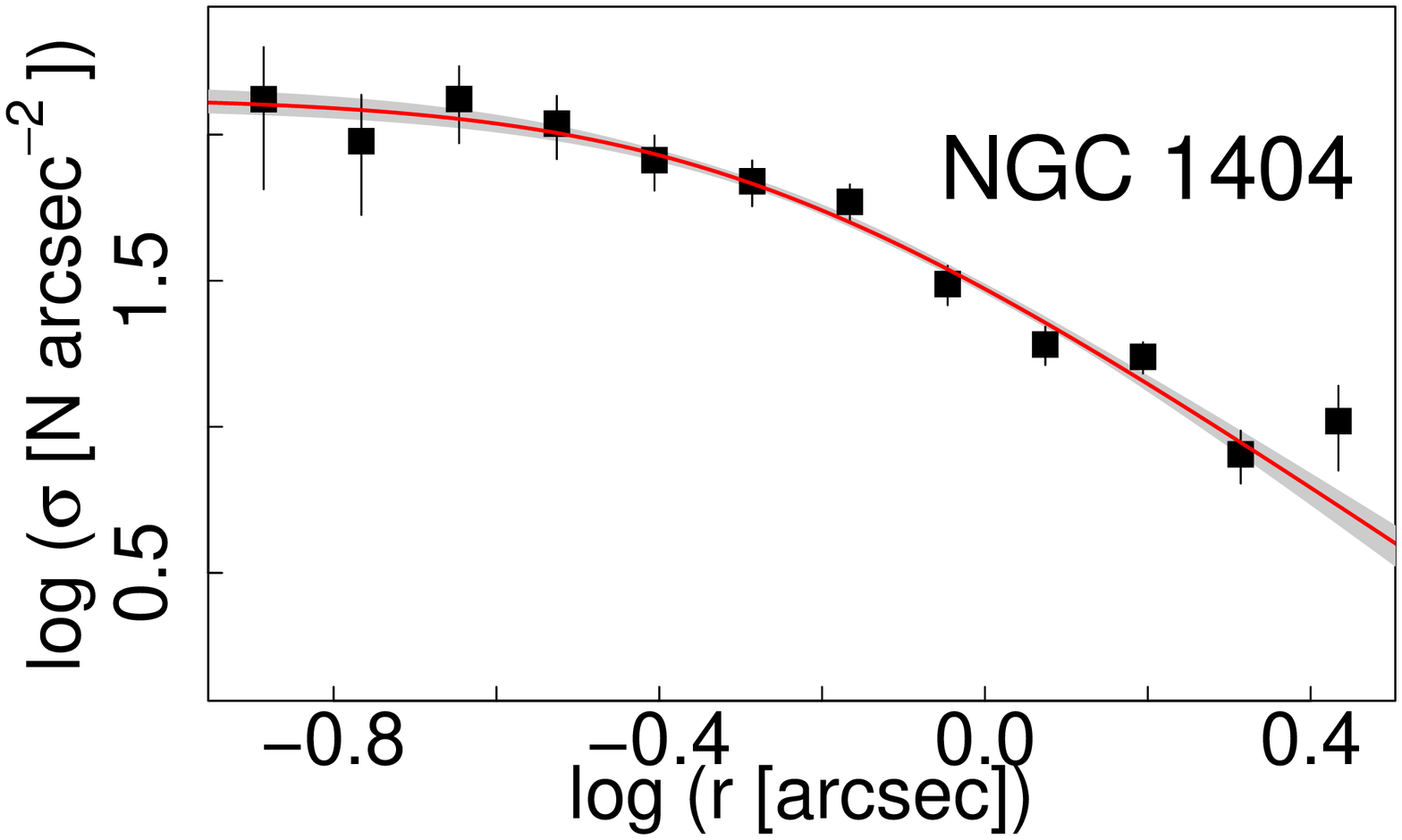}
\includegraphics[width=55mm]{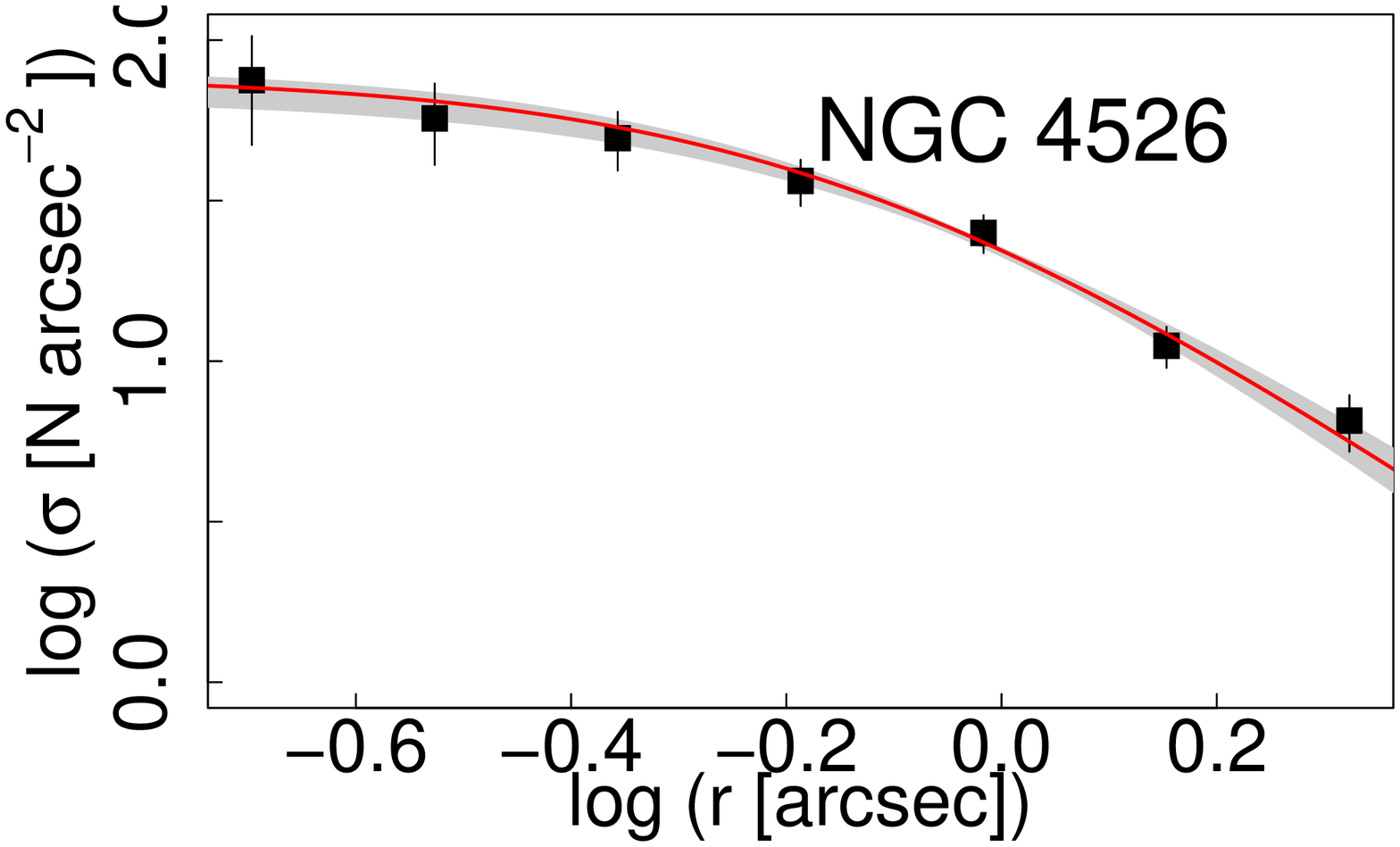}
\includegraphics[width=55mm]{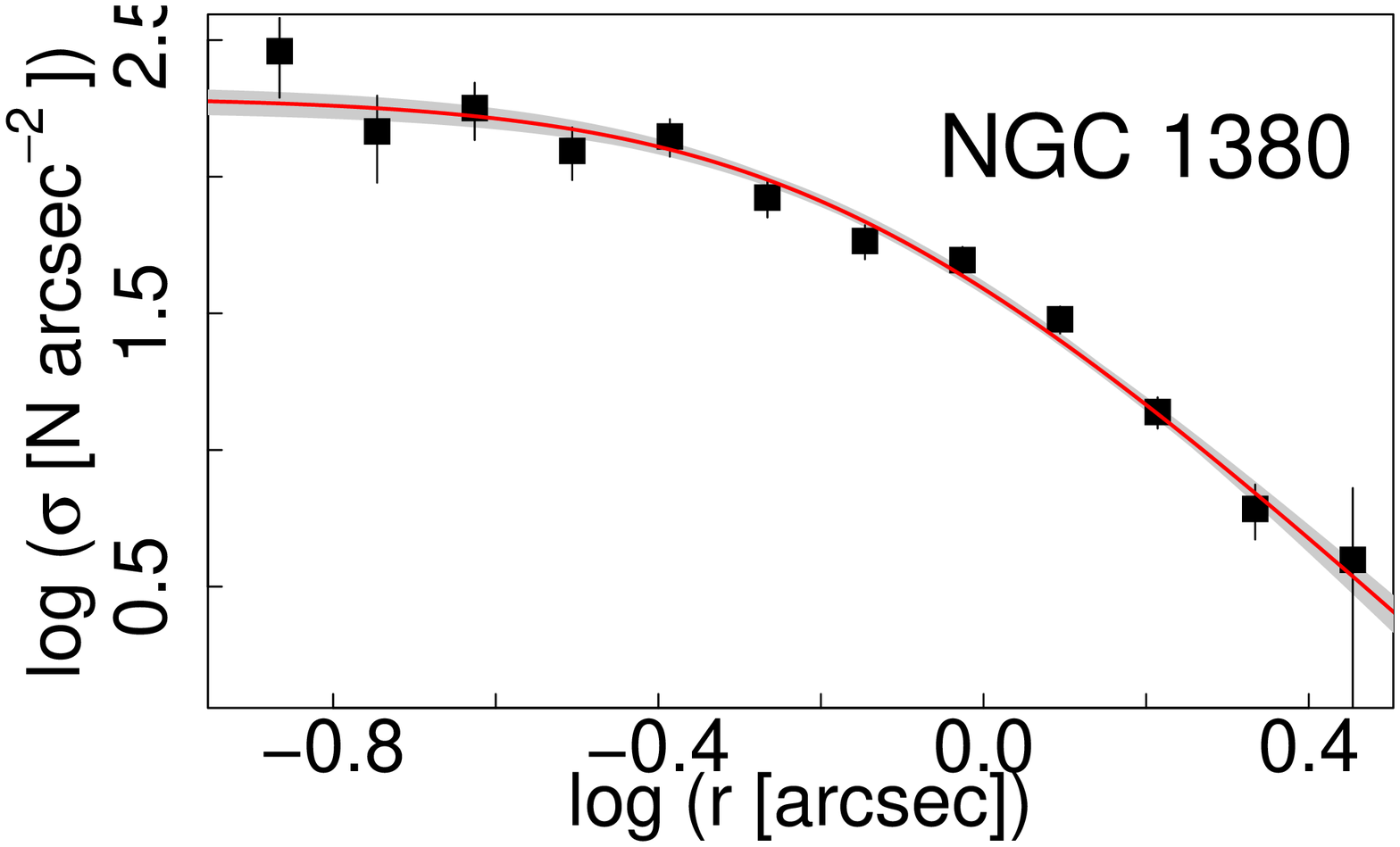}\\    
\includegraphics[width=55mm]{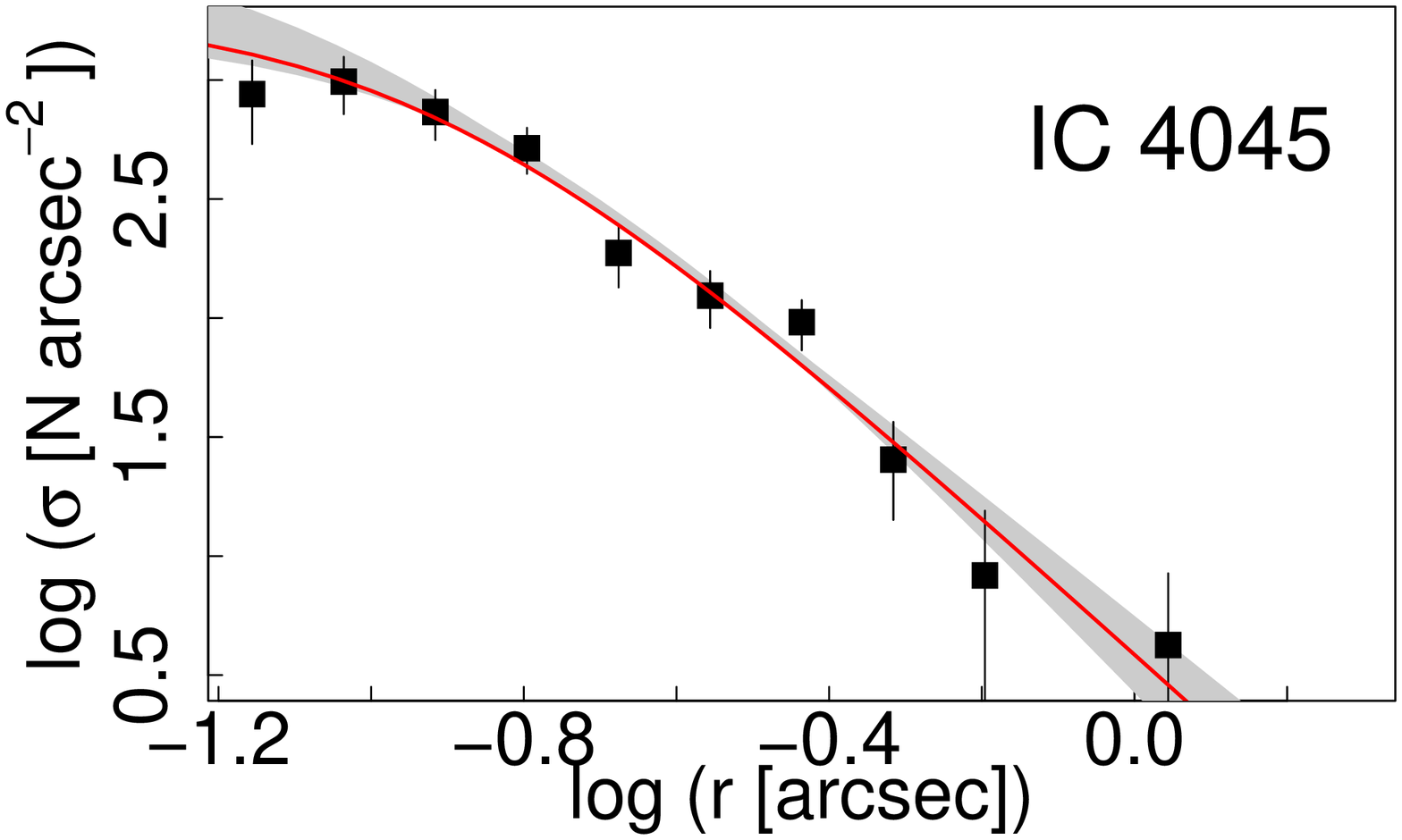}   
\includegraphics[width=55mm]{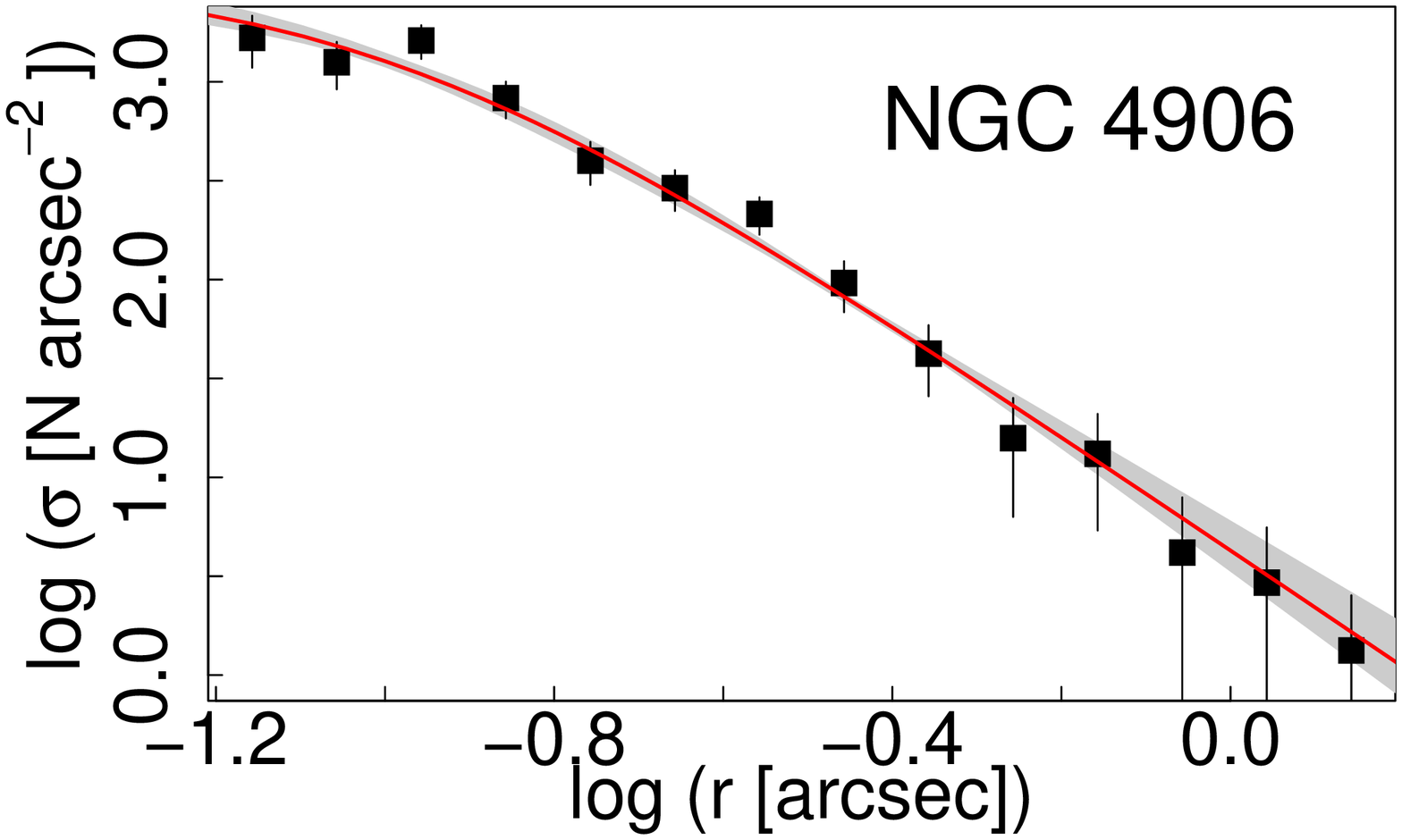}    
\includegraphics[width=55mm]{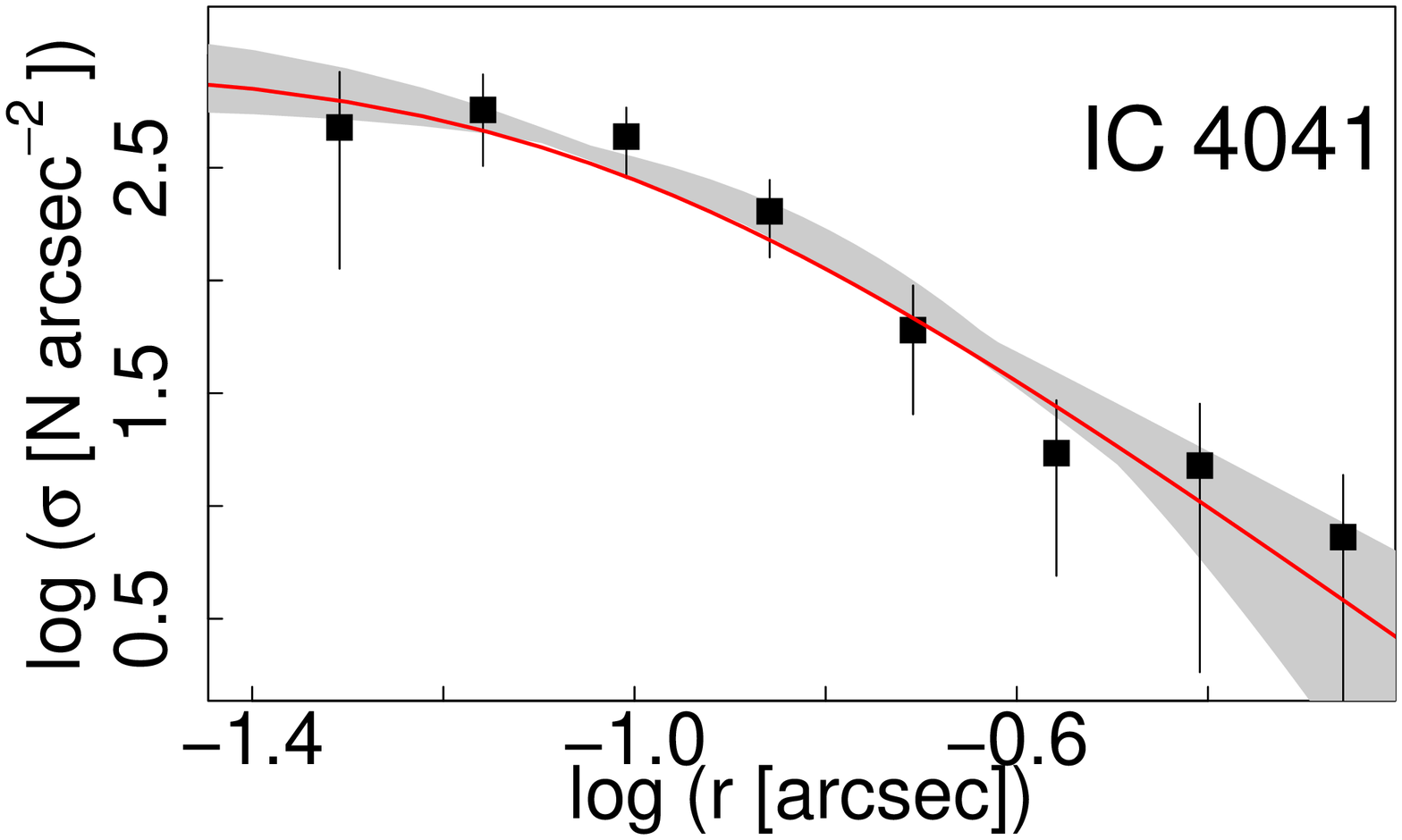}\\    
\includegraphics[width=55mm]{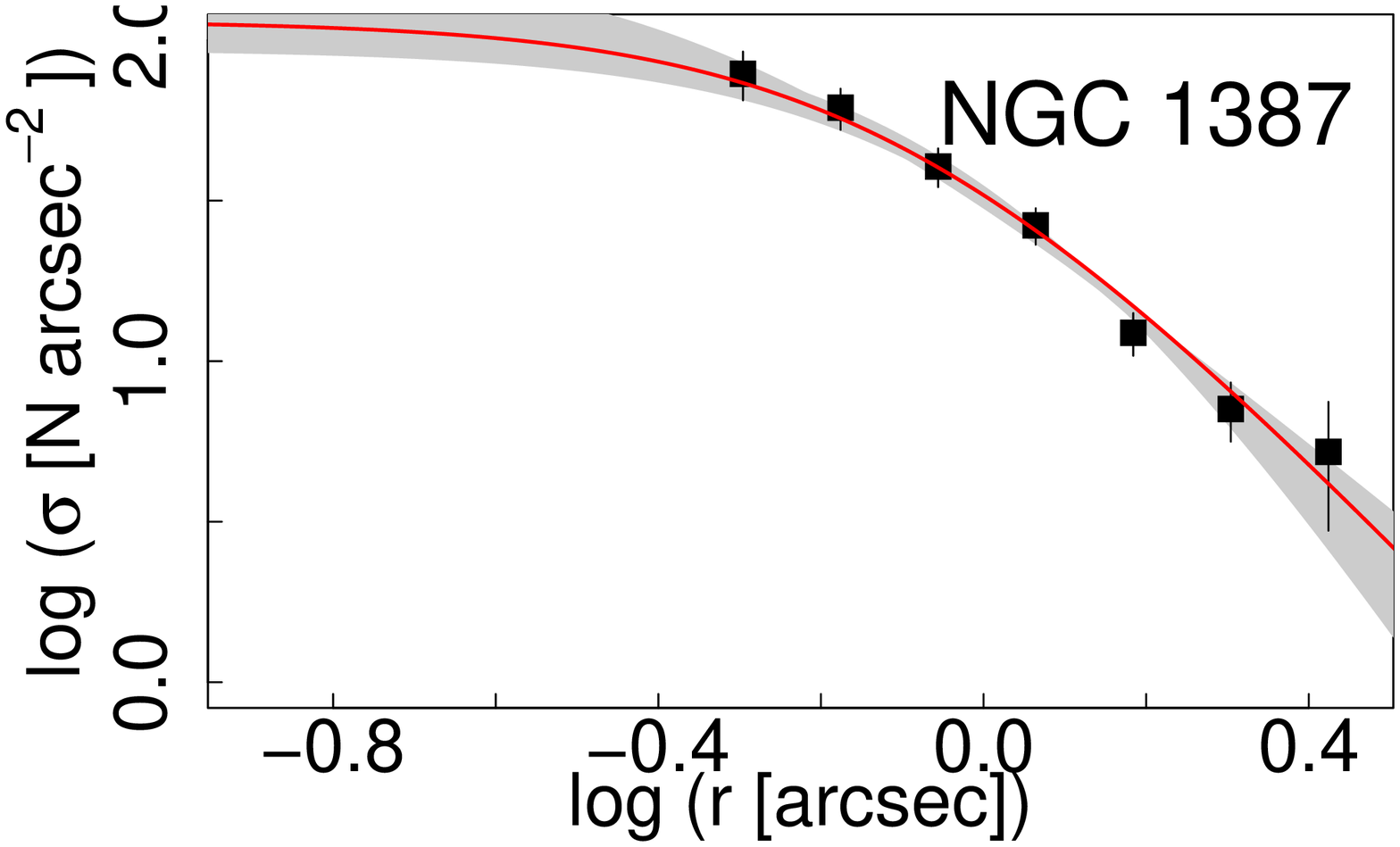} 
\includegraphics[width=55mm]{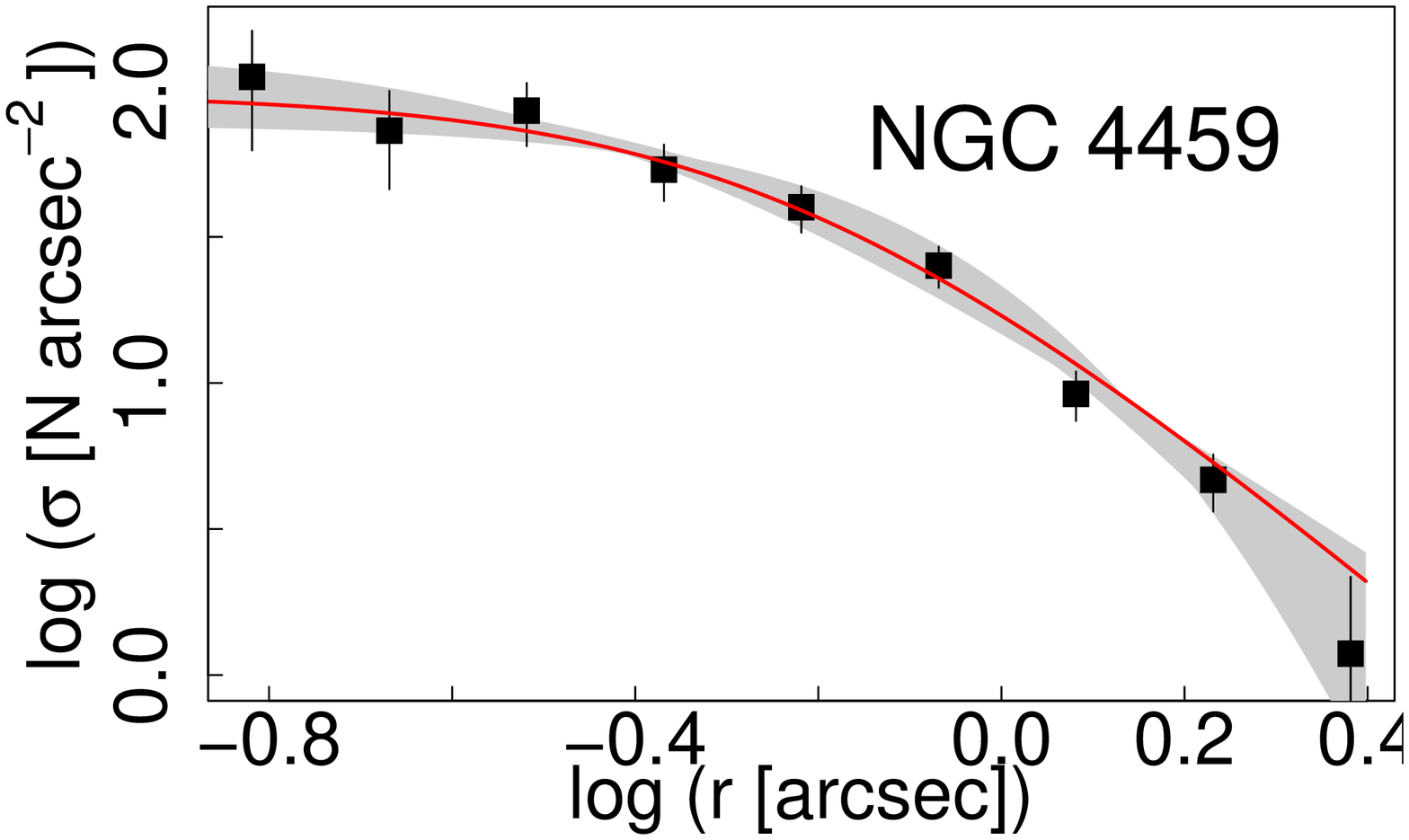}    
\includegraphics[width=55mm]{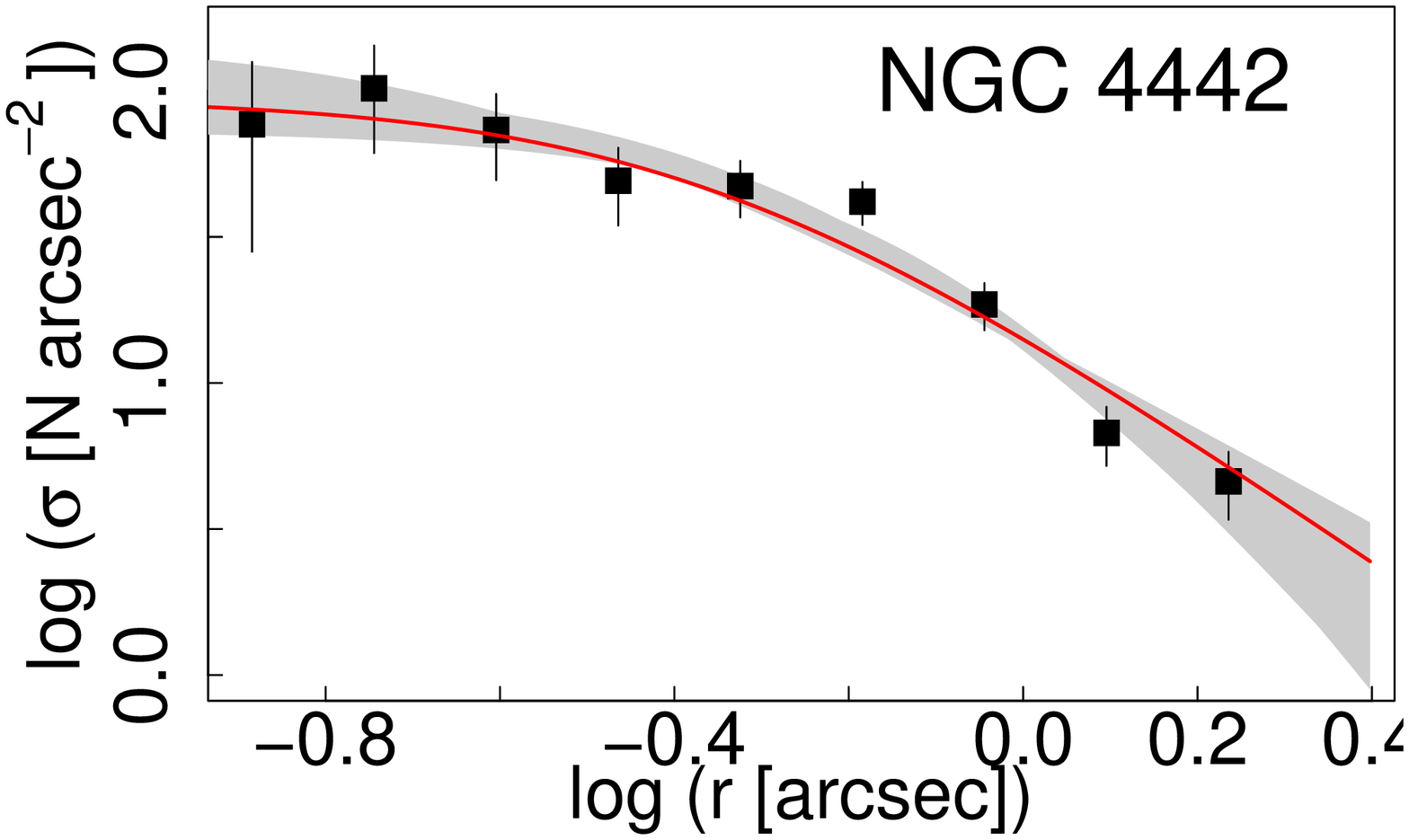}\\    
\includegraphics[width=55mm]{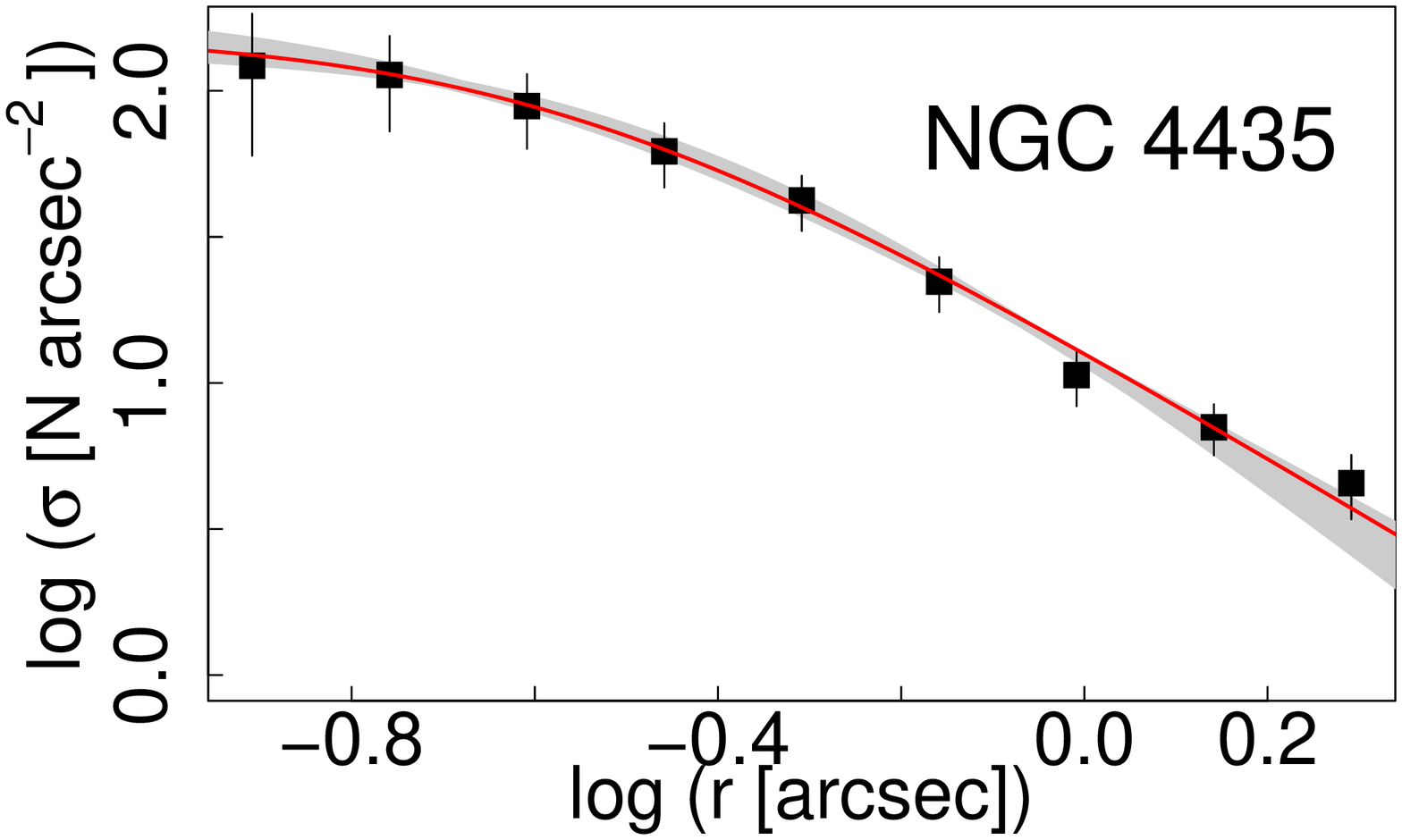}    
\includegraphics[width=55mm]{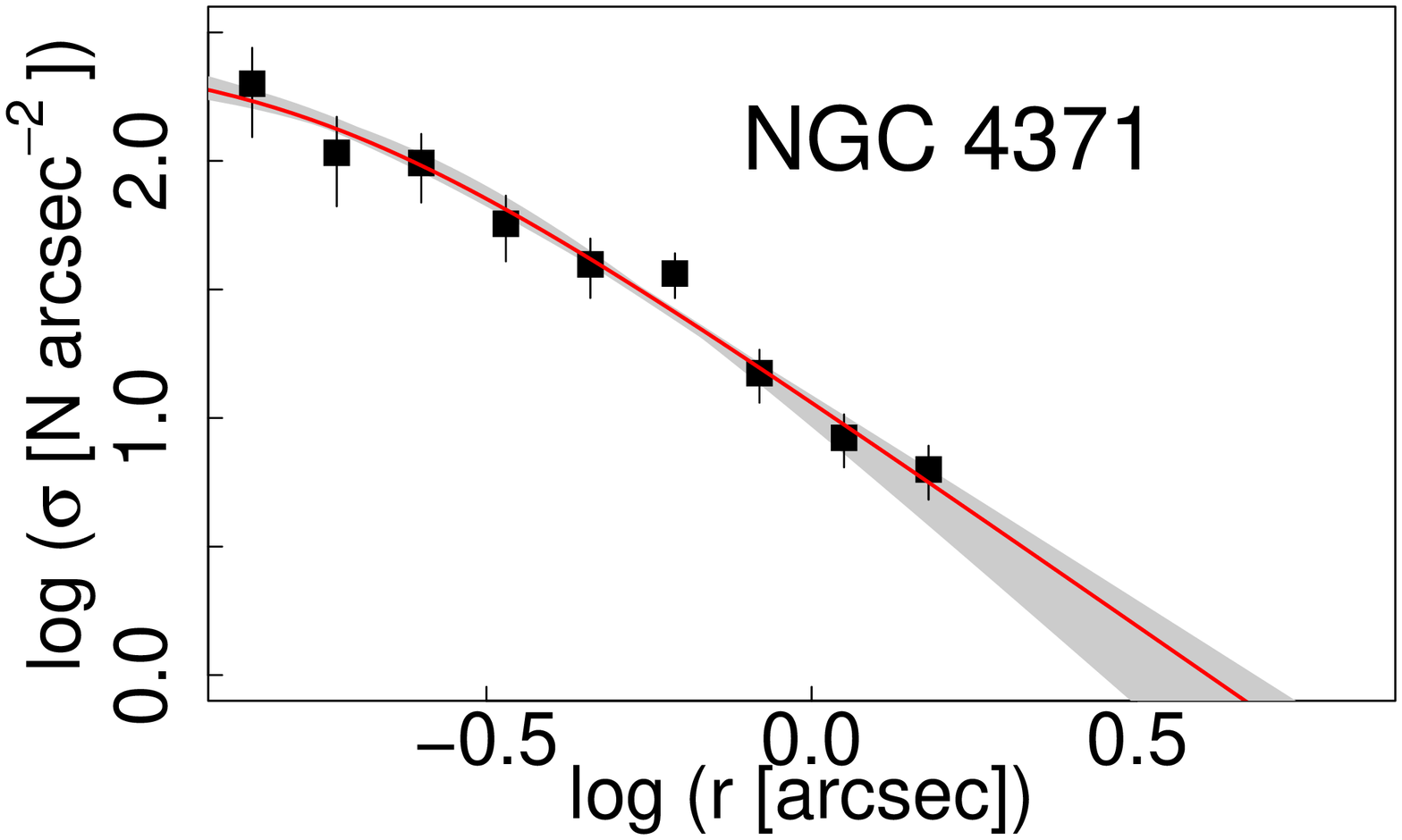}    
\includegraphics[width=55mm]{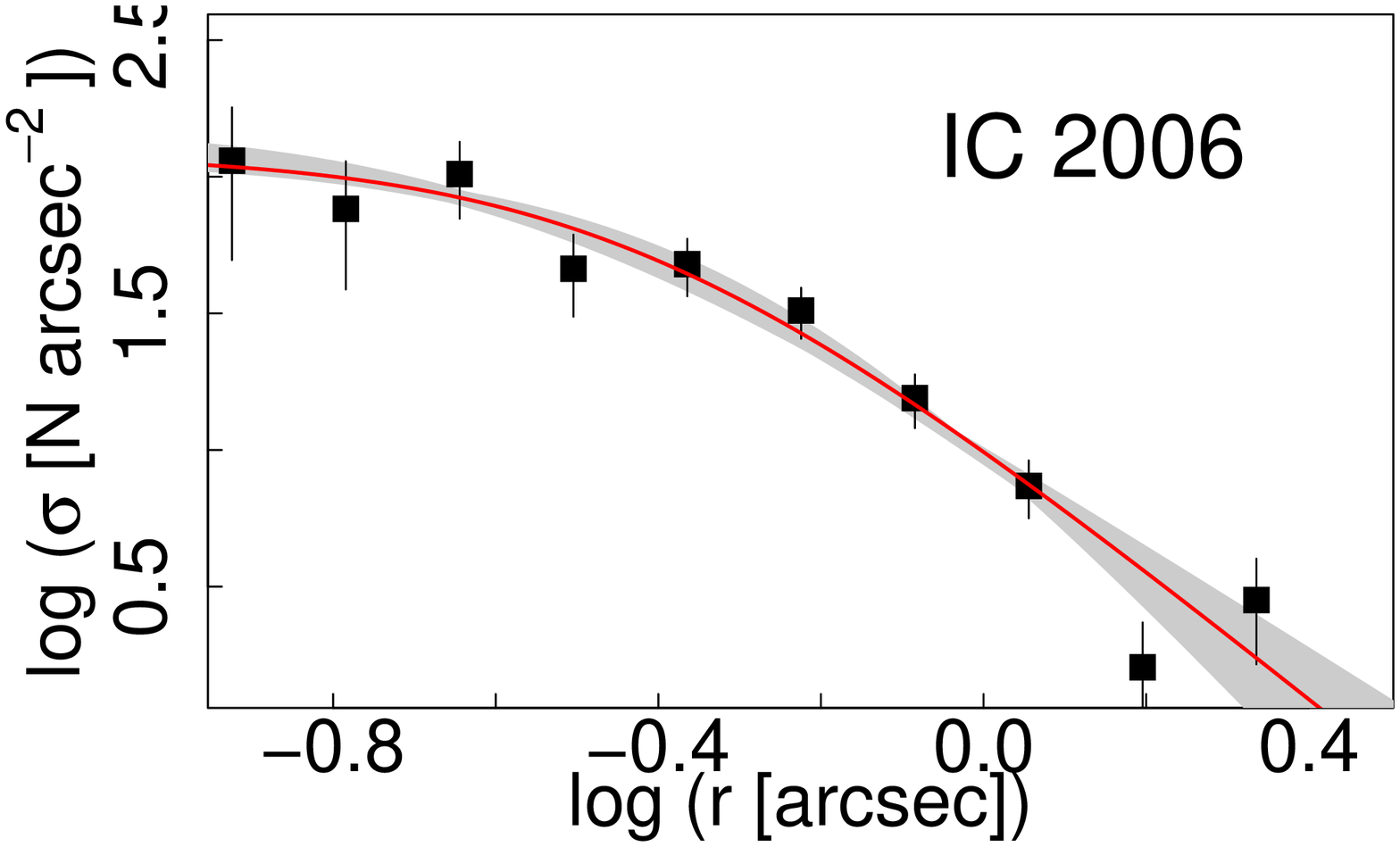}\\
\includegraphics[width=55mm]{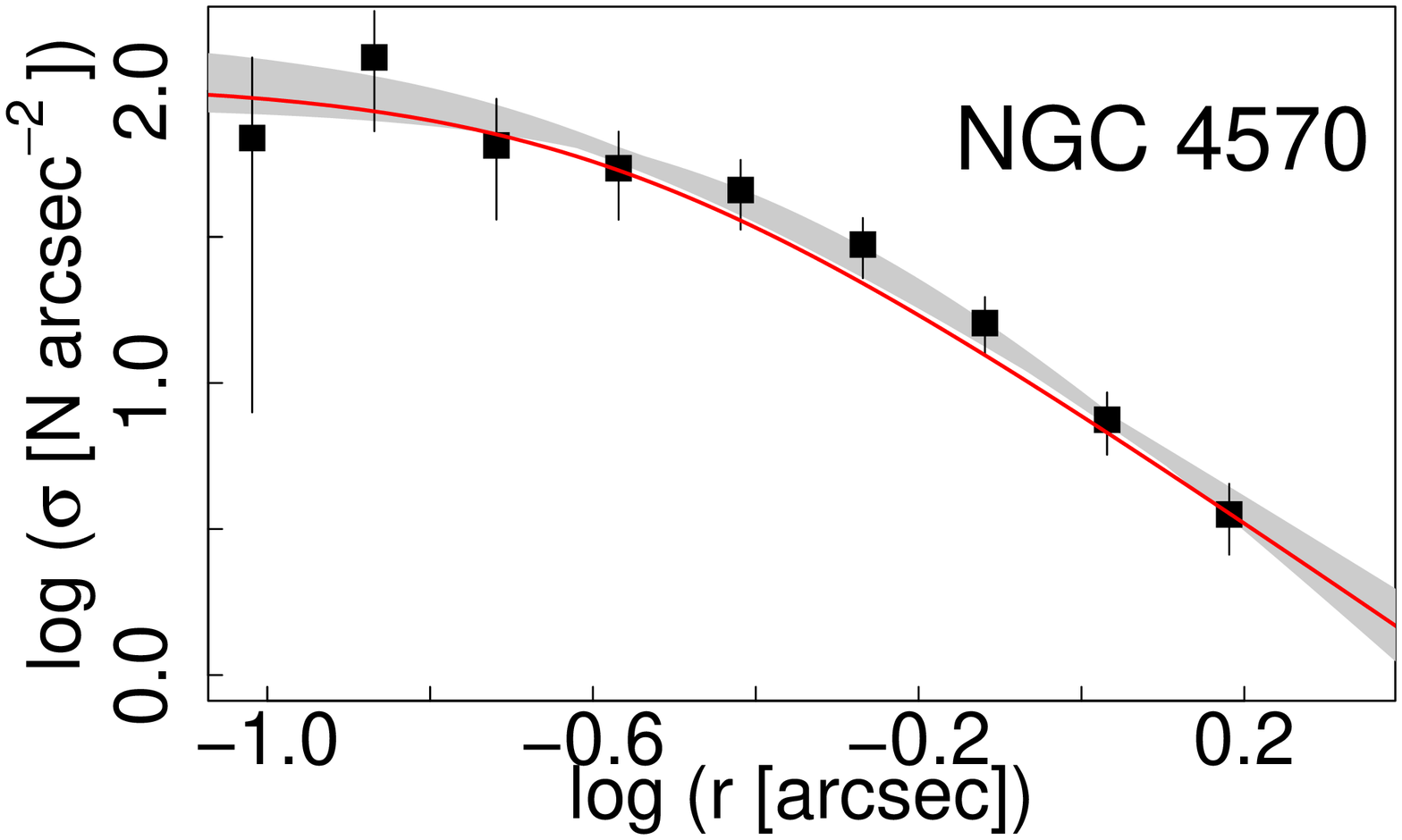}
\includegraphics[width=55mm]{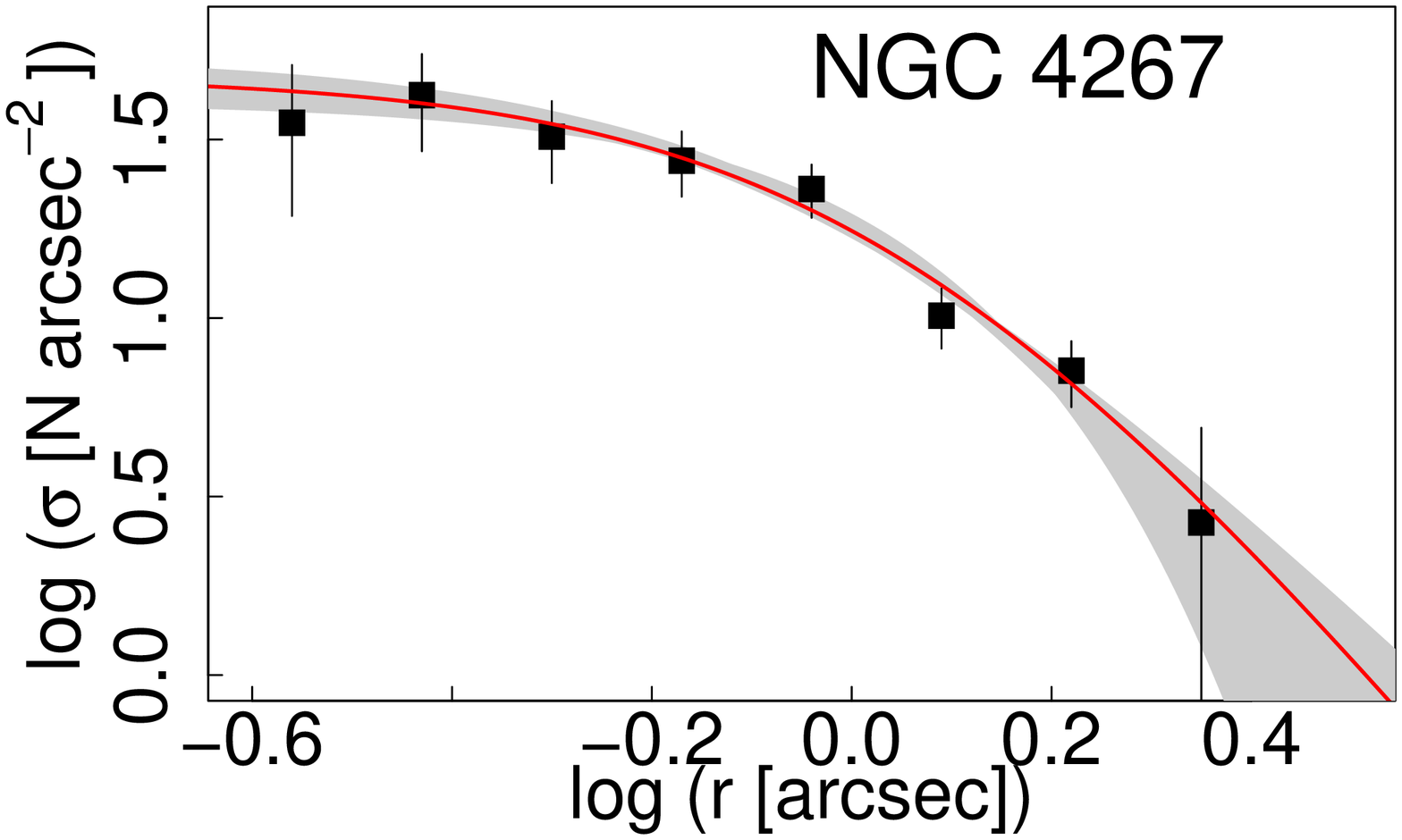}
\includegraphics[width=55mm]{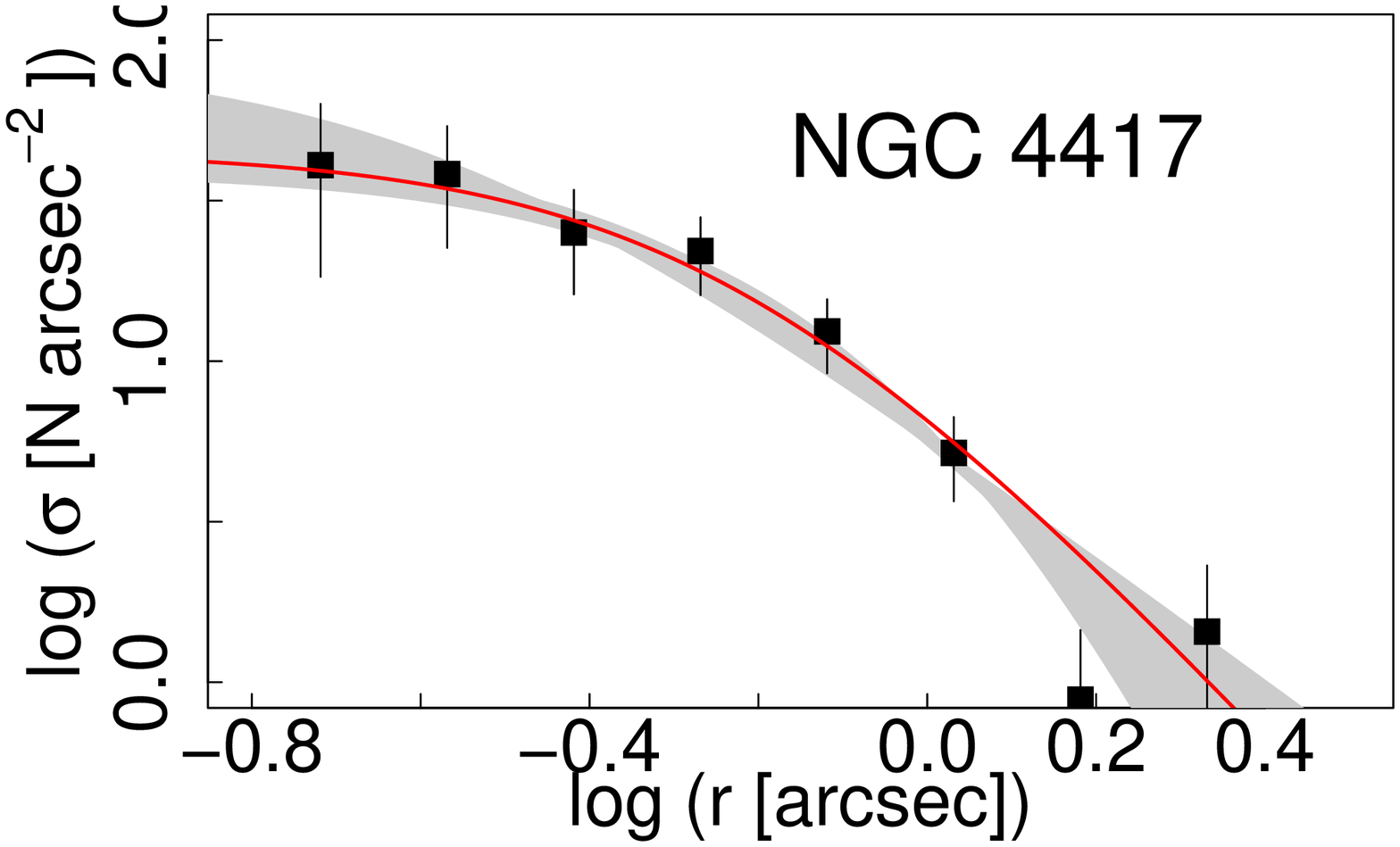}
\\  
\includegraphics[width=55mm]{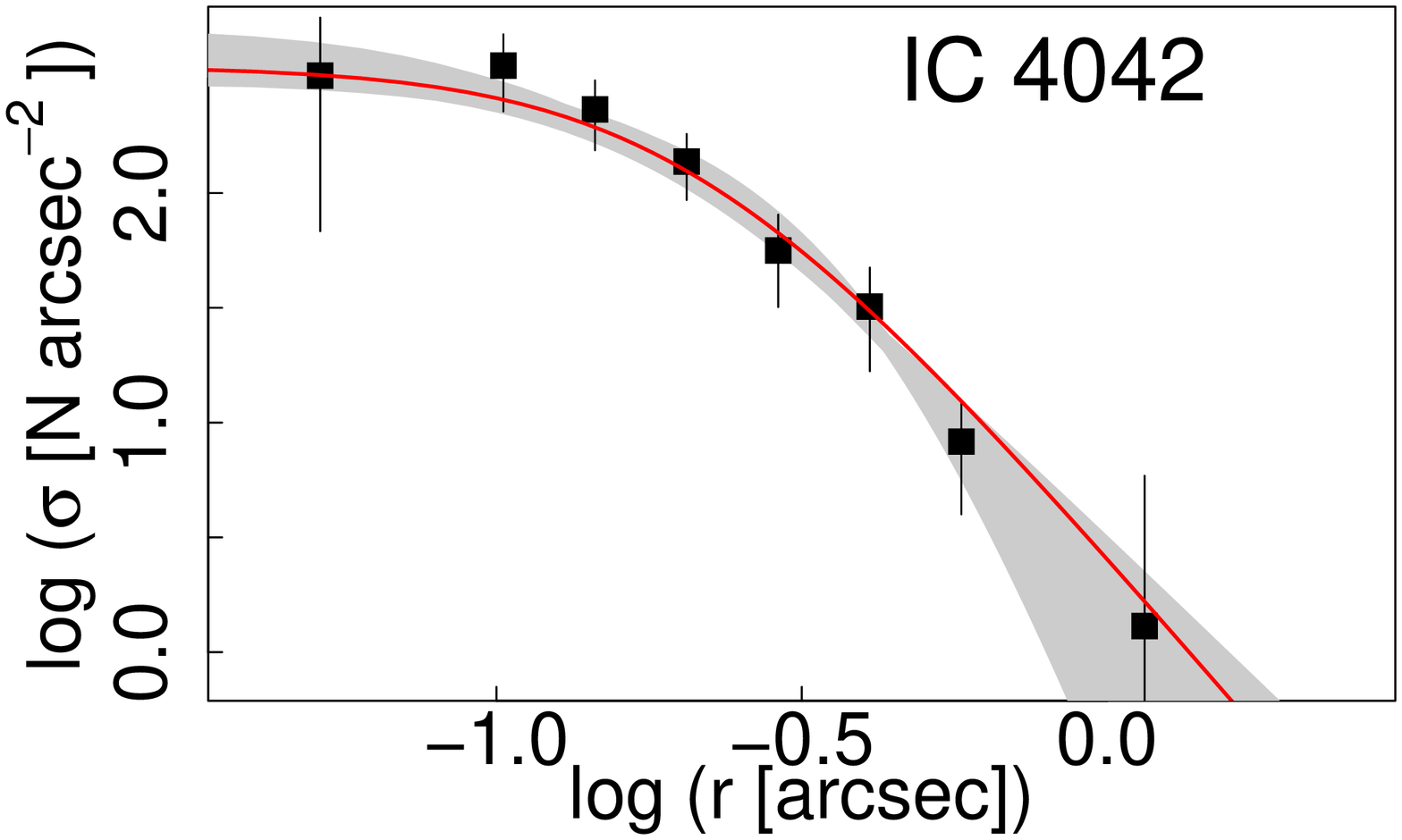}    
\includegraphics[width=55mm]{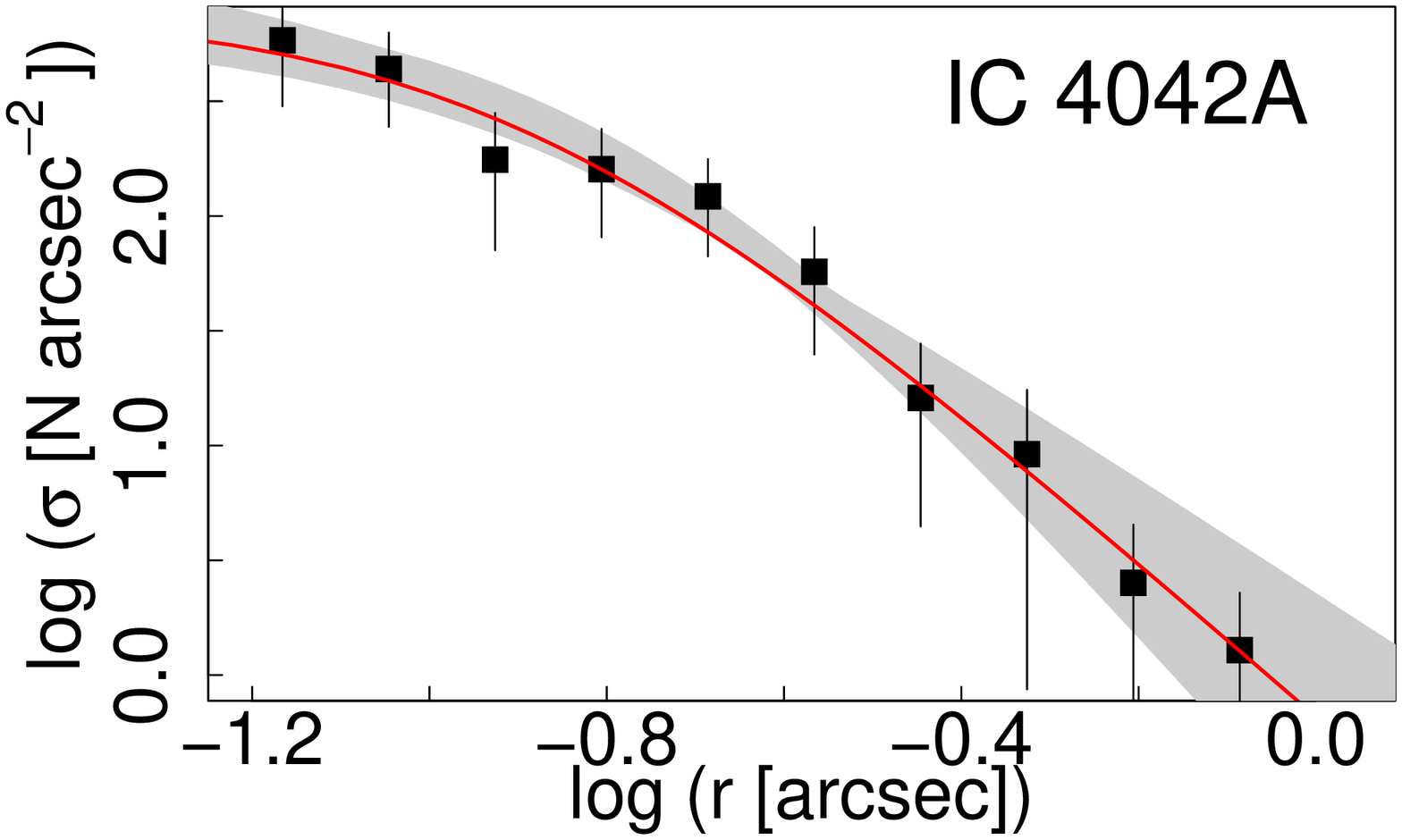}    
\includegraphics[width=55mm]{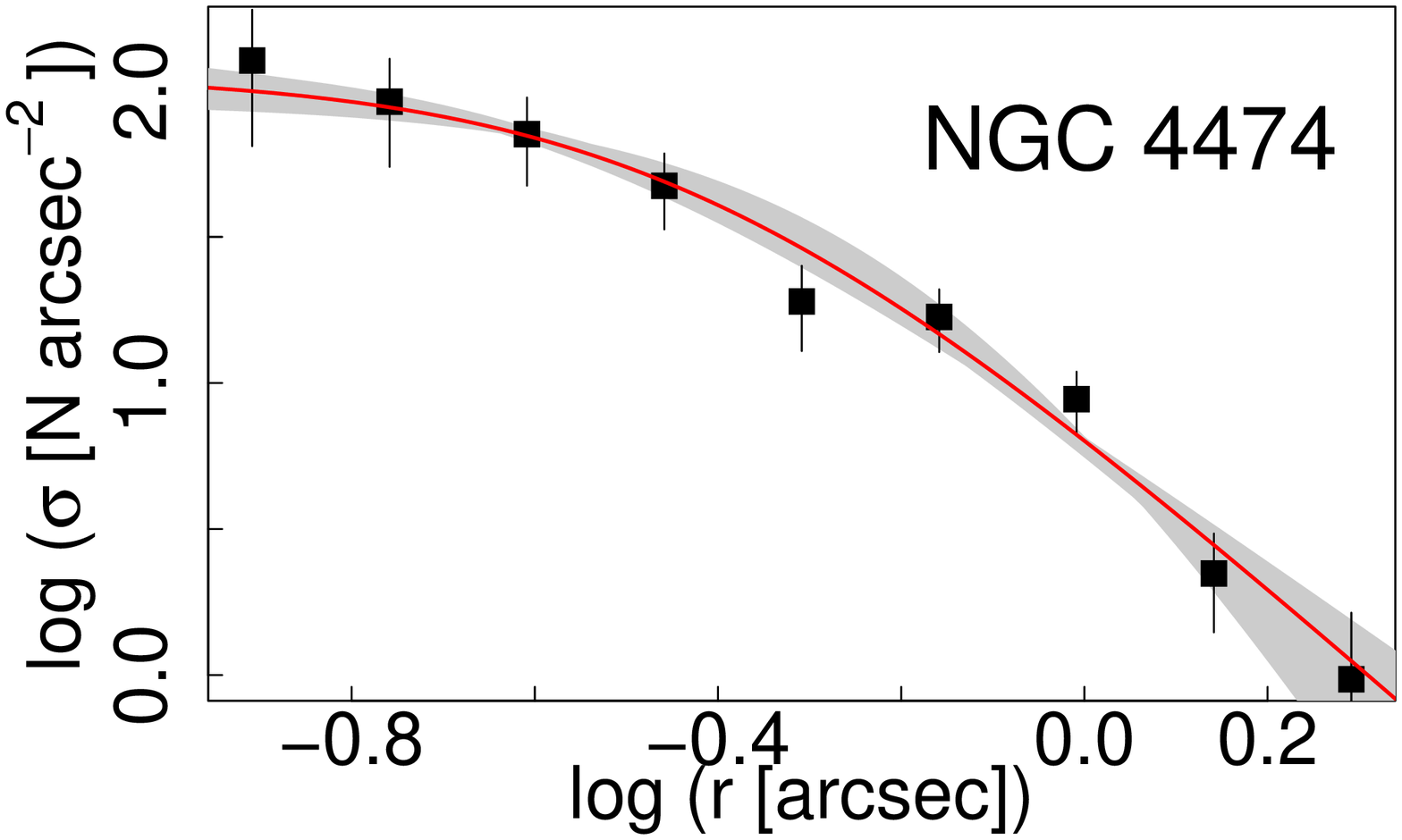} \\  
\includegraphics[width=55mm]{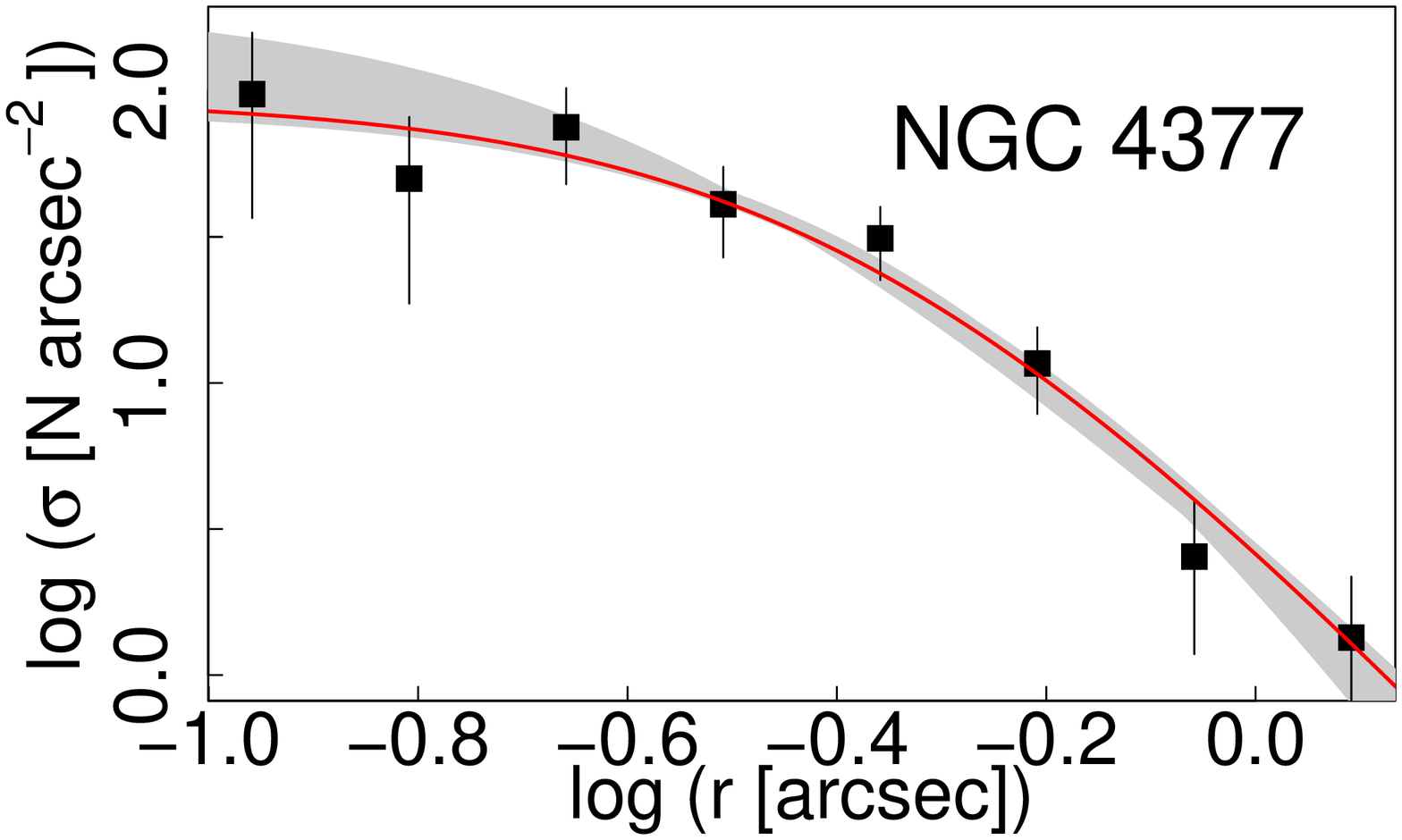}
\includegraphics[width=55mm]{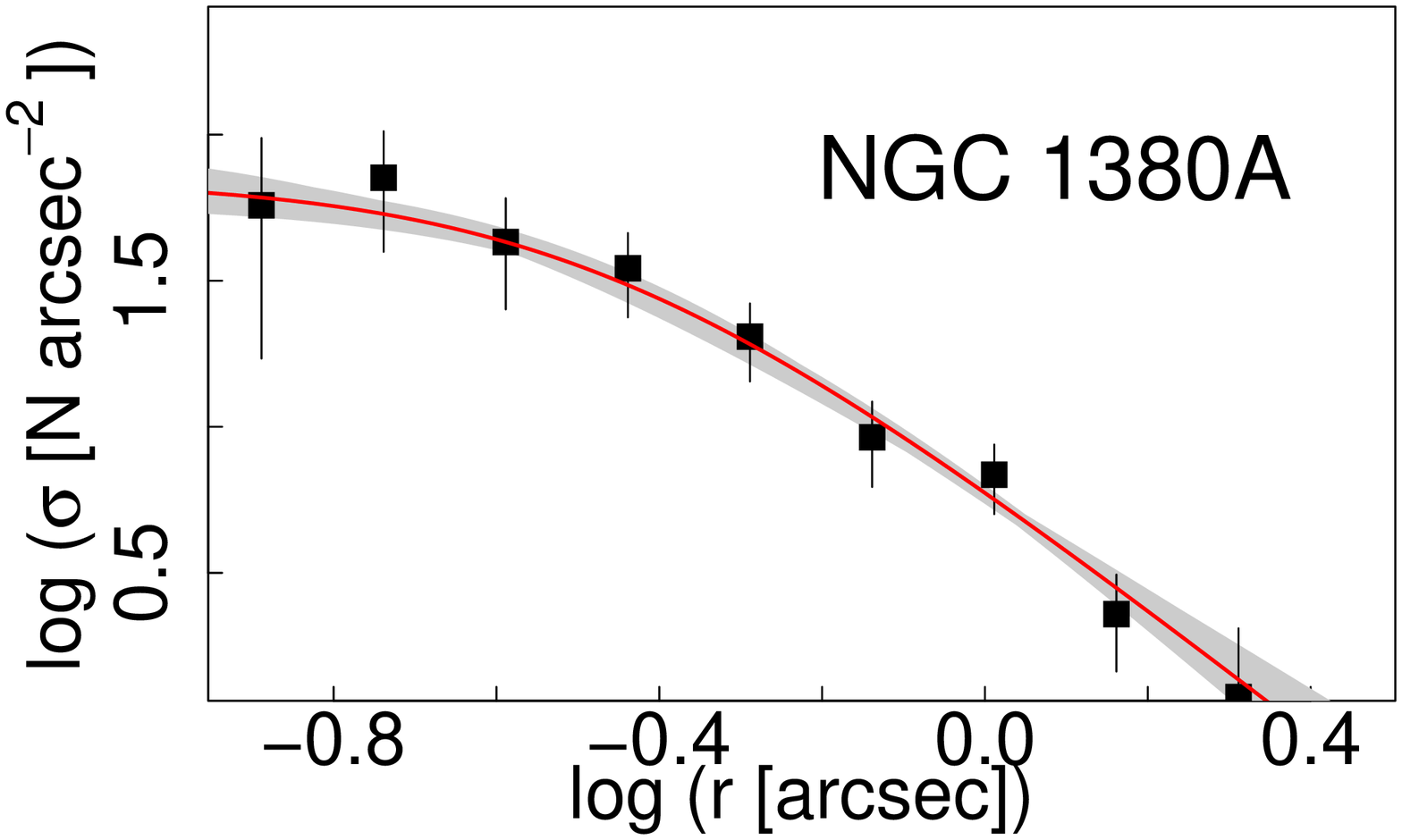}
\includegraphics[width=55mm]{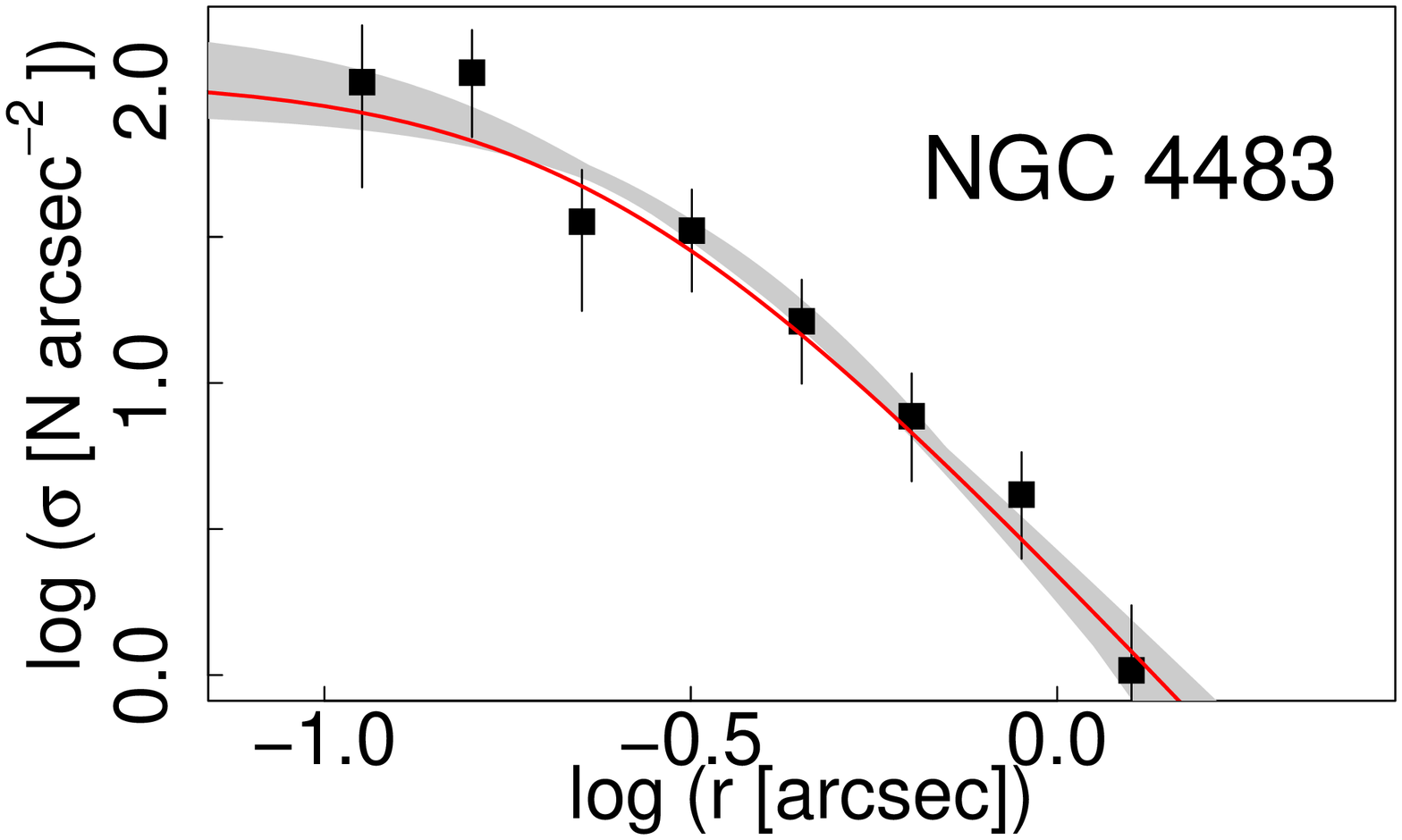}  
\\   
\label{hubprof}    
\end{figure*}

\begin{figure*}    
\includegraphics[width=55mm]{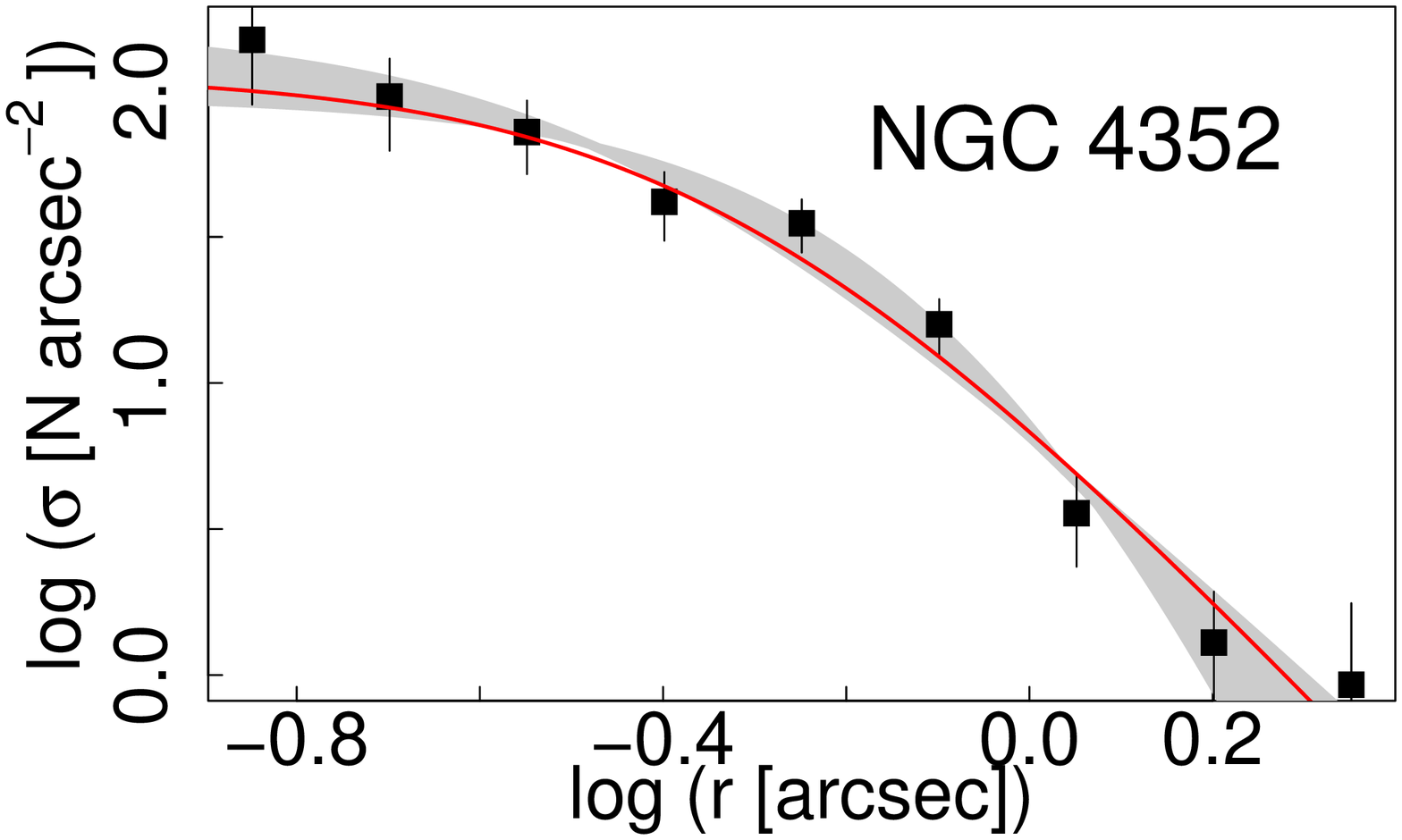}
\includegraphics[width=55mm]{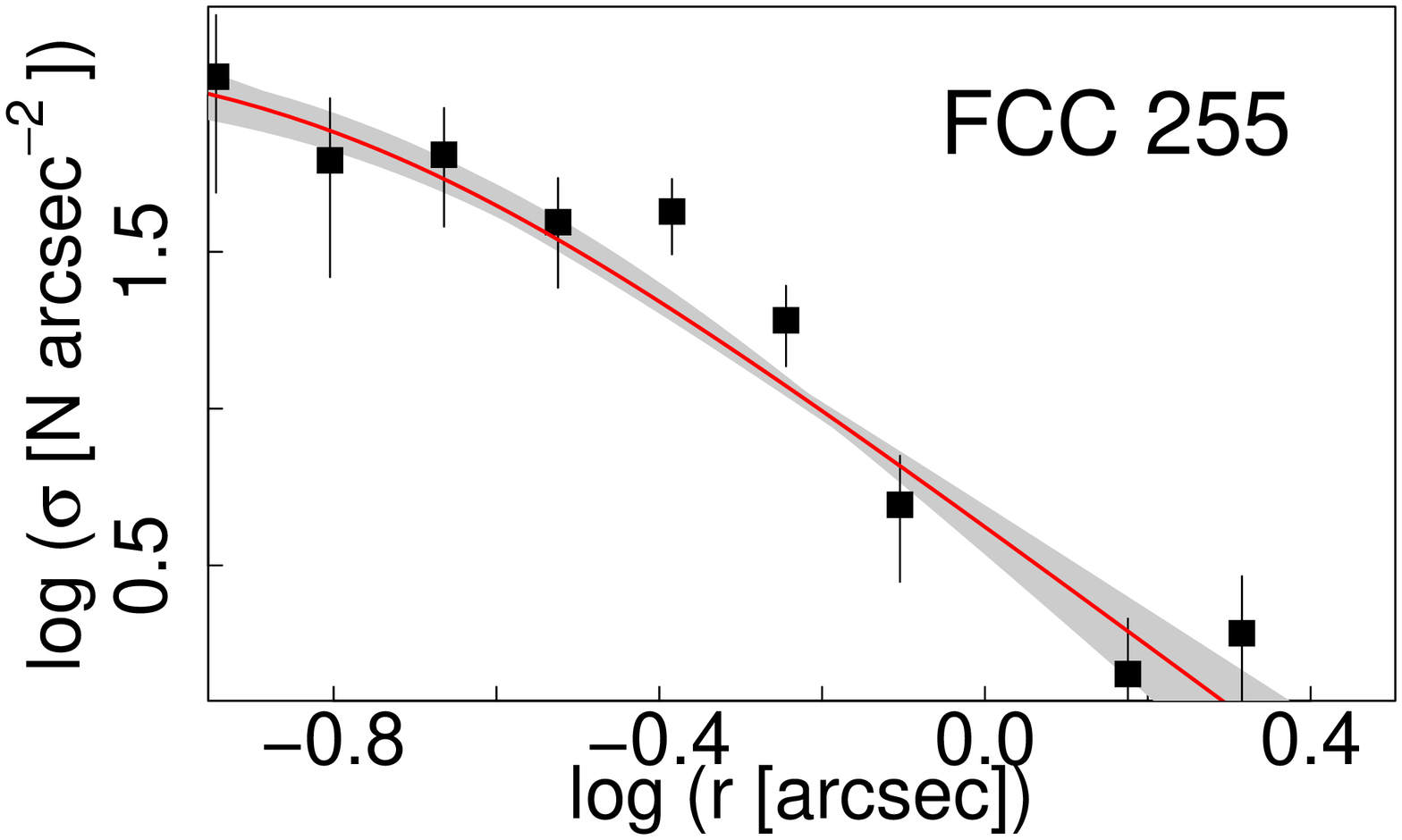}
\includegraphics[width=55mm]{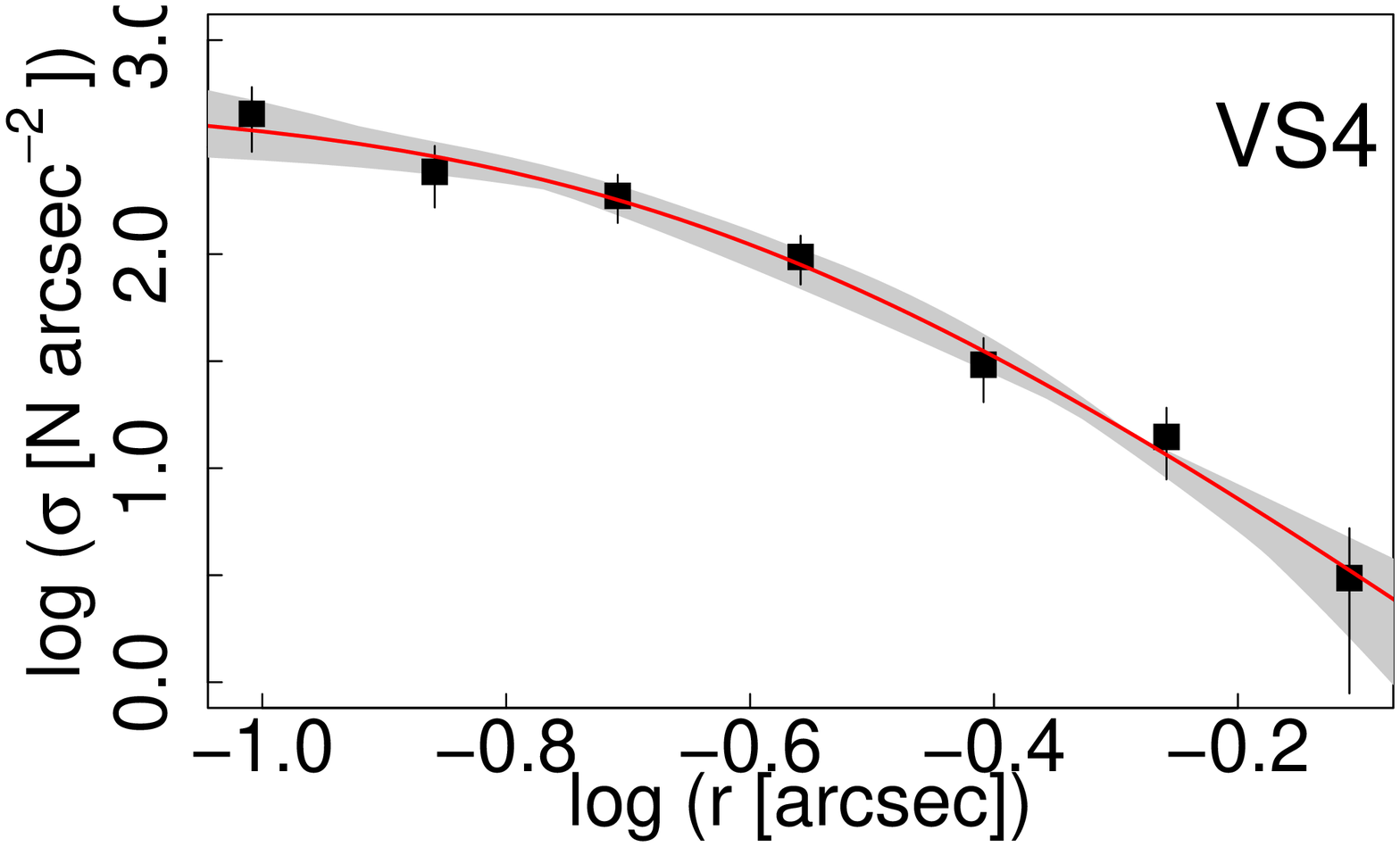}\\    
\includegraphics[width=55mm]{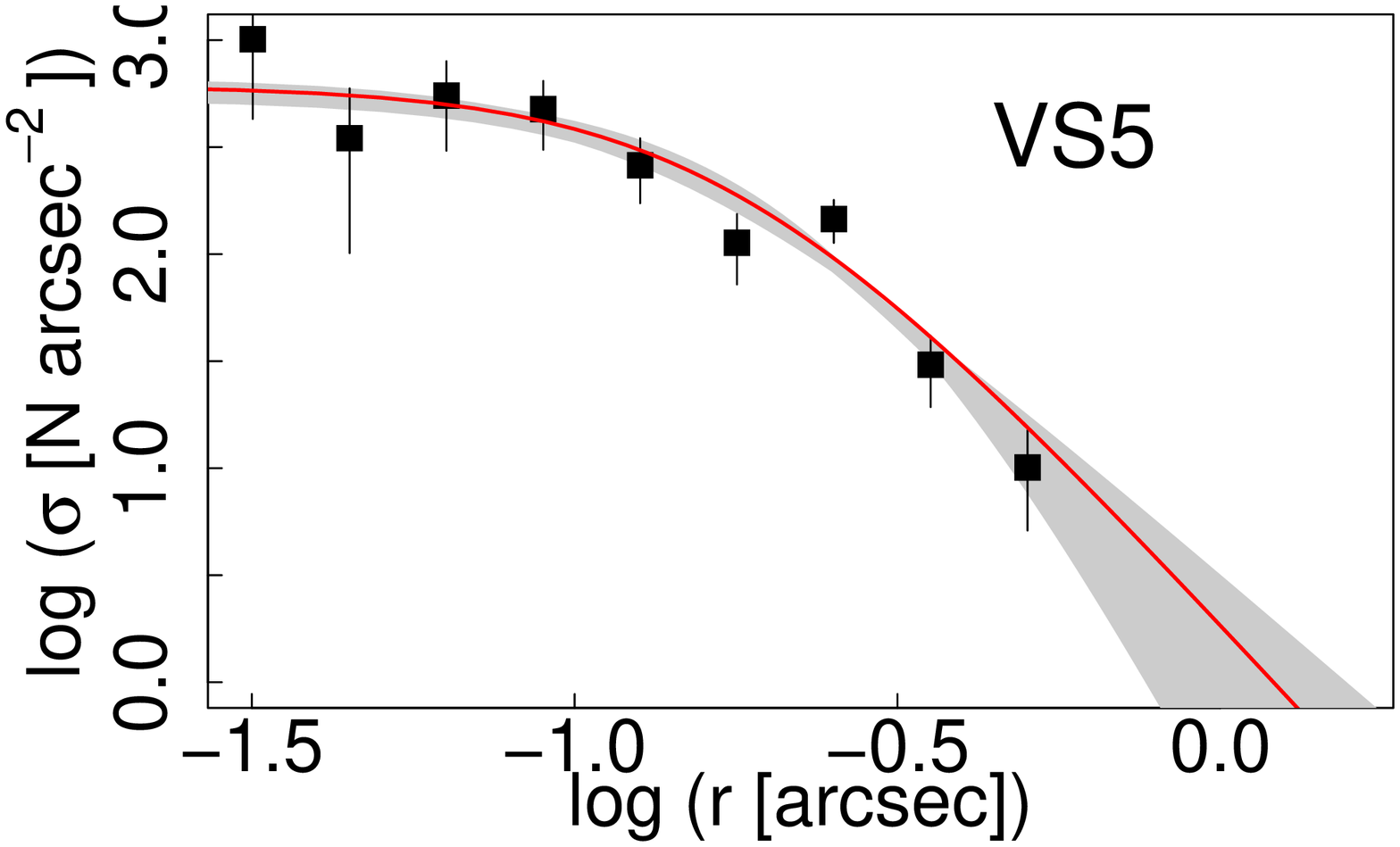}  
\includegraphics[width=55mm]{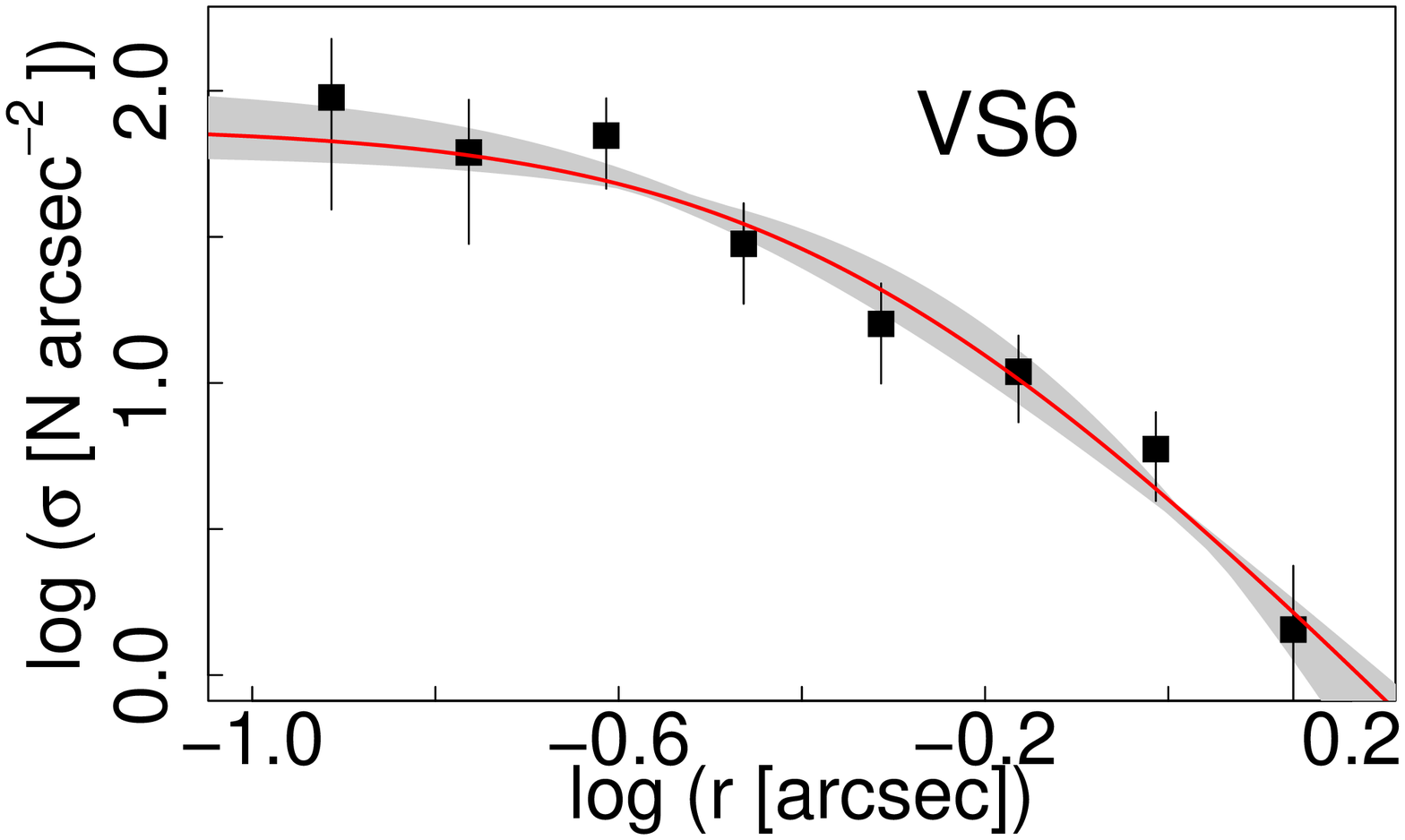}  
\includegraphics[width=55mm]{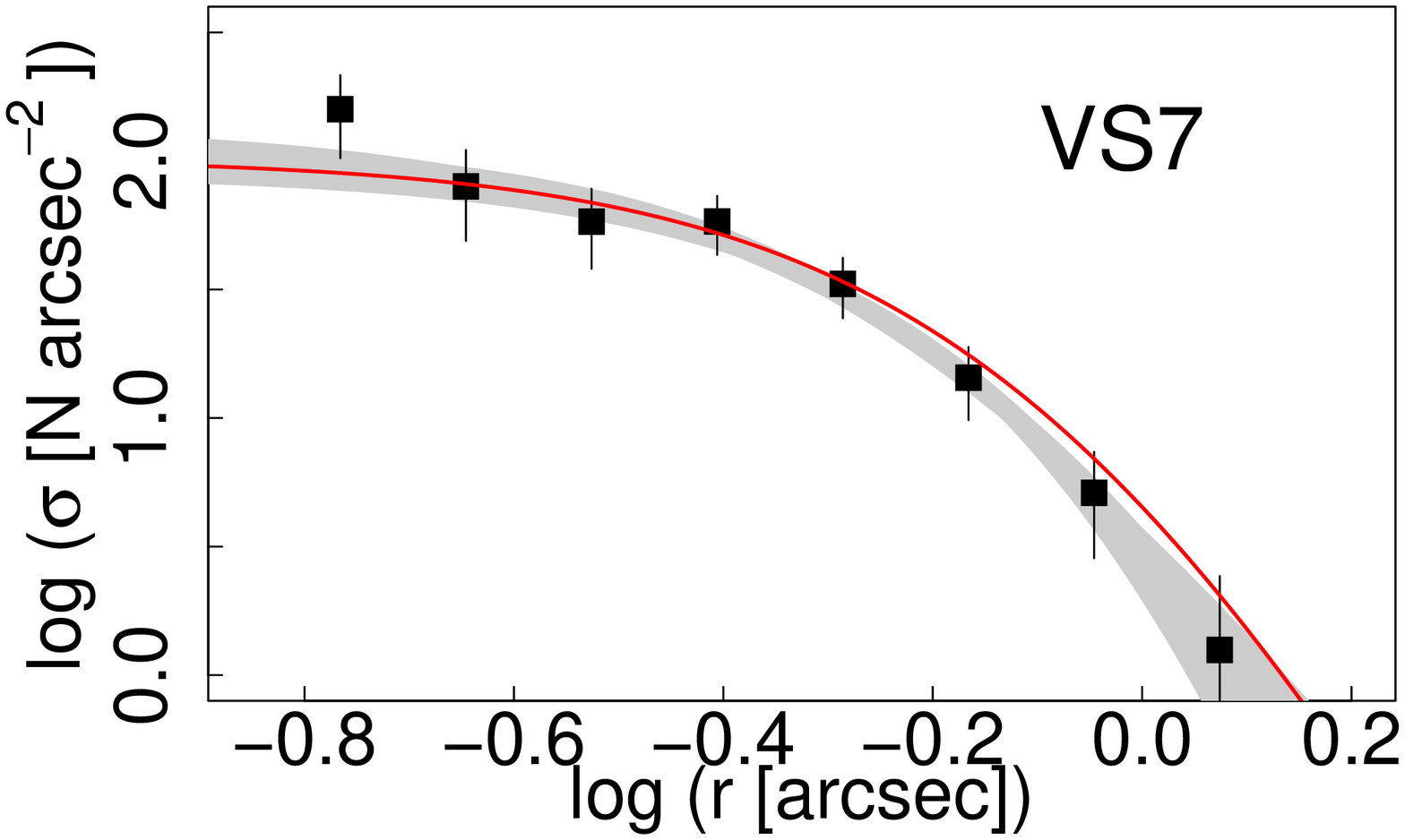}\\  
\contcaption{The stacked GCSs from Virgo dwarf  galaxies are indicated with the acronym VS\#.}    
\label{hubprof2}    
\end{figure*}

\subsection{Radial profiles of globular cluster systems}
\label{sec.radprof}
 In this section, we describe the fits for  the radial projected distributions of our sample of 23 GCSs of intermediate luminosity ETGs from the Virgo, Fornax, and Coma clusters, plus other 4 ones obtained by stacking Virgo dwarfs.

The radial profile of a GCS can be fitted using a number of mathematical expressions, such as a power-law \citep[e.g.][]{esc15,sal15}, de Vaucouleurs law \citep[e.g.][]{fai11}, S\'ersic model, \citep[e.g][]{ush13,kar14} or modified Hubble profile \cite[e.g][]{bin87,bas17}. Following Paper\,I, we use the latter one, which has provided accurate fits for most GCSs

\begin{equation}
n(r) = a \left( 1+ \left(\frac{r}{r_0} \right)^2 \right)^{-b}
\end{equation}

\noindent and behaves as a power-law with an exponent $2\,b$ at large galactocentric distances. In the inner regions, the flattening is  ruled by the core radius $r_0$.  We numerically integrate the modified Hubble profile to calculate the projected effective radius of the GCS ($r_{\rm eff,GCS}$), so that we can compare with studies based on S\'ersic profiles.

 The projected GC density distributions are built
 using concentric circular annuli, and are corrected by differential completeness and contamination, which we explain in the following paragraphs. Regarding the shape of the GCS with respect to the galaxy, \citet{wan13} study the spatial alignment of the GCS hosted by bright galaxies in the Virgo cluster and, for those with noticeable elongation, find that GCs are preferentially aligned along the major axis of the host galaxy. The limited population of GCs in the galaxies from our sample prevents us from obtaining accurate measurements of ellipticity and position angle from the GCSs. In order to constrain possible changes in the results as a consequence of using concentric circular rings instead of elliptical ones, GCSs with elongated spatial distributions are simulated and fitted in the same manner than the observed ones.  We consider four test cases of GCSs, ranging from 150 to 1000 members, the typical populations for the GCSs in the sample fitted in this paper. The parameters for their Hubble profiles, $r_0$, $b$ and $r_{\rm L}$, are obtained from the relations derived in Paper\,I as a function of $N_{\rm GC}$, and six values for ellipticity, ranging from 0 to 0.5, were applied to generate the elongated GCSs. Then, 100  samples are simulated with Monte Carlo for each combination of ellipticity and set of parameters of the radial profile.
The resulting parameters of the radial distribution built through circular annuli are fitted with a Hubble profile, in order to compare them with the original ones. The results indicate mild differences, but negligible in comparison with the typical dispersion due to small numbers statistics, and the intrinsic dispersion found in the literature for the scaling relations (e.g., Paper\,I and references therein). This proves the suitability of using concentric circular annuli, instead of elliptical ones.

In the case of Coma galaxies completeness functions are derived individually (see Section\,\ref{compl}), and the contamination density is obtained from adjacent fields, considering that the spatially extended GCSs associated to both dominant gEs, plus the intra-cluster GC population, represent the largest sources of contamination \citep[see][]{pen11}.
For Virgo and Fornax galaxies, we use model galaxies (NGC\,4621 for Virgo and NGC\,1340 for Fornax) to estimate the completeness, as described in Section\,\ref{datlit}. The contamination is derived from ACS fields from the respective clusters, containing dwarf galaxies with no significant GCSs \citep{pen08,liu19}, and located at comparable projected distance from each cluster central galaxy.

The radial binning is constant on a logarithmic scale for all profiles, with its size varying according to the number of GCs detected in each galaxy, but with a typical value of $\Delta log_{10} r = 0.1$, with $r$ in arcmin. To account for uncertainties caused by noise, the bin breaks were shifted in small amounts ten times, and the final parameters (listed in Table \ref{hubpar2}) result from the weighed mean of the parameters from each individual run.

In the case of the dwarf galaxies, typically fainter than $M_V \approx -18$ and with only a few dozen members in their GCSs, it is not possible to fit individual radial profiles without a significant scatter. Stacking GCSs associated to galaxies with similar luminosity and stellar masses \citep{pen08}, allows us to fit mean radial distributions and minimise the scatter. Four GCSs are obtained in this way, listed in Table\,\ref{hubpar2} and presented in the four last panels of Fig.\,\ref{hubprof}. They are labelled with the acronym VS$\#$ (for ``Virgo stacked'' subsamples) and consecutive numbers starting with VS\,4, as the first three ones have already been presented in Paper\,I. The first stack, VS\,4, corresponds to galaxies VCC\,21, VCC\,1499, VCC\,1539, VCC\,1489, VCC\,1661 and VCC\,230, presenting $V$ absolute magnitudes between $-16.83$ and $-15.61$, and stellar masses ${\rm M_{\star} \approx 0.14-0.69 \times 10^9\,M_{\odot}}$; in the second stack, VS\,5, the  galaxies  VCC\,1833, VCC\,571, VCC\,1075, VCC\,1440, VCC\,1407 and VCC\,1185, present $M_V$ in the range $-17.32$ and $-16.72$, and ${\rm M_{\star} \approx 0.86-1.24 \times 10^9\,M_{\odot}}$; in VS\,6 the stacked galaxies, VCC\,1355, VCC\,1695, VCC\,1545 and VCC\,1828, present $M_V$ in the range $-17.51$ and $-16.78$, and ${\rm M_{\star} \approx 1.26-1.82 \times 10^9\,M_{\odot}}$; and the last group, VS\,7, corresponds to six galaxies with $M_V$ between $-18.98$ and $-17.51$ and ${\rm M_{\star} \approx 2.22-3.68 \times 10^9\,M_{\odot}}$, these are VCC\,2048, VCC\,856, VCC\,140 and VCC\,1861. The stacked galaxies have intermediate values of the parameter $\Sigma_{10}$, ranging from $0.1$ to $1.2$. Their projected distances to M\,87 span $0.5-5.6$\,deg.
Hence, none of the dwarfs used to build these ``stacked'' GC profiles resides in the core nor the outskirts of the Virgo cluster, where different physical processes might lead to a paucity of GCs \citep{pen08}. However, the intrinsic scatter in the environmental parameters of the galaxies in each ``stack'' prevents us from including them in the analysis of environmental dependence.

In Fig.\,\ref{hubprof} we present the Hubble profiles fitted to the sample of 23 GCSs of intermediate luminosity ETGs, plus the 4 ``stacks'' from Virgo dwarfs, corrected by contamination and completeness. The grey regions cover the variations in the obtained function caused by the shifting of the bin breaks, while the red solid curve shows the Hubble modified profile that results from the weighed mean of the parameters.

Apparent magnitudes in several bands, colour excess, distance, parameters obtained from the fits of the GC projected distributions, and other physical properties are listed in Table\,\ref{hubpar2}, making a total of 27 GC profiles. That is, the intermediate luminosity galaxy sample that includes 23 individual ETGs from Virgo (12 galaxies), Fornax (6 galaxies), and Coma (5 galaxies) clusters, plus the 4 ``stacked profiles'' associated to dwarf galaxies in the Virgo cluster. The sample spans absolute magnitudes from $M_B\approx-16.7$ to $M_B\approx-22.4$. Column ${\rm r_L}$ presents the projected extension of the GCS obtained from our profiles, which is calculated as the galactocentric distance for which the contamination-corrected projected density falls to $30$\,per cent of the contamination level. This criterion has been used previously to define the GCS extension \citep[e.g.][]{bas17,cas17,cas19}, as well as in Paper\,I.
Column ${\rm r_{eff,GCS}}$ corresponds to the effective radius of the GCS, which encloses half of the GCs. It is obtained from the numerical integration of the fitted density profile, hence it depends on ${\rm r_L}$, $r_0$ and $b$. The numerical integration of the S\'ersic radial profile up to ${\rm r_L}$ gives as a result the number of GCs more luminous than the completeness limit, i.e. $z_0=24$\,mag for Virgo and Fornax galaxies, and $I_0=26.5$\,mag for Coma ones. The fraction of faint GCs below the completeness limit is calculated from the GC luminosity function (GCLF), leading to the total population of GCs ($N_{\rm GCs}$). For Virgo and Fornax GCSs, the parameters of the GCLF are obtained from \citet{vil10}, the populations derived in this manner agree with those published in the literature \citep[and listed in Table\,\ref{hubpar2}, from][]{pen08,liu19}.
For Coma galaxies, the completeness limit agrees with the turn-over magnitude of the GCSs when assuming that the distance modulus for Coma is $(m-M)=35$\,mag. Then, the doubling of the integrated value leads to $N_{\rm GCs}$. It is worth noting that the extrapolation of radial profiles forced by the limitations of the FOV might lead to uncertainties larger than those estimated in the case of the most extended GCSs in our sample. The last column in Table\,\ref{hubpar2} corresponds the central velocity dispersion of the host galaxy, obtained from the HyperLeda web page\footnote{http://leda.univ-lyon1.fr} \citep{mak14}. 

\subsection{Scaling relations for GCSs}

 This section is devoted to analysing the scaling relations of GCSs in ETGs as a function of several parameters, and particularly the possibility of an environmental dependence already underlying in the relations derived in Paper\,I. The sample contains 100 GCS profiles, including the 27 cases analysed in the previous section, plus 67 systems included in Paper\,I, and 6 newly added galaxies from the literature. The latter ones are listed in Table\,\ref{tab.otros}, which shows their apparent magnitude in several bands, colour excess, distance, parameters from the fits of the GC projected distributions, and central velocity dispersion.

The GCS and the stellar population of the host galaxy are intrinsically related through physical processes that affect both components.

\begin{landscape}
\begin{table}   
\begin{minipage}{220mm}   
\begin{center}   
\caption{Galaxies analysed in this paper, listed in decreasing $B$-band luminosity.  Membership to a galaxy cluster, namely Virgo (V), Fornax (F) or Coma (C). Magnitudes (col. 3 to 6) were obtained from NED and reddening corrections from the recalibration by \citet{sch11}. Distances correspond to SBF measurements listed in NED, typically from \citet{tul13} or \citet{bla09}. The parameters $a$, ${\rm r_0}$ and $b$ (col. 9 to 11) correspond to the modified Hubble profiles fitted to the GCS radial profiles. ${\rm r_L}$, ${\rm r_{eff,GCS}}$ and ${\rm N_{GCs}}$ represent the projected extension of the GCS, its effective radius and the total GC population, respectively, obtained as indicated in the text (Section\,4.3). Central velocity dispersion (${\rm \sigma_0}$) was taken from HyperLeda database.}   
\label{hubpar2}   
\begin{tabular}{@{}lcccccccccccccc@{}}   
\hline
\multicolumn{1}{@{}c}{Name}&\multicolumn{1}{@{}c@{}}{Cluster}&\multicolumn{1}{c}{$B$}&\multicolumn{1}{c}{$V$}&\multicolumn{1}{c}{$J$}&\multicolumn{1}{c}{$K$}&\multicolumn{1}{c}{$E_{(B-V)}$}&\multicolumn{1}{c}{$D$}&\multicolumn{1}{c}{$a$}&\multicolumn{1}{c}{${\rm r_0}$}&\multicolumn{1}{c}{$b$}&\multicolumn{1}{c}{${\rm r_L}$}&\multicolumn{1}{c}{${\rm r_{eff,GCS}}$}&\multicolumn{1}{c}{${\rm N_{GCs}}$}&\multicolumn{1}{c}{${\rm \sigma_0}$}\\   
\multicolumn{1}{@{}c}{}&\multicolumn{1}{c}{}&\multicolumn{1}{c}{mag}&\multicolumn{1}{c}{mag}&\multicolumn{1}{c}{mag}&\multicolumn{1}{c}{mag}&\multicolumn{1}{c}{mag}&\multicolumn{1}{c}{Mpc}&\multicolumn{1}{c}{}&\multicolumn{1}{c}{arcmin}&\multicolumn{1}{c}{}&\multicolumn{1}{c}{kpc}&\multicolumn{1}{c}{kpc}&\multicolumn{1}{c}{}&\multicolumn{1}{c}{km\,s$^{-1}$}\\ 
\hline
NGC\,1404 & F & 10.97 & 10.00 & 7.76 & 6.82 & 0.010 & 22.20 & $2.13\pm0.02$ & $0.50\pm0.05$ & $0.94\pm0.07$ & $96.9\pm 12.7 $ & $20.3\pm 4.2 $ & $311\pm8^2$ & $229.7\pm 3.8 $ \\
NGC\,4526 & V & 10.66 & 9.70 & 7.45 & 6.47 & 0.020 & 16.90 & $1.89\pm0.03$ & $0.71\pm0.08$ & $1.16\pm0.10$ & $51.4\pm 7.8 $ & $9.6\pm 1.9 $ & $388\pm117^1$ & $224.6\pm 9.4 $ \\
NGC\,1380 & F & 10.87 & 9.93 & 7.77 & 6.86 & 0.015 & 17.60 & $2.29\pm0.02$ & $0.66\pm0.05$ & $1.36\pm0.08$ & $41.2\pm 3.3 $ & $6.3\pm 0.4 $ & $504\pm77^2$ & $214.5\pm 4.6 $ \\
IC\,4045 & C & 14.96 & 13.94 & 11.85 & 10.90 & 0.011 & 115.00 & $3.29\pm0.08$ & $0.12\pm0.02$ & $1.45\pm0.08$ & $26.8\pm 2.3$ & $6.0\pm 0.7$ & $161\pm11^3$ & $216.5\pm3.8$\\
NGC\,4906 & C & 15.09 & 14.10 & 12.15 & 11.24 & 0.011 & 115.00 & $3.50\pm0.10$ & $0.10\pm0.01$ & $1.41\pm0.06$ & $28.1\pm 1.3$ & $5.3\pm 0.3$ & $200\pm11^3$ & $168.8\pm4.2$\\
IC\,4041 & C & 15.30 & 14.35 & 12.58 & 11.66 & 0.010 & 115.00 & $2.96\pm0.08$ & $0.09\pm0.02$ & $1.51\pm0.15$ & $13.0\pm 2.0$ & $4.0\pm 0.7$ & $36\pm11^3$ & $132.7\pm3.3$\\
NGC\,1387 & F & 11.68 & 10.69 & 8.44 & 7.52 & 0.011 & 19.90 & $2.06\pm0.03$ & $0.81\pm0.09$ & $1.34\pm0.11$ & $47.8\pm 12.6 $ & $8.6\pm 1.8 $ & $299\pm31^2$ & $167.3\pm 11.6 $ \\
NGC\,4459 & V & 11.32 & 10.37 & 8.10 & 7.15 & 0.040 & 16.10 & $1.99\pm0.03$ & $0.60\pm0.09$ & $1.32\pm0.13$ & $32.1\pm 10.1 $ & $5.5\pm 1.4 $ & $218\pm28^1$ & $171.8\pm 4.8 $ \\
NGC\,4442 & V & 11.38 & 10.44 & 8.21 & 7.29 & 0.020 & 15.30 & $1.98\pm0.04$ & $0.44\pm0.10$ & $1.04\pm0.15$ & $42.7\pm 17.1 $ & $8.2\pm 4.1 $ & $178\pm30^1$ & $179.2\pm 3.9 $ \\
NGC\,4435 & V & 11.74 & 10.80 & 8.42 & 7.30 & 0.026 & 16.70 & $2.21\pm0.04$ & $0.24\pm0.03$ & $0.88\pm0.04$ & $61.7\pm 11.4$ & $13.6\pm 3.3 $ & $345\pm80^1$ & $155.0\pm 4.0 $ \\
NGC\,4371 & V & 11.79 & 10.81 & 8.60 & 7.70 & 0.032 & 17.00 & $3.02\pm0.07$ & $0.14\pm0.04$ & $0.82\pm0.05$ & $57.9\pm 14.1 $ & $9.9\pm 4.7 $ & $200\pm41^1$ & $128.7\pm 2.2 $ \\
IC\,2006 & F & 12.21 & 11.29 & 9.40 & 8.48 & 0.010 & 20.30 & $2.09\pm0.04$ & $0.38\pm0.05$ & $1.20\pm0.10$ & $30.7\pm 7.2 $ & $5.5\pm 1.6 $ & $122\pm6^2$ & $120.6\pm 1.9 $ \\
NGC\,4570 & V & 11.84 & 10.90 & 8.60 & 7.70 & 0.019 & 17.10 & $2.03\pm0.05$ & $0.26\pm0.05$ & $0.96\pm0.08$ & $39.3\pm 9.2 $ & $8.0\pm 2.7 $ & $139\pm23^1$ & $186.5\pm 5.2 $ \\
NGC\,4267 & V & 11.86 & 10.93 & 8.72 & 7.84 & 0.042 & 15.80 & $1.68\pm0.04$ & $1.13\pm0.25$ & $1.73\pm0.42$ & $26.7\pm 6.3 $ & $6.0\pm 0.8 $ & $179\pm17^1$ & $149.6\pm 4.6 $ \\
NGC\,4417 & V & 12.00 & 11.10 & 9.05 & 8.17 & 0.022 & 16.00 & $1.65\pm0.04$ & $0.59\pm0.10$ & $1.44\pm0.18$ & $19.6\pm 5.6 $ & $4.2\pm 1.0 $ & $72\pm10^1$ & $134.9\pm 2.4 $ \\
IC\,4042 & C & 15.28 & 14.27 & 11.79 & 10.76 & 0.011 & 99.10 & $2.58\pm0.03$ & $0.20\pm0.03$ & $1.58\pm0.10$ & $16.1\pm 0.9$ & $6.0\pm 0.3$ & $59\pm11^3$ & $150.6\pm14.1$\\
IC\,4042A & C & 16.30 & 15.34 & 12.90 & 12.10 & 0.011 & 103.00 & $2.81\pm0.08$ & $0.13\pm0.02$ & $1.60\pm0.15$ & $13.8\pm 1.5$ & $4.5\pm 0.6$ & $45\pm10^3$ & -- \\
NGC\,4474 & V & 12.38 & 11.50 & 9.62 & 8.70 & 0.037 & 15.60 & $2.06\pm0.03$ & $0.38\pm0.04$ & $1.39\pm0.10$ & $18.2\pm 4.0 $ & $2.9\pm 0.4 $ & $116\pm24^1$ & $89.8\pm 5.2 $ \\
NGC\,4377 & V & 12.76 & 11.88 & 9.73 & 8.83 & 0.033 & 17.80 & $1.97\pm0.04$ & $0.43\pm0.06$ & $1.92\pm0.22$ & $11.4\pm 1.1 $ & $2.2\pm 0.2 $ & $74\pm32^1$ & $128.4\pm 5.5 $ \\
NGC\,1380A & F & 13.31 & 12.41 & 10.52 & 9.57 & 0.013 & 20.00 & $1.85\pm0.04$ & $0.34\pm0.05$ & $1.09\pm0.09$ & $28.0\pm 6.5 $ & $6.1\pm 2.1 $ & $67\pm9^2$ & $52.6\pm 2.8 $ \\
NGC\,4483 & V & 13.14 & 12.23 & 10.11 & 9.29 & 0.017 & 16.70 & $2.04\pm0.05$ & $0.24\pm0.04$ & $1.34\pm0.12$ & $13.3\pm 2.3 $ & $2.2\pm 0.1 $ & $72\pm18^1$ & $100.0\pm 9.6 $ \\
NGC\,4352 & V & 13.49 & 12.58 & 10.63 & 9.87 & 0.023 & 18.70 & $2.06\pm0.04$ & $0.48\pm0.07$ & $1.68\pm0.16$ & $18.3\pm 3.1 $ & $3.2\pm 0.1 $ & $114\pm12^1$ & $67.6\pm 5.8 $ \\
FCC\,255 & F & 13.86 & -- & 11.33 & 10.43 & 0.006 & 20.00 & $2.17\pm0.15$ & $0.16\pm0.05$ & $0.96\pm0.10$ & $27.3\pm 6.2 $ & $5.2\pm 2.0 $ & $69\pm9^2$ & $46.0\pm 7.5 $ \\
VS\,4$^{a}$ & V & -- & -- & -- & -- & -- & 16.80 & $2.72\pm0.06$ & $0.23\pm0.03$ & $2.03\pm0.21$ & $5.3\pm 2.2 $ & $1.1\pm 0.1 $ & $20\pm7^3$ & -- \\
VS\,5$^{a}$ & V & -- & -- & -- & -- & -- & 16.80 & $2.77\pm0.05$ & $0.14\pm0.02$ & $1.35\pm0.13$ & $7.4\pm 3.3 $ & $1.3\pm 0.2 $ & $23\pm9^3$ & -- \\
VS\,6$^{a}$ & V & -- & -- & -- & -- & -- & 16.80 & $1.92\pm0.05$ & $0.36\pm0.05$ & $1.43\pm0.13$ & $9.4\pm 2.7 $ & $2.5\pm 0.2 $ & $25\pm9^3$ & -- \\
VS\,7$^{a}$ & V & -- & -- & -- & -- & -- & 16.80 & $2.01\pm0.03$ & $0.90\pm0.13$ & $3.92\pm0.75$ & $7.2\pm 1.0 $ & $2.2\pm 0.1 $ & $35\pm8^3$ & -- \\
\hline
\end{tabular}
\end{center}
$^{\rm a}$ Stacked galaxies from the Virgo cluster.
In these cases, distance of the Virgo cluster is assumed as a representative distance (see text for further details).\\
{\bf References:} $^1$\citet{pen08}, $^2$\citet{liu19}, $^3$This paper, 
\end{minipage}   
\end{table}   
\end{landscape}

 \begin{table*} 
\begin{minipage}{180mm}   
\begin{center}   
  \caption{Galaxies from the literature, listed in decreasing $B$-band luminosity. Magnitudes (col. 2 to 5) are obtained from NED and reddening corrections from the recalibration by \citet{sch11}. Distance correspond to SBF measurements listed in NED, typically from \citet{tul13}. Parameter $b$ corresponds to the exponent of the Hubble modified profile (analogue to half of the power-law slope). ${\rm r_L}$,  ${\rm r_{eff,GCS}}$ and ${\rm N_{GCs}}$ represent the projected extension of the GCS, its effective radius and the total population of GCs, respectively. Central velocity dispersion (${\rm \sigma_0}$) was taken from the HyperLeda database.}    
\label{tab.otros}   
\begin{tabular}{@{}l@{}c@{}c@{}c@{}c@{}c@{}c@{}cc@{}c@{}cc@{}}
\hline   
\multicolumn{1}{@{}c}{Name}&\multicolumn{1}{c}{$B$}&\multicolumn{1}{c}{$V$}&\multicolumn{1}{c}{$J$}&\multicolumn{1}{c}{$K$}&\multicolumn{1}{c}{$E_{(B-V)}$}&\multicolumn{1}{c}{$D$}&\multicolumn{1}{c}{$b$}&\multicolumn{1}{c}{${\rm r_L}$}&\multicolumn{1}{c}{${\rm r_{eff,GCS}}$}&\multicolumn{1}{@{}c}{${\rm N_{GCs}}$}&\multicolumn{1}{c@{}}{${\rm \sigma_0}$}\\   
 &\multicolumn{1}{c}{mag}&\multicolumn{1}{c}{mag}&\multicolumn{1}{c}{mag}&\multicolumn{1}{c}{mag}&\multicolumn{1}{c}{mag}&\multicolumn{1}{c}{Mpc}& &\multicolumn{1}{c}{kpc}&\multicolumn{1}{c}{kpc}& &\multicolumn{1}{c@{}}{km\,s$^{-1}$}\\   
\hline   
NGC\,4874 & 12.63 & 11.68 & 9.85 & 8.86 & 0.008 & 99.50 & $--$ & $200.0^1$ & $67.7^1$ & $23000\pm700^1$ & $271.9\pm 4.3 $ \\
NGC\,1316 & 9.42 & 8.53 & 6.44 & 5.59 & 0.019 & 20.80 & $--$ & $78.6^2$ & $--$ & $1500\pm--^2$ & $223.1\pm 3.3 $ \\
NGC\,6876 & 11.76 & 10.80 & 8.70 & 7.70 & 0.039 & 50.90 & $0.90\pm0.10^3$ & $125.0^3$ & $30.5^3$ & $9500\pm2500^3$ & $233.3\pm 16.1 $\\
NGC\,3610 & 11.70 & 10.84 & 8.84 & 7.91 & 0.009 & 34.80 & $1.42\pm0.08^4$ & $40.5^4$ & $9.6^4$ & $500\pm110^4$ & $168.3\pm 3.3 $\\
NGC\,3613 & 11.82 & 10.89 & 8.93 & 8.00 & 0.011 & 30.10 & $1.15\pm0.19^5$ & $70.9^5$ & $17.2^5$ & $2075\pm130^5$ & $212.5\pm 4.3 $ \\
NGC\,4546 & 11.30 & 10.32 & 8.31 & 7.39 & 0.029 & 14.00 & $--$ & $50.1^6$ & $3.2^6$ & $390\pm60^6$ & $195.9\pm 5.4 $ \\
\\
\hline
\end{tabular}
\end{center}
{\bf References:} $^1$\citet{pen11}, $^2$\citet{ric12a}, $^3$\citet{enn19}, $^4$\citet{bas17}, $^5$\citet{debo20}, $^6$\citet{esc20} 
\end{minipage}   
\end{table*}

 Then, analysing scaling relations for several GCS properties as a function of the stellar mass of the host galaxy (${\rm M_{\star}}$) becomes a natural step. This latter parameter is calculated, for all  galaxies in the sample, as the mean of the values derived from the luminosity in $J$ and $K$ bands, using the mass-to-light ratios ($M/L$) from \citet{bel03} and the $(B-V)$ colours, adopting a Salpeter initial mass function (see  Tables\,\ref{hubpar2}, \ref{tab.otros}, and Paper\,I). In the case of the stacked GCSs from Virgo dwarfs, we use as stellar mass the average of the masses of the dwarfs included in each stack, and the distance of the Virgo cluster as a representative distance.

\begin{figure}
\includegraphics[width=85mm]{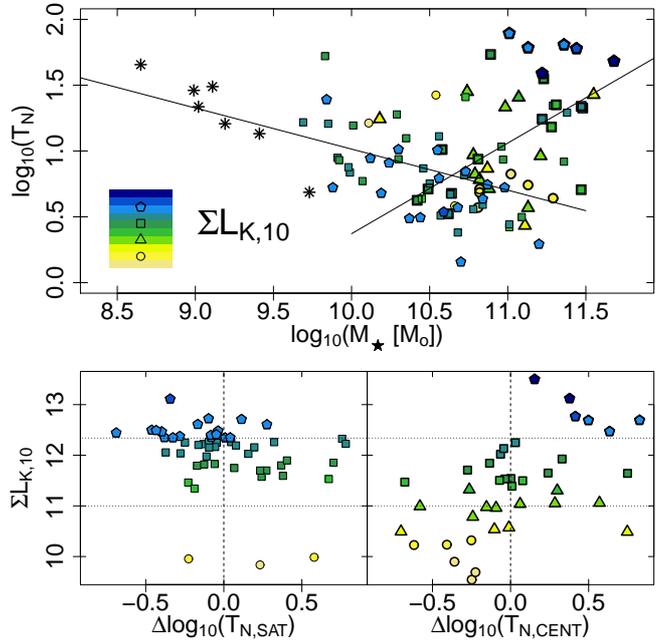}\\    
\caption{
{\bf Upper panel:} Richness of the GCS, represented by the logarithm of the parameter ${\rm T_N}$, as a function of the logarithm of the ${\rm M_{\star}}$ of the host galaxy. Different symbols and colour gradient represent increasing values of the environmental density parameter $\Sigma L_{K,10}$, from yellow circles to blue pentagons. Framed symbols highlight central galaxies and asterisks correspond to the stacked GCS profiles of low-mass galaxies from the Virgo cluster. The solid lines correspond to  linear fits to satellite and central galaxies. {\bf Lower panels:} Environmental density $\Sigma L_{K,10}$ as a function of residuals of the linear relations for satellite (left panel) and central galaxies (right panel). The dashed vertical lines represent null residuals, and the horizontal dotted lines correspond to values of $\Sigma L_{K,10}$ used to split the sample for statistical analysis (see Section\,\ref{sec.tn}).}
\label{mestTN}
\end{figure}

\subsubsection{Richness of the GCSs}
\label{sec.tn}

 We have stated in the Introduction that several studies in the literature point to a paucity of GCs in satellite galaxies located in dense environments, as well as in bright galaxies in the field. Although our sample was built to analyse the scaling relations derived from the GCS radial profiles and this restricts its size at the low mass regime, it is straightforward to present the richness of the GCSs in our sample versus the stellar mass of the host galaxy. The upper panel in Fig.\,\ref{mestTN} shows the richness of the GCSs represented by the logarithm of the parameter ${\rm T_N}$, as a function of the logarithm of ${\rm M_{\star}}$. \citet{zep93} define the parameter ${\rm T_N}$ as the ratio between the number of GCs (${\rm N_{GCs}}$) and the stellar mass of the host galaxy, in units of $10^9\,{\rm M_{\odot}}$. The different symbols and colour gradients represent increasing values of the density parameter $\Sigma L_{K,10}$ (see Section\,\ref{sec.dens}), from yellow circles to blue pentagons. Framed symbols highlight central galaxies and asterisks correspond to the stacked profiles of low-mass galaxies from the Virgo cluster. The evolution for satellites and centrals is distinctive, the parameter ${\rm T_N}$ increases for larger ${\rm M_{\star}}$ in the latter ones, but is inversely proportional for satellites. As expected, a Kendall test \citep{ken38} reveals significant correlation at the 95\,per cent confidence for satellites and centrals separately, but results are not conclusive for the entire sample. The solid lines represent linear fits to the satellite and central samples separately, that result  in slopes of $-0.31\pm0.07$ and $0.7\pm0.15$, respectively. This change of trend leads to a minimum richness of GCSs that corresponds to galaxies with ${\rm M_{\star}}\approx 5\times 10^{10}\,{\rm M_{\odot}}$. \citet{har13} indicate that such behaviour is due to the increasing efficiency of the star formation in galaxies that reaches a maximum at about that stellar mass \citep[e.g.][]{leg19}. 
 
 On the other hand, in central galaxies the merger history also plays a relevant role in the increase of ${\rm T_N}$. It also agrees with the pivot mass for several scaling relations introduced in Paper\,I and revisited in this Section. The lower left panel in Fig.\,\ref{mestTN} shows the density parameter $\Sigma L_{K,10}$ for satellites, as a function of the residuals from the bilinear fit in the upper panel. The vertical dashed line corresponds to null residuals, and is included for comparison purposes. The colour gradient and symbols represents increasing values for the parameter $\Sigma L_{K,10}$, as already explained. The number of satellites below $\Sigma L_{K,10}=11$ is negligible (NGC\,7332, NGC\,1400 and NGC\,7457), and we focus on density ranges above this limit. Satellites in denser environments present mainly negative residuals, on the contrary to satellites in intermediate density environments. We select $\Sigma L_{K,10}=12.3$ to separate both samples, with the horizontal dotted lines representing those limits in the corresponding panel. The samples skewness is calculated from the adjusted Fisher–Pearson standardised moment \citep{joa98}, resulting $5.9\pm0.4$ and $-1.5\pm0.5$ for the samples with the intermediate and large values of density parameters, respectively. A commonly used criterion to recognise skewed samples is to calculate the ratio between the skewness and its error, assuming that it follows a normal distribution \citep[e.g.][]{cra97}. Then, absolute values for this ratio larger than 2 correspond to a 95\,per cent confidence level, and indicate that the sample is significantly skewed. In both cases, the criterion is fulfilled, pointing that satellites in very dense environments present poorer GCSs than their analogues in intermediate environments, probably as a consequence of stripping processes. This result supports previous statements for particular clusters by \citet{pen08} and \citet{liu19}. 
 
  The right lower panel is analogue, but for central galaxies. The symbols and colour gradient follow the same coding than previous panels. Unlike satellites, the centrals present negative skewness, $-6.2\pm0.8$, for galaxies below $\Sigma L_{K,10}=11$, and $5.8\pm0.4$ for galaxies in intermediate density environments. The centrals in the denser environments also present positive residuals, although they are just a few.  The correlation between richness and environmental density is also present comparing $\Sigma L_{K,10}$ versus the logarithm of ${\rm T_N}$. In this case, the Kendall test leads to significant correlation at the 99\,per cent. These point to the relevance of the environment in the build up of rich GCSs for central galaxies. Central galaxies in dense environments are supposed to have experienced a rich merger history in comparison with galaxies in the field, besides secular processes like tidal stripping of GCs from satellites.

\begin{figure}
\includegraphics[width=85mm]{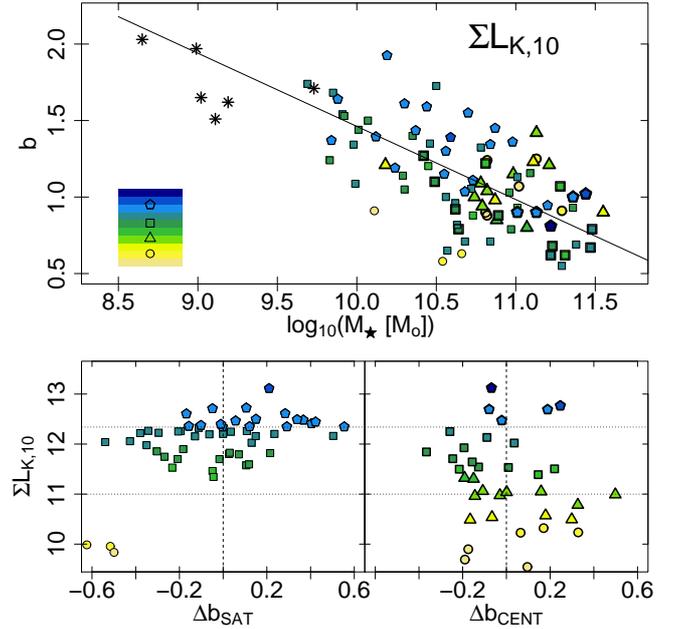}\\ 
\caption{
{\bf Upper panel:} Exponent $b$ of the modified Hubble profile fit to the GCS, as a function of the logarithm of the host stellar mass (${\rm M_{\star}}$). The colour gradient and the different symbols represent increasing values of the density parameter $\Sigma L_{K,10}$, from yellow circles to blue pentagons. The solid line corresponds to a linear fit to the data (Equation\,\ref{eq:mb}). Framed symbols highlight central galaxies and asterisks correspond to the stacked low-mass galaxies from the Virgo cluster. {\bf Lower panels:} Environmental density $\Sigma L_{K,10}$ versus  residuals from the linear fit for satellites (left panel) and central galaxies (right panel). Both panels follow the same symbol coding as the upper panel. The dashed vertical lines represent null residuals, and the horizontal dotted lines correspond to values of $\Sigma L_{K,10}$ used to split the sample for statistical analysis (see Section\,\ref{sec.b}).}
\label{mestb}
\end{figure}

\subsubsection{The exponent $b$ of the modified Hubble profile}
\label{sec.b}

The environmental processes that affect the halo of satellite galaxies are also supposed to have effects on the radial distribution of GCs, which have proven to be useful as a tracer population of the halo kinematics \citep[e.g.][]{sch12,ric15}. The exponent $b$ of the modified Hubble profile (or alternatively, the power-law profile) provides a direct estimation of a GCS compactness. It is available for a large number of GCSs, making the comparison easier. The upper panel in Fig.\,\ref{mestb} shows the $b$ parameter as a function of the logarithm of ${\rm M_{\star}}$. The colour gradient and symbols represent the same as in Fig.\,\ref{mestTN}.
This enlarged sample confirms our results from Paper\,I, where the $b$ parameter inversely correlates with ${\rm M_{\star}}$. Less massive galaxies present steeper radial distributions than their giant counterparts. The solid line corresponds to a linear fit to the data, leading to

\begin{equation}
b = 6.2\pm0.6 -0.48\pm0.06 \times X
\label{eq:mb}
\end{equation}

\noindent with ${\rm X}$ being ${\rm log_{10}(M_{\star})}$. The lower left panel in Fig.\,\ref{mestb} shows the density parameter $\Sigma L_{K,10}$ versus the residuals of the linear fit for satellites. The symbols and colour gradient follow the same symbol coding as the upper panel. The dashed line represents null residuals and is included for comparison purposes. The horizontal dotted lines represent values of the density parameter $\Sigma L_{K,10}=11$ and $\Sigma L_{K,10}=12.3$, used to split the sample in intermediate and dense environments. The skewness estimator for these groups result  $-0.8\pm0.4$ and $1.6\pm0.5$, respectively. In both cases, the criterion based on the ratio between the estimator and its error (see Section\,\ref{sec.tn}) hints at skewed samples. To support differences between the sample of satellites in intermediate and dense environments, a Mann-Whitney-Wilcoxon test \citep[][hereafter MWW test]{man47} is applied to the residuals of both groups, showing differences in their  distributions at the 95\,per cent confidence. These results point to GCSs in dense environments being steeper, probably due to processes of stripping that affect both, the halo of  galaxies and the populations residing in them.

The lower right panel in Fig.\,\ref{mestb} is analogue, but for central galaxies. In this case, galaxies in the range $\Sigma L_{K,10}<11$ present mainly positive residuals, and skewness estimator $1.4\pm0.6$. For galaxies in the range $11< \Sigma L_{K,10}<12.3$, this estimator has a value of $-1.1\pm0.5$. In both cases, these values indicate skewed samples. A MWW test results in differences at the 95\,per cent confidence between these samples, that roughly represent field/isolated galaxies, and central galaxies in groups. The number of central galaxies in dense environments is small, which prevents us from performing a statistical analysis. The results for central galaxies suggest that galaxies in the field present steeper GCSs than those in groups/clusters. This can be interpreted in the context of the two-phases scenario for GCS formation \citep{for11}, with galaxies in sparser environments lacking satellites to supply GCs for their outer haloes.

\subsubsection{Extension of the GCSs}
\label{sec.rL}

\begin{figure}    
\includegraphics[width=85mm]{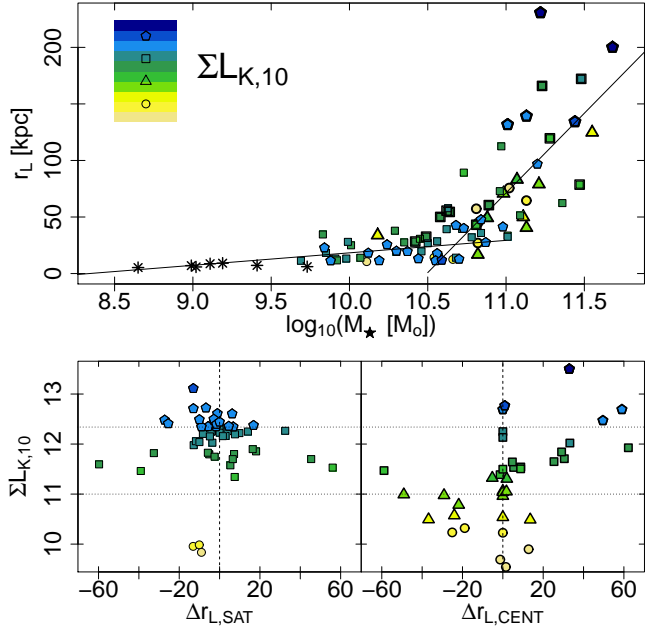}\\    
\caption{
{\bf Upper panel:} Extension of the GCS, as a function of the logarithm of ${\rm M_{\star}}$, with different symbols and colour gradient representing increasing values of the environmental parameter $\Sigma L_{K,10}$, from yellow circles to blue pentagons. Solid lines represent the bilinear relation fitted to the data (Equation\,\ref{eq:mrl}), in agreement with Paper\,I. Framed symbols discriminate central galaxies and asterisks correspond to the stacked GC profiles of low-mass galaxies from the Virgo cluster. {\bf Lower panels:} Environmental density $\Sigma L_{K,10}$ versus residuals from the bilinear relation for satellite (left panel) and central galaxies (right panel). Both panels follow the same symbol coding than the upper panel. The horizontal dotted lines represent ranges of $\Sigma L_{K,10}$ used to split the samples for statistical analysis.}
\label{MrL}
\end{figure}

The stripping of loosely bound GCs from satellites due to the interaction with the central galaxy is expected to affect the extension of their GCS. This is supported by the study of GCSs of satellites close to massive galaxies \citep[e.g.][]{bas06b,weh08}, but also from the existence of spatially extended intra-cluster components in dense environments \citep[e.g.][]{lon18,mad18}. The upper panel in Fig.\,\ref{MrL} shows the projected extension of the GCS in kpc (${\rm r_L}$) as a function of the logarithm of ${\rm M_{\star}}$ of the host galaxy. The symbols and colour gradient follow the same coding as previous figures.

It was already shown in Paper\,I that a different behaviour is noticeable for GCSs across the whole range of stellar mass  of host galaxies, presenting a pivot mass at ${\rm M_{\star}}\approx 5 \times 10^{10}\,{\rm M_{\odot}}$. Moreover, the extension of the GCS presents a larger dispersion for massive galaxies, probably due to the complexity in determining its value for extended GCSs that usually exceed their FOV. 

Following the results from Paper\,I, we fit a bilinear relation of the form:

\begin{eqnarray}
\left.\begin{aligned}
{\rm r_L} =&-92\pm20 + 11\pm2.5 \times {\rm X}, & {\rm M_{\star} \lesssim 5\times 10^{10}\,M_{\odot}}\\
 &-1500\pm275 + 141\pm25 \times {\rm X}, & {\rm M_{\star} \gtrsim 5\times 10^{10}\,M_{\odot}}
\end{aligned}\right.
\label{eq:mrl}
\end{eqnarray}

\noindent with X representing ${\rm log_{10}(M_{\star})}$. These relations are in agreement with those derived in Paper\,I for a smaller sample. They also agree with the results from \citet{kar14} for massive galaxies, if the differences in ${\rm M_{\star}}$ from the relations by \citet{zep93} and \citet{bel03} are taken into account \citep[see][]{kar16}. The lower left panel shows the parameter $\Sigma L_{K,10}$ for satellites, as a function of residuals from the bilinear relation. The symbols and colour gradient follow the same coding than in the upper panel. The horizontal dotted lines show the two selected samples, $11<\Sigma L_{K,10}<12.3$ and $\Sigma L_{K,10}>12.3$. The skewness estimator for them are $-1.5\pm0.5$ and $0.5\pm0.4$, respectively. Then, from the criterion based on the ratio between the skewness and its error (see Section\,\ref{sec.tn}), only the sample from dense environments is significantly skewed. A MWW test shows differences between both samples at the 90\,per cent confidence. The lower right panel is analogue for central galaxies. The skewness estimator results $-1.5\pm0.7$ for galaxies with $\Sigma L_{K,10}<11$, and $1.9\pm0.6$ for those presenting $11<\Sigma L_{K,10}<12.3$. In both cases, these values lead to skewed samples. The MWW test suggests differences at the 99\,per cent confidence. The results in the current section are in agreement with those from Section\,\ref{sec.b}, with satellites in dense environment presenting less extended systems, probably due to large fractions of mass loss, that also affect the extension of the GCSs. The scenario for central galaxies points to the relevance of the merging history to build up extended (and populated) GCSs, with galaxies in the field presenting less extended GCSs.

\begin{figure}    
\includegraphics[width=85mm]{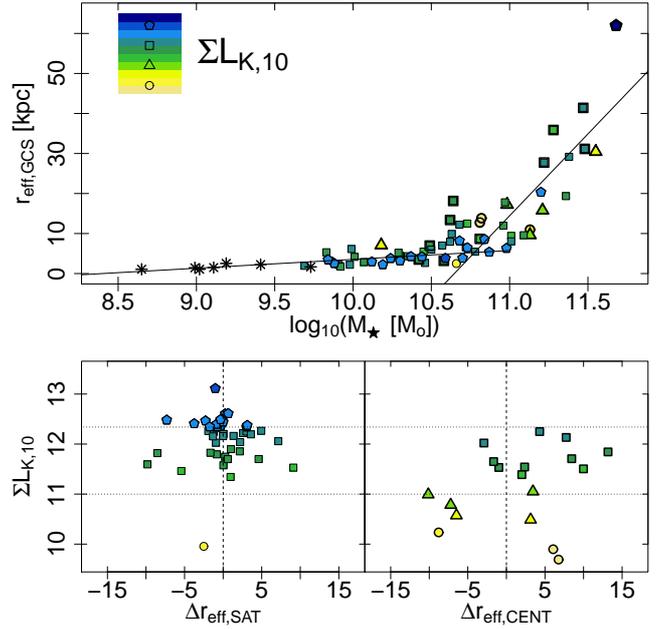}\\    
\caption{
{\bf Upper panel:} Effective radius of the GCS (${\rm r_{eff,GCS}}$) as a function of the logarithm of ${\rm M_{\star}}$. The different symbols and colour gradient represent increasing values of the environmental parameter $\Sigma L_{K,10}$,  from yellow circles to blue pentagons. Solid lines represent the bilinear relation fitted to the data (Equation\,\ref{eq:mreff}). Framed symbols discriminate central galaxies and asterisks correspond to the stacked low-mass galaxies from the Virgo cluster. {\bf Lower panels:} show the parameter $\Sigma L_{K,10}$ versus residuals from the bilinear relation for satellites (left panel) and centrals (right panel). Both panels follow the same symbol coding than the upper one. The horizontal dotted lines represent ranges of $\Sigma L_{K,10}$ used to split the samples for statistical analysis.}
\label{Mreff}
\end{figure}  

\subsubsection{Effective radius of the GCSs}
\label{sec.reff}

The effective radius of the GCS (${\rm r_{eff,GCS}}$) is commonly related to the size of the GCS. The upper panel in Fig.\,\ref{Mreff} shows ${\rm r_{eff,GCS}}$ versus the logarithm of ${\rm M_{\star}}$. The symbols and colour gradient follow the same coding as previous figures.

The behaviour of ${\rm r_{eff,GCS}}$ with ${\rm M_{\star}}$ is similar to that found for ${\rm r_L}$ in the previous subsection, with a break in the relation at a host stellar mass of $\approx 5\times 10^{10}\,{\rm M_{\odot}}$, becoming steeper for more massive galaxies. In comparison, ${\rm r_{eff,GCS}}$ presents a lower dispersion, as it is directly obtained by fitting S\'ersic profiles and more accurately calculated than the extension of  GCSs in massive galaxies. It is worth emphasising that the number of systems with ${\rm r_{eff,GCS}}$ measurements is smaller than those with estimations of ${\rm r_L}$, particularly for massive galaxies. Solid lines correspond to the bilinear relation fitted to the entire sample:

\begin{eqnarray}
\left.\begin{aligned}
{\rm r_{eff,CGS}} =&-18\pm4.7 + 2.2\pm0.5 \times {\rm X}, & {\rm M_{\star} \lesssim 5\times 10^{10}\,M_{\odot}}\\
 &-435\pm57 + 40.9\pm5.2 \times {\rm X}, & {\rm M_{\star} \gtrsim 5\times 10^{10}\,M_{\odot}}
\end{aligned}\right.
\label{eq:mreff}
\end{eqnarray}

\noindent with $X$ being the $log_{10}({\rm M_{\star}})$. The lower left panel presents the parameter $\Sigma L_{K,10}$ as a function of residuals from the bilinear fit for satellites. The symbols and colour range follow the same symbol coding as the upper panel. The horizontal dotted lines separate satellites with $11<\Sigma L_{K,10}<12.3$ and $\Sigma L_{K,10}>12.3$. The skewness estimator for these two groups are $0.2\pm0.4$ and $-1.8\pm0.6$, respectively. They indicate that satellites from dense environments are a skewed sample, based on the criterion explained in Section\,\ref{sec.tn}. A MWW test supports significant differences for both distributions at the 90\,\,per cent confidence. The lower right panel is analogue for central galaxies. In this case, the size of the sample prevents us from going further in the analysis, although central galaxies in the range $11<\Sigma L_{K,10}<12.3$ seem to be positive skewed. These results show that satellites in high density environments present smaller ${\rm r_{eff,GCS}}$ at fixed stellar mass, being consistent with the effects of environmental processes already stated in previous sections.

\begin{figure}    
\includegraphics[width=85mm]{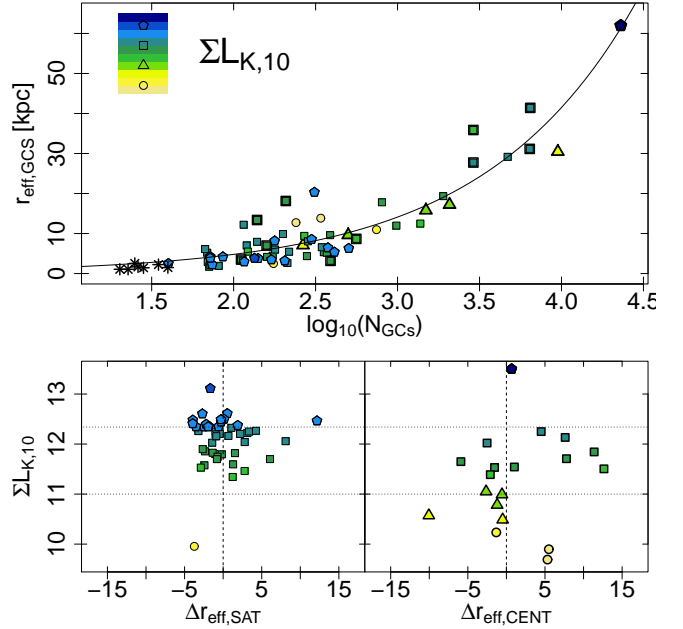}
\centering
\caption{{\bf Upper panel:} Effective radius of the GCS (${\rm r_{eff,GCS}}$), as a function of the logarithm of the population of GCs (${\rm N_{GCs}}$). The different symbols and colour gradient represent increasing values of the environmental parameter $\Sigma L_{K,10}$, from yellow circles to blue pentagons. Framed symbols discriminate central galaxies and asterisks correspond to the stacked GC profiles for Virgo dwarfs. Solid curve represents a power-law fitted to the data (Equation\,\ref{eq:reffngc}). {\bf Lower panels:} parameter $\Sigma L_{K,10}$ versus  residuals from the power-law fit for satellites (left panel) and centrals (right panel). Both panels follow the same symbol coding than the upper one. The horizontal dotted lines represent ranges of $\Sigma L_{K,10}$ used to split the samples for statistical analysis.}
\label{fig:reff2}
\end{figure}  

\subsubsection{Scaling relations involving other parameters}

In Paper\,I we also analysed scaling relations for parameters of the radial profile versus the logarithm of the population of GCs ($\rm N_{GCs}$). Unlike relations based on the ${\rm M_{\star}}$ (see Section\,\ref{sec.rL} and \ref{sec.reff}), ${\rm r_L}$ and ${\rm r_{eff,GCS}}$ of the GCS versus $\rm N_{GCs}$ evolve smoothly. This is expected, as the pivot mass at $5 \times 10^{10}\,{\rm M_{\odot}}$ is common to these three parameters. It is relevant to point out that both scaling relations, i.e. ${\rm r_L}$ versus $\rm N_{GCs}$ and ${\rm r_{eff,GCS}}$ versus $\rm N_{GCs}$, suggest an environmental dependence for satellites. We only present this latter relation, shown in the upper panel of Fig.\,\ref{fig:reff2}. The symbols and colour gradient follow the same coding as previous figures. Unlike Paper\,I, now we fit a power-law to the sample, resulting in

\begin{equation}
{\rm r_{eff,GCS}} = (0.55\pm0.1) \times {\rm N_{GCs}}^{0.47\pm0.02}
\label{eq:reffngc}
\end{equation}

The lower left panel in Fig.\,\ref{fig:reff2} shows the parameter $\Sigma L_{K,10}$ as a function of residuals from the power-law for satellites, following the same symbol coding than the upper panel.
This implies that the processes that satellites experience in dense environments have a larger impact on the spatial size of the GCSs than on their population. The horizontal dotted lines separate satellites presenting $11<\Sigma L_{K,10}<12.3$ and $\Sigma L_{K,10}>12.3$. The skewness estimator for these two groups are $0.4\pm0.4$ and $-1.5\pm0.6$, respectively. These indicate that satellites from dense environments behave as a skewed sample, based on the criterion explained in Section\,\ref{sec.tn}. A MWW test is not conclusive, with only 75\,\,per cent confidence for significant differences between both distributions. The lower right panel is analogue, but for central galaxies. The size of the sample prevent us from further analysis.

The formation and evolution of a GCS and the stellar population of the host galaxy are intrinsically connected, through the experience of processes that modelled their current properties. In this sense, several studies \citep[e.g.][]{kar16} have compared the ${\rm r_{eff,GCS}}$ with the effective radius of the host galaxy (${\rm r_{eff,gal}}$). In Paper\,I we fitted a linear relation between these parameters, in agreement with \citet{for17} and \citet{hud18}. In this work, we update the ${\rm r_{eff,gal}}$, replacing those from \citet{fab89} based on de Vaucouleurs profiles with values fitted from S\'ersic profiles calculated by us or available in the literature.

\begin{figure}    
\includegraphics[width=85mm]{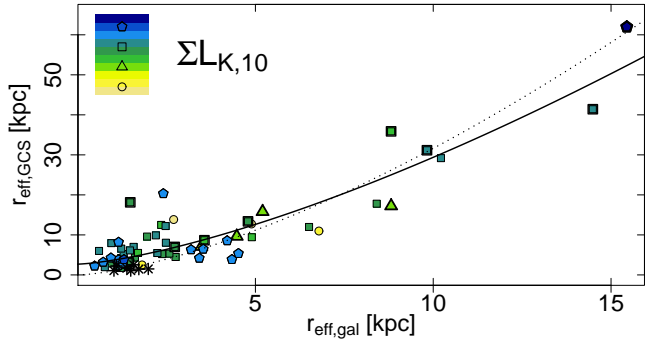}
\centering
\caption{
Effective radius of the GCS (${\rm r_{eff,GCS}}$), as a function of the effective radius of host galaxy (${\rm r_{eff,gal}}$). Different symbols and colour gradient represent increasing values of the environmental parameter $\Sigma L_{K,10}$,  from yellow circles to blue pentagons.  Framed symbols highlight central galaxies and asterisks show the stacked GC profiles from Virgo low-mass galaxies. The dotted line shows a power-law with exponent $1.4$, derived from previous scaling relations (see text for further details). A power-law plus an additive constant was also fitted to the data (Equation\,\ref{eq:reffreff2}), represented by the solid line.}
\label{fig:reffreff}
\end{figure}

Fig.\,\ref{fig:reffreff} presents ${\rm r_{eff,GCS}}$ as a function of ${\rm r_{eff,gal}}$. The symbols and colour gradient follow the same coding as previous figures.
We find no evidence of environmental dependence in this relation. From this enlarged sample, we realise that a linear fit results in systematic residuals for galaxies with small 
${\rm r_{eff,gal}}$. From Equation\,\ref{eq:reffngc}, plus the correlation between the mass enclosed in a GCS (${\rm M_{GCS}}$) and the halo mass (${\rm M_h}$) from \citet{har15}, and the ${\rm r_{eff,gal}}$-to-${\rm M_h}$ relations from the literature \citep[e.g.][]{kra13,rod21}, the ${\rm r_{eff,gal}}$ and the ${\rm r_{eff,GCS}}$ come to be related by a power-law with exponent $\approx 1.4$ (shown with a dotted line in Fig.\,\ref{fig:reffreff}). Thus, we propose a power-law plus a zero point, leading to the following relation,

\begin{equation}
{\rm r_{eff,GCS}} = 2.7\pm0.6 + {\rm r_{eff,gal}}^{1.43\pm0.02}
\label{eq:reffreff2}
\end{equation}

\noindent which is represented in Fig.\,\ref{fig:reffreff} with a solid line. This fit is in agreement with the previously derived slope. Although a more complete sample is desirable, the apparent lack of environmental dependence indicates that both parameters are mostly regulated by the physical processes that the galaxy experience in dense environment.

\vspace{16pt}
In summary, we would like to stress that at fixed stellar mass, the richness of a GCS depends on the environment where its host galaxy resides. Satellites present poor GCSs in dense environments, and for centrals the environmental density and the richness correlate. Analysing the slope of the radial profiles shows that GCSs are steeper at denser environments in the case of satellite galaxies, with central galaxies presenting the opposite behaviour. At fixed stellar mass, the GCSs of satellite galaxies are less extended and present lower effective radii in denser environments. On the contrary, the GCSs of central galaxies in the field are typically less extended than their analogues in denser environments, pointing out the relevance of the environment in the merger history. In the relations of the effective radius and the extension of the GCS versus its GC population, a trend with the environment seems to be present. The effective radius of the GCS as a function of the effective radius of the host galaxies does not evidence any environmental dependence. In the scaling relation between the effective radii of the host galaxy and its GCS, we find that fitting power-law rather than a linear function is a physically motivated choice.

\section{Discussion}
\label{sec.dis}

\subsection{The pivot mass}
In Section\,\ref{sec.tn} we analyse, for our sample, the richness of the CGSs versus the stellar mass of the host galaxy. This relation has a breaking point at ${\rm M_{\star} \approx 5\times 10^{10} M_{\odot}}$, with less massive galaxies than this pivot mass hosting poorer GCSs
with increasing mass. We find a similar breaking point in the scaling relations involving the extension (Section\,\ref{sec.rL}) and effective radius of the GCSs (Section\,\ref{sec.reff}). This behaviour has already been noticed in the literature and interpreted as a consequence of changes in the star formation efficiency of the galaxy stellar population, rather than a relative variation of the GC population \citep[e.g.][]{geo10,har13}. This is supported by studies of the star formation efficiency and the  stellar-to-halo mass ratio through redshift \citep[e.g.][]{lea12,gir20}, but recent results point to changes in the time-scale and in the efficiency of the environmental quenching for satellites as a function of their mass \citep{kaw17,cor19}. On the other hand, \citet{mie14} propose that the U-shaped relation between GCS richness and galaxy luminosity depends on the rate of GC  disruption at early epochs, based on models from \citet{broc14}. However, that work evaluates model galaxies in isolated conditions that do not evolve, hence it does not consider environmental processes that might play a relevant role. Other recent theories like \citet{cho19}, discuss the fractions of accreted stars and GCs as a function of the host galaxy mass. This dependence could have an impact on the behaviour change in the relations.

For galaxies with stellar masses above the pivot mass, typically central galaxies, mergers constitute the main processes ruling their evolution in the last Gyrs \citep[e.g.][]{jim11,xu12}. \citet{rod16} study the fraction of {\it in situ} and {\it ex situ} stars in simulated galaxies from Illustris, and indicates that {\it in situ} stellar formation dominates in galaxies up to ${\rm M_{\star} \approx 10^{11} M_{\odot}}$, with {\it ex situ} contribution increasing for larger masses. Moreover, they state that minor mergers contribute with only $\approx 20$\,\,per cent of the {\it ex situ} component. On the contrary, minor mergers are expected to be a major contribution to the build up of GCSs in massive galaxies, and particularly for their outskirts, leading to extended and rich GCSs \citep{for18b}. In this sense, \citet{kru15} suggest that the pivot mass corresponds to the point where the GC population becomes dominated by the metal poor GCs accreted from stripped dwarf galaxies. 

These results are in agreement with the pivot mass being present in several scaling relations from this work and from Paper\,I, as well as with steep behaviour in the high-mass regime.

\subsection{Richness of the GCSs}
Regarding the environmental dependence proposed in this paper for the relative richness of the GCSs (Section\,\ref{sec.tn} and Fig.\,\ref{mestTN}), it has been largely accepted that the environment plays a main role in the evolutionary history of galaxies and their current morphology \citep[e.g][]{dre80,bro98,tem11}. After the infall, a satellite galaxy experiences different environmental processes that produce a significant mass loss from its halo \citep{gan10,dra20}, but also the removal of the hot gas supply, that leads to the quenching of its star formation \citep{pen10,wet13,dar16}. In the first pericentric passage of a satellite, its halo typically loses $\approx 20-30$\,\,per cent of its mass \citep{rhe17,vdb18}, and this fraction increases for radial or tightly bound orbits, due to stronger tidal forces from the host potential \citep{ogi19}. For ancient satellites, the successive passages at the pericenter produces a mass loss of $\approx 75$\,\,per cent in a time-scale of several Gyr \citep{nie19}. 

Focusing on the effect of tidal stripping in GCSs belonging to satellites in cluster-like environments, \citet{ram18} propose a particle tagging technique on a dark matter simulation to mimic the GCSs of satellite galaxies. They find that satellites have lost $\approx 60$\,percent of their GCs at $z=0$, mainly blue (metal-poor) GCs (more spatially extended). They also indicate that the GC stripping preferentially occurs when the satellite crosses the core of the cluster. 

For massive central galaxies, the usual high GCS richness and its correlation with the environmental density naturally emerges from the two-phases scenario for the build up of GCSs  \citep{for11,for18b}, and the importance of minor mergers to increase the relative richness of a GCS  \citep{kru15,cho19}. Although it has been already stated that minor mergers contribute little to the stellar mass growth of the galaxy, the relative richness of their GCSs  \citep[see][]{pen08,geo10,liu19} reflects their relevance in the build up of GCSs in massive galaxies. Moreover, \citet{kru15} suggest that field galaxies should experience low merging rates, leading to poorer GCSs. This is caused by the combination of low survival rates to disruption at early phases, plus the scarce contribution of accreted GCs. In some cases, ellipticals in these environments even  present less massive dark matter haloes  \citep[e.g.][]{coc09,lan15}, though a variety of results have been obtained \citep[e.g.][]{ric15}.

All of the above support the results from Section\,\ref{sec.tn}, pointing to the existence of environmental dependencies for satellites and centrals in the scaling relation involving the GCS richness and the stellar mass of the host galaxy.

\subsection{Parameters of the GCs radial profile}
In Sections\,\ref{sec.b} to \ref{sec.reff} we explore the environmental dependence for several parameters of the GCS radial profile versus the stellar mass. The results show that satellites in dense environments present more concentrated and less extended GCSs. As a halo population, it is straightforward to consider the evolution of subhaloes in high density environments. It is already stated that low-mass haloes in regions of high tidal forces experience a significant mass loss due to tidal stripping \citep[e.g.][]{ogi19}. These haloes typically present higher NFW concentrations than those in lower density regions, caused by the steepening of their outer density profiles via preferential removal of material from the outer regions \citep{lee17,lee18,dra20}. 

In the very low-mass regime, \citet{sha21} analyse a sample of analogues to Fornax dSph from the E\-MOSAIC simulation, and indicate that satellite galaxies have more concentrated GC distributions than their field analogues. The existence of large populations of intracluster GCs in cluster environments (e.g. Fornax \citep{dab16,pot18}, Virgo \citep{dur14}, Coma \citep{pen11,mad18}, Perseus \citep{har20}, etc) reinforce the relevance of stripping processes in satellite galaxies. Although this intracluster component is typically dominated by blue GCs, $\approx 20-25$\,per cent are metal rich GCs \citep{pen11,lon18}. This implies that the contribution to the intracluster component comes from stripped galaxies with a wide range of stellar masses. \citet{ram20} point to similar results from numerical analysis. For nine massive clusters from the Illustris simulation, they obtain that the main contributors to the intracluster component are galaxies with ${\rm M_{\star} \gtrsim 10^{10} M_{\odot}}$, that survive at $z=0$ as cluster satellites. Then, it is expected in very dense environments to detect intermediate mass galaxies with stripped GCSs (as well as dark matter haloes), less extended than analogues in other environments.

Regarding central galaxies, \citet{amo19} uses N-body simulations for Virgo-like systems and finds that minor mergers with low-mass galaxies play a main role in the build up of extended GCSs in massive galaxies. This is also supported by observational studies with a large FOV that point to the two-phases scenario \citep{for11,par13,lee16,cas17}. This is in agreement with the differences in centrals galaxies as a function of the environmental density, considering that massive galaxies in the field are not likely to experience a large number of mergers.

Both the radial evolution of stripped dark matter haloes and the large contribution of blue GCs to the intracluster component are in agreement with the dependence of the parameters of the GCSs radial profiles on environment for satellites. The differences with the environment for centrals are a consequence of the main role of minor mergers in the build up of populated and extended GCSs. 

To summarise, the results obtained for the galaxy sample analysed in this paper agree with previous findings from Paper\,I, including the presence of a pivot mass in several scaling relations. We remind that the main goal of the present work is to provide evidence for the role of the environment in the build up of the GCSs in ETGs.

\section{Conclusions}
\label{sec.sum}
We have analysed the projected radial distribution of GCs in 23 intermediate luminosity ETGs from Virgo, Fornax and Coma clusters, plus 4 ``stacked'' GC profiles that were built on the basis of 20 Virgo dwarfs separated in four groups with similar stellar mass. We also included the parameters of the GCS radial profiles associated to 6 ETGs taken from the literature. This sample was supplemented with that presented in Paper\,I (i.e. galaxies analysed by us plus those with published parameters), obtaining a total enlarged sample of $100$ GCS  (Appendix Tables\,\ref{morfodens} and \ref{morfodens2}), where the stacked galaxies are counted as single ones.

Based on projected density estimators, we explore the role of the environment in shaping the radial distribution of the GCSs. The results point to differences in the scaling relations for ETGs as a function of the environmental density, but also to distinct behaviours for central and satellite ones. Such behaviours can be explained by the evolution of galaxies in a hierarchical scenario and its impact on stripping and accretion processes, including minor and major mergers. We summarise here our main results:
\begin{itemize}
\item The parameter ${\rm T_N}$ as a function of ${\rm M_{\star}}$ shows a distinctive behaviour for central and satellite galaxies, with a turning point at ${\rm M_{\star}} \approx 5\times 10^{10}\,M_{\odot}$. For larger masses, ${\rm T_N}$ increases towards larger ${\rm M_{\star}}$, but presents an  opposite trend for lower masses, typically associated to satellites. As a secondary effect, we found a correlation with environment for both satellites and centrals.
\item In Paper\,I we showed that the exponent of the modified Hubble profile $b$, inversely correlates with the host galaxy stellar mass. In the present work, we confirmed this result for a larger sample and found a relation with the environment, with satellite galaxies presenting steeper GCS radial profiles in denser environments and the opposite in the case of central galaxies.
\item The segmented relations between the extension and effective radius of a GCS versus the stellar mass of the host galaxy,  already reported in Paper\,I, also show an underlying dependence with the environment.
Satellite galaxies in high density environments present less extended and
more compact GCSs at fixed stellar mass. The opposite is found in the case of central galaxies.
\item We have reproduced the correlation between the effective radius of the GCS and that of its host galaxy. This time, we fitted a power-law instead of a linear relation. It was compared with that expected from equivalent scaling relations, taken from the literature. 
\end{itemize}

\section*{Acknowledgments}
We thank the useful comments of the referee, which
helped to improve this paper. We are grateful to Francisco Azpilicueta and Ricardo Salinas for their constructive comments. This work was funded with grants from Consejo Nacional de Investigaciones Cient\'{\i}ficas y T\'ecnicas de la Rep\'ublica Argentina, Agencia Nacional de Promoci\'on Cient\'{\i}fica y Tecnol\'ogica, and Universidad Nacional de La Plata (Argentina).
This research has made use of the NASA/IPAC Extragalactic Database (NED) which is operated by the Jet Propulsion Laboratory, California Institute of Technology, under contract with the National Aeronautics and Space Administration.
Based on observations made with the NASA/ESA Hubble Space Telescope, obtained from the data archive at the Space Telescope Science Institute. STScI is operated by the Association of Universities for Research in Astronomy, Inc. under NASA contract NAS 5-26555. 

\section*{Data Availability}
All raw data from ACS/HST can be found in the archive.

\bibliography{biblio}
\bibliographystyle{mn2e}

\appendix
\section{Environmental parameters}

\begin{table}
\begin{center}
\caption{Galaxies analysed in this paper and in Paper\,I, listed in decreasing B-band luminosity. Effective radii were taken from the mentioned references and environmental density parameter, $\Sigma
L_{K,10}$, was calculated as explained in section \ref{sec.dens}}   
\label{morfodens}
\tabcolsep=1mm
\begin{tabular}{@{}l@{}c@{}cc@{}c@{}ccc@{}}
\hline
\multicolumn{1}{@{}c}{Name}&\multicolumn{1}{c}{${\rm r_{eff,gal}}$}&\multicolumn{1}{c}{${\rm r_{eff,gal}}$}&\multicolumn{1}{c}{$\Sigma_{10}$}&\multicolumn{1}{c}{$\Sigma
L_{K,10}$}\\
\multicolumn{1}{c}{}&\multicolumn{1}{c}{{\rm arcsec}}&\multicolumn{1}{c}{{\rm kpc}}&\multicolumn{1}{c}{${\rm log(Mpc^{-2})}$}&\multicolumn{1}{c}{${\rm log(\frac{erg}{s\,Mpc^2})}$}\\   
\hline
NGC\,1404 & 22.4$^1$ & 2.4 & 1.2 & 12.5 \\
NGC\,4526 & 23.8$^1$ & 2.0  & 0.7 & 11.8 \\
NGC\,1380 & 37.4$^1$ & 3.2  & 1.2 & 12.5 \\
IC\,4045 & 4.4$^2$ & 2.5 & 1.4 & 12.3 \\
NGC\,4552 & 84.7$^3$ & 6.5  & 0.6 & 11.8 \\
NGC\,4906 & 8.1$^2$ & 4.5 & 1.6 & 12.4 \\
NGC\,3818 & 27.6$^1$ & 4.9  & -1.1 & 9.9 \\
NGC\,1340 & 39.5$^4$ & 3.6  & 0.2 & 11.4 \\
IC\,4041 & 7.8$^2$ & 4.4 & 1.6 & 12.4 \\
NGC\,4621 & 116.2$^3$ & 8.4 & 0.6 & 11.7 \\
NGC\,4473 & 47.7$^5$ & 3.5 & 1.0 & 12.3 \\
NGC\,1387 & 43.7$^1$ & 4.2 & 1.6 & 12.6 \\
NGC\,1439 & 38.4$^1$ & 4.8 & 0.5 & 11.7 \\
NGC\,4459 & 28.8$^3$ & 2.2 & 1.2 & 12.2 \\
NGC\,4442 & 15.6$^3$ & 1.2 & 1.1 & 12.4 \\
NGC\,1426 & 24.4$^1$ & 2.7 & 0.5 & 11.5 \\
NGC\,7173 & 9.8$^1$ & 1.5 & 0.7 & 11.8 \\
NGC\,4435 & 15.1$^1$ & 1.2 & 1.0 & 12.2 \\
NGC\,4371 & 26.7$^3$ & 2.2 & 1.1 & 12.3 \\
IC\,2006 & 17.2$^1$ & 1.7 & 0.1 & 11.3 \\
NGC\,4570 & 11.5$^5$ & 1.0  & 0.9 & 12.2 \\
NGC\,4267 & 7.8$^3$ & 0.6 & 1.0 & 12.2 \\
NGC\,4033 & 12.2$^1$ & 1.3  & 0.3 & 11.5 \\
NGC\,4417 & 12.1$^3$ & 0.9 & 1.2 & 12.5 \\
NGC\,1351 & 25.5$^4$ & 2.4 & 1.0 & 11.9 \\
NGC\,4564 & 15.8$^3$ & 1.2 & 1.1 & 12.3 \\
NGC\,1339 & 16.9$^4$ & 1.6 & 0.4 & 11.6 \\
NGC\,1172 & 33.9$^1$ & 3.5 & -0.4 & 10.5 \\
NGC\,3377 & 53.5$^1$ & 2.8 & 0.5 & 11.7 \\
IC\,4042 & 4.0$^2$ & 1.9 & 1.6 & 13.1 \\
IC\,4042A & 6.9$^2$ & 3.4 & 1.3 & 12.5 \\
NGC\,4434 & 10.6$^3$ & 1.15 & 1.0 & 12.4 \\
NGC\,4660 & 9.9$^3$ & 0.7 & 1.1 & 12.4 \\
NGC\,4474 & 16.8$^3$ & 1.3 & 1.2 & 12.4 \\
NGC\,4377 & 5.5$^3$ & 0.5 & 1.2 & 12.3 \\
NGC\,1419 & 10.9$^4$ & 1.2 & 0.5 & 11.7 \\
NGC\,1336 & 28.2$^1$ & 2.6 & 0.7 & 11.8 \\
NGC\,4387 & 10.7$^3$ & 0.9 & 0.7 & 11.8 \\
NGC\,1380A & 15.2$^1$ & 1.5 & 1.1 & 12.2 \\
NGC\,4458 & 17.3$^3$ & 1.4 & 0.9 & 12.2 \\
NGC\,4483 & 14.0$^3$ & 1.1 & 1.2 & 12.2 \\
NGC\,4623 & 14.5$^3$ & 1.2 & 0.7 & 11.8 \\
NGC\,4352 & 15.6$^3$ & 1.4 & 0.8 & 12.2 \\
NGC\,4515 & 9.5$^3$ & 0.8 & 0.8 & 12.0 \\
NGC\,1380B & -- & --  & 1.2 & 12.6 \\
NGC\,1428 & -- & --  & 1.0 & 12.4 \\
FCC\,255 & 13.4$^1$ & 1.3 & 0.9 & 12.2 \\
VS\,1 & 13.5$^{\rm a}$  & 1.1 & -- & -- \\
VS\,2 & 24.7$^{\rm a}$  & 2.0 & -- & -- \\
VS\,3 & 21.5$^{\rm a}$  & 1.7 & -- & -- \\
VS\,4 & 18.4$^{\rm a}$  & 1.5 & -- & -- \\
VS\,5 & 12.6$^{\rm a}$  & 1.0 & -- & -- \\
VS\,6 & 19.4$^{\rm a}$  & 1.6 & -- & -- \\
VS\,7 & 13.5$^{\rm a}$  & 1.1 & -- & -- \\
\\
\hline
\end{tabular}
\end{center}
$^{\rm a}$ Calculated as the mean of the effective radii of individual galaxies.\\
{\bf References:} $^1$This paper or Paper\,I, $^2$\citet{hoy11}, $^3$\citet{fer06b}, 
$^4$\citet[][please note that in this paper de Vaucouleurs profile was  used, instead of S\'ersic one]{fab89}, $^5$\citet{korm09}
\end{table}   

\begin{table}
\begin{center}
\caption{Galaxies from the literature compiled in this paper and in Paper\,I, listed in decreasing B-band luminosity. Effective radii were taken from the mentioned references and environmental density parameter, $\Sigma
L_{K,10}$, was calculated as explained in section \ref{sec.dens}}   
\label{morfodens2}
\tabcolsep=1mm
\begin{tabular}{@{}l@{}c@{}cc@{}c@{}ccc@{}}
\hline
\multicolumn{1}{@{}c}{Name}&\multicolumn{1}{c}{${\rm r_{eff,gal}}$}&\multicolumn{1}{c}{${\rm r_{eff,gal}}$}&\multicolumn{1}{c}{$\Sigma_{10}$}&\multicolumn{1}{c}{$\Sigma
L_{K,10}$}\\
\multicolumn{1}{c}{}&\multicolumn{1}{c}{{\rm arcsec}}&\multicolumn{1}{c}{{\rm kpc}}&\multicolumn{1}{c}{${\rm log(Mpc^{-2})}$}&\multicolumn{1}{c}{${\rm log(\frac{erg}{s\,Mpc^2})}$}\\ 
\hline
NGC\,4874 & 32.0$^1$ & 15.4 & 4.2 & 15.2 \\
NGC\,1316 & 69.5$^2$ & 7.0 & 0.2 & 11.5 \\
NGC\,6876 & 99.0$^3$ & 24.4 & -0.7 & 10.6 \\
NGC\,1407 & 71.9$^4$ & 9.8 & 0.9 & 12.0 \\
NGC\,4486 & 81.3$^4$ & 6.6 & 1.2 & 12.7 \\
NGC\,1395 & 45.4$^4$ & 5.2 & 0.8 & 11.9 \\
NGC\,4594 & -- & -- & 0.6 & 11.6 \\
NGC\,4649 & 128.2$^5$ & 10.2 & 1.2 & 12.3 \\
NGC\,4406 & 202.7$^5$ & 16.1 & -- & -- \\
NGC\,4374 & 52.5$^4$ & 4.7 & 1.2 & 12.2 \\
NGC\,3962 & 34.4$^4$ & 6.1 & -0.7 & 10.2 \\
NGC\,5813 & 57.5$^4$ & 8.8 & 0.2 & 11.5 \\
NGC\,720 & 39.5$^4$ & 5.2 & -0.6 & 10.8 \\
NGC\,3610 & 26.5$^6$ & 4.5 & -0.3 & 11.0 \\
NGC\,3311 & -- & -- & 1.5 & 12.8 \\
NGC\,2768 & 63.1$^4$ & 6.8 & -0.9 & 10.2 \\
NGC\,4636 & 89.1$^4$ & 6.4 & 0.6 & 11.6 \\
NGC\,4365 & 128.1$^5$ & 14.5 & 1.5 & 12.1 \\
NGC\,3923 & 53.3$^4$ & 5.5 & 0.2 & 11.5 \\
NGC\,6411 & 26.7$^4$ & 6.7 & -1.2 & 9.5 \\
NGC\,4762 & 43.7$^4$ & 4.9 & 0.4 & 11.5 \\
NGC\,1399 & 42.4$^4$ & 3.6 & 1.7 & 13.1 \\
NGC\,7507 & 31.4$^4$ & 3.7 & -0.7 & 10.5 \\
NGC\,3613 & 60.5$^7$ & 8.8 & -0.2 & 11.0 \\
NGC\,4494 & 49.0$^4$ & 4.0 & 0.1 & 11.3 \\
NGC\,2865 & 11.7$^4$ & 2.2 & -0.1 & 11.0 \\
NGC\,3268 & 154.9$^8$ & 27.6 & 1.5 & 12.5 \\
NGC\,3258 & 55.6$^8$ & 9.4 & 1.5 & 12.7 \\
NGC\,5866 & 36.3$^4$ & 2.7 & -1.4 & 9.7 \\
NGC\,6861 & 22.8$^4$ & 3.2 & 0.3 & 11.3 \\
NGC\,821 & 39.8$^4$ & 4.5 & -0.9 & 10.3 \\
NGC\,3115 & 36.1$^4$ & 1.7 & -0.6 & 10.5 \\
NGC\,1052 & 36.9$^4$ & 3.4 & -0.3 & 11.0 \\
NGC\,3379 & 39.8$^4$ & 2.2 & 0.8 & 12.0 \\
NGC\,5128 & -- & -- & -0.2 & 11.0 \\
NGC\,4278 & 31.6$^4$ & 2.4 & 0.4 & 11.5 \\
NGC\,1379 & 42.4$^4$ & 4.2 & 1.7 & 12.7 \\
NGC\,1427 & 32.9$^4$ & 3.1 & 1.1 & 12.3 \\
NGC\,7332 & 17.4$^4$ & 1.8 & -1.3 & 10.0 \\
NGC\,4754 & 31.6$^4$ & 2.5 & 0.4 & 12.0 \\
NGC\,1374 & 30.0$^4$ & 2.8 & 1.6 & 12.7 \\
NGC\,4546 & 21.1$^9$ & 1.4 & 0.4 & 11.6 \\
NGC\,2271 & -- & -- & -0.2 & 11.0 \\
NGC\,1400 & 37.8$^4$ & 3.0 & -1.1 & 10.0 \\
NGC\,3384 & 32.3$^4$ & 1.6 & 0.6 & 12.0 \\
NGC\,7457 & 36.3$^4$ & 2.1 & -1.2 & 9.8 \\
\\
\hline
\end{tabular}
\end{center}
{\bf References:} $^1$\citet{vea17}, $^2$\citet{ric12a}, $^3$\citet{enn19},
$^4$\citet[][we are aware that in this paper de Vaucouleurs profiles were used, instead of S\'ersic ones]{fab89}, $^5$\citet{korm09}, $^6$\citet{bas17}, $^{7}$\citet{debo20}, $^{8}$\citet{cal18}, $^{9}$\citet{esc20}
\end{table}  

\label{lastpage}
\end{document}